%% file: 000_main.tex
\documentclass[acmsmall,screen,nonacm]{acmart}

\renewcommand\footnotetextcopyrightpermission[1]{}

\setcopyright{acmlicensed}
\copyrightyear{2018}
\acmYear{2018}
\acmDOI{XXXXXXX.XXXXXXX}

\acmJournal{TACO}
\acmVolume{37}
\acmNumber{4}
\acmArticle{111}
\acmMonth{8}

\usepackage[acronym,shortcuts,nohypertypes={acronym,notation}]{glossaries}
\usepackage[inline]{enumitem}
\usepackage{amsmath}
\usepackage{graphicx}
\usepackage{subcaption}
\usepackage{color}
\usepackage{fontawesome}
\usepackage{makecell}
\usepackage{multirow}
\usepackage{wrapfig}
\usepackage[shortcuts]{extdash}
\usepackage{todonotes}

\setlist[itemize]{nosep,leftmargin=1em,labelwidth=*,align=left}

\usepackage{changepage}
\usepackage{hyphenat}
\usepackage{tikz}
\usetikzlibrary{fit,decorations.pathreplacing,calc,tikzmark}
\usepackage{cleveref}
\usepackage{tabularx}

\definecolor{applegreen}{rgb}{0.55, 0.71, 0.0}
\definecolor{carnationpink}{rgb}{1.0, 0.65, 0.79}
\definecolor{lightgray}{rgb}{0.83, 0.83, 0.83}

\newcommand{\successmark}{\faCheck{}}

\usepackage{listings}
\usepackage{xcolor}

\makeatletter
\let\old@lstKV@SwitchCases\lstKV@SwitchCases
\def\lstKV@SwitchCases#1#2#3{}
\makeatother
\usepackage{lstlinebgrd}
\makeatletter
\let\lstKV@SwitchCases\old@lstKV@SwitchCases

\lst@Key{numbers}{none}{%
    \def\lst@PlaceNumber{\lst@linebgrd}%
    \lstKV@SwitchCases{#1}%
    {none:\\%
        left:\def\lst@PlaceNumber{\llap{\normalfont
            \lst@numberstyle{\thelstnumber}\kern\lst@numbersep}\lst@linebgrd}\\%
        right:\def\lst@PlaceNumber{\rlap{\normalfont
            \kern\linewidth \kern\lst@numbersep
            \lst@numberstyle{\thelstnumber}}\lst@linebgrd}%
    }{\PackageError{Listings}{Numbers #1 unknown}\@ehc}}
\makeatother

\definecolor{applegreen}{rgb}{0.55, 0.71, 0.0}
\definecolor{carnationpink}{rgb}{1.0, 0.65, 0.79}
\definecolor{lightgray}{rgb}{0.83, 0.83, 0.83}
\definecolor{codegray}{rgb}{0.5,0.5,0.5}
\definecolor{backcolour}{rgb}{0.95,0.95,0.92}

\lstdefinestyle{cstyle}{
    language=C,
    backgroundcolor=\color{backcolour},
    commentstyle=\color{codegray},
    keywordstyle=\color{blue},
    numberstyle=\tiny\color{codegray},
    stringstyle=\color{red},
    basicstyle=\scriptsize\ttfamily,
    breakatwhitespace=false,
    breaklines=true,
    captionpos=b,
    keepspaces=true,
    numbers=left,
    numbersep=3pt,
    showspaces=false,
    showstringspaces=false,
    showtabs=false,
    tabsize=2,
    xleftmargin=1mm,
    escapeinside={(*@}{@*)},  
    columns=flexible,
    mathescape=false,
    texcl=false,
}

\lstdefinestyle{pythonstyle}{
    language=Python,
    backgroundcolor=\color{backcolour},
    commentstyle=\color{codegray},
    keywordstyle=\color{blue},
    numberstyle=\tiny\color{codegray},
    stringstyle=\color{red},
    basicstyle=\footnotesize\ttfamily,
    breakatwhitespace=false,
    breaklines=true,
    captionpos=b,
    keepspaces=true,
    numbers=left,
    numbersep=3pt,
    showspaces=false,
    showstringspaces=false,
    showtabs=false,
    tabsize=2,
    xleftmargin=1mm,
    escapeinside={},
    columns=flexible,
}

\lstdefinestyle{asmstyle}{
    language={[x86masm]Assembler},
    backgroundcolor=\color{backcolour},
    commentstyle=\color{codegray}\itshape,
    keywordstyle=\color{blue},
    numberstyle=\tiny\color{codegray},
    stringstyle=\color{red},
    basicstyle=\scriptsize\ttfamily,
    breakatwhitespace=false,
    breaklines=true,
    captionpos=b,
    keepspaces=true,
    numbers=left,
    numbersep=3pt,
    showspaces=false,
    showstringspaces=false,
    showtabs=false,
    tabsize=2,
    xleftmargin=1mm,
    escapeinside={||},
    columns=flexible,
    morecomment=[l]{//},
    morecomment=[l]{\#},
    identifierstyle=\color{black},
    morekeywords={push, pop, mov, load, store, add, sub, mul, div, call, ret, jmp, je, jne, jg, jl, jge, jle, cmp, lea, inc, dec, if, end, loop},
    alsoletter={:},
    emph={[1]Rcs, Rtmp1, Rtmp2, xaddr, hot_func},
    emphstyle={[1]\color{violet}},
}

\graphicspath{{./figures/}}

\begin{document}

\title{A Magnified View into Heterogeneous-ISA Thread Migration Performance without State Transformation}

\author{Nikolaos Mavrogeorgis}
\orcid{0009-0003-2109-1407}
\affiliation{%
  \institution{University of Edinburgh}
  \country{United Kingdom}
}
\email{nikos.mavrogeorgis@ed.ac.uk}

\author{Christos Vasiladiotis}
\orcid{0000-0001-7936-2183}
\affiliation{%
  \institution{University of Edinburgh}
  \country{United Kingdom}
}
\email{c.vasiladiotis@ed.ac.uk}

\author{Pei Mu}
\orcid{0000-0003-2781-8032}
\affiliation{%
  \institution{University of Edinburgh}
  \country{United Kingdom}
}
\email{pei.mu@ed.ac.uk}

\author{Amir Khordadi}
\orcid{0009-0002-2003-8455}
\affiliation{%
  \institution{University of Edinburgh}
  \country{United Kingdom}
}
\email{amir.khordadi@ed.ac.uk}

\author{Björn Franke}
\orcid{0000-0002-1219-8523}
\affiliation{%
  \institution{University of Edinburgh}
  \country{United Kingdom}
}
\email{b.franke@ed.ac.uk}

\author{Antonio Barbalace}
\orcid{0000-0003-1641-0779}
\affiliation{%
  \institution{University of Edinburgh}
  \country{United Kingdom}
}
\email{antonio.barbalace@ed.ac.uk}

\renewcommand{\shortauthors}{Mavrogeorgis et al.}

\input{001_definitions.tex}

\input{002_abstract.tex}


\maketitle

\input{010_introduction.tex}

\input{020_background_motivation.tex}

\input{030_example.tex}

\input{040_design.tex}

\input{050_implementation.tex}

\input{070_evaluation.tex}

\input{080_relatedwork.tex}

\input{090_conclusion.tex}

\bibliographystyle{ACM-Reference-Format}
\bibliography{099_references}

\end{document}

%% file: 001_definitions.tex

\newcommand{\tcircle}[1]{\textcircled{\raisebox{-0.9pt}{#1}}}
\newcommand{\asm}[1]{\texttt{#1}}
\newcommand{\code}[1]{\texttt{#1}}



\newcommand{\tool}{\textsc{Unifico}}
\newcommand{\popcorn}{\textsc{Popcorn}}


\newcommand{\pcarch}{\textsc{x86}}
\newcommand{\armarch}{\textsc{ARM}}
\newcommand{\pcback}{\textsc{x86}}
\newcommand{\armback}{\textsc{ARM}}
\newcommand{\riscv}{\textsc{RISC-V}}

\newcommand{\cpp}{\textsc{C++}}
\newcommand{\cproglang}{\textsc{C}}
\newcommand{\python}{\textsc{Python}}
\newcommand{\linux}{\textsc{Linux}}
\newcommand{\llvm}{\textsc{LLVM}}
\newcommand{\gcc}{\textsc{GCC}}
\newcommand{\clang}{\textsc{clang}}
\newcommand{\tablegen}{\textsc{TableGen}}
\newcommand{\gdb}{\textsc{gdb}}
\newcommand{\biglittle}{\textsc{big.LITTLE}}
\newcommand{\smartnic}{\textsc{SmartNIC}}
\newcommand{\smartssd}{\textsc{SmartSSD}}
\newcommand{\genz}{\textsc{Gen\-/Z}}
\newcommand{\opencapi}{\textsc{OpenCAPI}}
\newcommand{\ccix}{\textsc{CCIX}}
\newcommand{\cxl}{\textsc{CXL}}
\newcommand{\criu}{\textsc{CRIU}}
\newcommand{\hetcriu}{\textsc{HetCRIU}}
\newcommand{\dapper}{\textsc{Dapper}}
\newcommand{\pcie}{\textsc{PCI}e}
\newcommand{\musl}{\textsc{musl libc}}
\newcommand{\perf}{\textsc{perf}}
\newcommand{\spec}{\textsc{SPEC CPU2017}}
\newcommand{\upmem}{\textsc{UPMEM}}
\newcommand{\dynamorio}{\textsc{DynamoRIO}}



\newacronym{npb}{NPB}{NAS Parallel Benchmarks}


\newacronym{loc}{LoC}{lines of code}
\newacronym{api}{API}{application programming interface}
\newacronym{abi}{ABI}{application binary interface}
\newacronym{cpu}{CPU}{central processing unit}
\newacronym{gpu}{GPU}{graphics processing unit}
\newacronym{tpu}{TPU}{tensor processing unit}
\newacronym{fpga}{FPGA}{field\-/programmable gate array}
\newacronym{isa}{ISA}{instruction set architecture}
\newacronym{smp}{SMP}{shared\-/memory programming}
\newacronym{nic}{NIC}{network interface card}
\newacronym{fp}{FP}{floating\-/point}
\newacronym{ir}{IR}{intermediate representation}
\newacronym{os}{OS}{operating system}
\newacronym{hpc}{HPC}{high\-/performance computing}
\newacronym{dma}{DMA}{direct memory access}
\newacronym{sota}{SotA}{state\-/of\-/the\-/art}
\newacronym{pim}{PIM}{processing\-/in\-/memory}
\newacronym{dse}{DSE}{design space exploration}
\newacronym{dbt}{DBT}{dynamic binary translation}
\newacronym{hsa}{HSA}{heterogeneous software architecture}
\newacronym{ssd}{SSD}{solid\-/state drive}
\newacronym{byoc}{BYOC}{``bring your own core''}
\newacronym{oss}{OSS}{open source software}
\newacronym{ram}{RAM}{random access memory}
\newacronym{aslr}{ASLR}{address space layout randomization}
\newacronym{ipc}{IPC}{instructions per cycle}
\newacronym{ndp}{NDP}{near\-/data processing}
\newacronym{risc}{RISC}{reduced instruction set computer}
\newacronym{cisc}{CISC}{complex instruction set computer}
\newacronym{bfs}{BFS}{breadth\-/first search}
\newacronym{rop}{ROP}{return\-/oriented programming}
\newacronym{elf}{ELF}{executable and linkable format}
\newacronym{melf}{MELF}{multi\-/variant ELF}
\newacronym{mite}{MITE}{macro instruction translation engine}
\newacronym{dsb}{DSB}{decoded stream buffer}
\newacronym{tma}{TMA}{top\-/down microarchitecture analysis}
\newacronym{pmu}{PMU}{performance monitoring unit}
\newacronym{mpki}{MPKI}{misses per kilo instructions}
\newacronym{ilp}{ILP}{instruction\-/level parallelism}
\newacronym{dsm}{DSM}{distributed shared memory}

%% file: 002_abstract.tex

\begin{abstract}
Heterogeneous-ISA processor designs have attracted considerable research interest.
However, unlike their homogeneous-ISA counterparts, explicit software support for bridging ISA heterogeneity is required.
The lack of a compilation toolchain ready to support heterogeneous-ISA targets has been a major factor hindering research in this exciting emerging area.
For any such compiler, ``getting right'' the mechanics involved in state transformation upon migration and doing this efficiently is of critical importance.
In particular, any runtime conversion of the current program stack from one architecture to another would be prohibitively expensive.
In this paper, we design and develop \textsc{Unifico}, a new multi-ISA compiler that generates binaries that maintain the \textit{same} stack layout during their execution on either architecture.
\textsc{Unifico} avoids the need for runtime stack transformation, thus eliminating overheads associated with ISA migration.
Additional responsibilities of the \textsc{Unifico} compiler backend include maintenance of a uniform ABI and virtual address space across ISAs.
\textsc{Unifico} is implemented using the LLVM compiler infrastructure, and we are currently targeting the \textsc{x86-64} and \textsc{ARMv8} ISAs.
We have evaluated \textsc{Unifico} across a range of compute-intensive NAS benchmarks and show its minimal impact on overall execution time, where less than $6\%$ ($10\%$) overhead is introduced on average for high-end (low-end) processors.
We also analyze the performance impact of \textsc{Unifico}'s key design features and demonstrate that they can be further optimized to mitigate this impact.
When compared against the state\hyp{}of\hyp{}the\hyp{}art \textsc{Popcorn} compiler, \textsc{Unifico} reduces binary size overhead from ${\sim}200\%$ to ${\sim}10\%$, whilst eliminating the stack transformation overhead during ISA migration.
\end{abstract}

\begingroup
\renewcommand\thefootnote{}\footnotetext{
    Extension of conference paper: Earlier version presented at \textsc{CC 2024}~\cite{mavrogeorgis2024unifico}.
    We extend prior work in the following directions:
    1. We present an analysis of \tool{}'s impact to native performance on \armarch{} and \pcarch, identifying the most critical features introduced for an identical stack layout (\Cref{subsec:perf-breakdown}).
    2. For the most impactful features (callsite alignment, register pressure, register allocation, and machine scheduling),
       we provide insights on why some applications are affected more than others, and present performance improvements (\Cref{subsec:callsite-alignment,subsec:reg-pressure,subsec:misched}).
    3. We demonstrate the performance portability of \tool{} across additional machines with different microarchitectures (\Cref{subsec:portability}).
    4. We extend a transformation framework based on \hetcriu{}~\cite{bapat2024dapper, xing2022} to support heterogeneous checkpoint-restore with and without transformation, and compare the end-to-end performance with \popcorn{}\hyp{}compiled binaries.
       We show that the transformation cost savings depend on the size of the transformed state, execution time of the application, and number of migrations (\Cref{subsec:migration-comparison}).

}
\addtocounter{footnote}{-1}
\endgroup

%% file: 010_introduction.tex
\section{Introduction}\label{sec:intro}

Heterogeneity in computing hardware (CPUs, GPUs, TPUs, and FPGAs) is common practice today in a multitude of deployments and configurations~\cite{kumar2005, zahran2016,jouppi2017}.
This has been driven by the ever\hyp{}increasing computational demands of workloads, whose data\hyp{}processing requirements boomed recently~\cite{liu2021a}.
Yet, processing (large amounts of) data across heterogeneous processing units poses several problems, like hindering programmability.

Classical software compilation, targeting a single \ac{isa} and the related programming models, e.g., shared memory programming, is not applicable as\hyp{}is to heterogeneous\hyp{}\ac{isa} platforms.
Instead, the current practice is to isolate (or mark) a set of functions to be run on a processing unit different from the main \ac{cpu}, compile them for the specific \ac{isa}, and offload them at runtime.
Bespoke programming frameworks exist to support the application programmer (e.g., OpenCL, CUDA), and include development and runtime environments~\cite{kim2017a}.

To improve the programmability of such platforms, different solutions have been introduced, like the reuse of data pointers across \acp{cpu} and heterogeneous processing units.
These required hardware memory management per processing unit, and motivated the introduction of coherent shared memory between the host \ac{cpu} and heterogeneous processing units, either on the same chip~\cite{amd-apu101-webpage}, or via the peripheral bus~\cite{opencapi-webpage,genz-webpage,ccix-webpage,cxl-webpage}.

\paragraph{Emerging heterogeneous\hyp{}\ac{isa} platforms.}
At the same time, the landscape of heterogeneous hardware computing is widening.
While classical heterogeneous\hyp{}\ac{isa} platforms, comprising a single general\hyp{}purpose \ac{cpu} and multiple special\hyp{}purpose processing units (GPUs, TPUs, and FPGAs), are widely available,
platforms with multiple general\hyp{}purpose diverse\hyp{}\ac{isa} processing units are emerging.
Similarly to classical heterogeneous hardware, emerging heterogeneous\hyp{}\ac{isa} platforms are also going to offer shared memory amongst \acp{isa}.

While academia proposed single\hyp{}chip, cache\hyp{}coherent heterogeneous\hyp{}\ac{isa} \acp{cpu}, e.g., BYOC~\cite{balkind2020}, which never reached the market, new peripheral interconnects like \cxl{}~\cite{cxl-webpage} promise to enable coherent shared memory between the main \ac{cpu} and the processing units of peripheral devices, such as SmartNICs~\cite{broadcom-stingray-smartnic-webpage,riscvsmartnic} or SmartSSDs~\cite{8671589,barbalace2021computational} (\armarch{}- or \riscv{}- based).
Also, \cxl{} will accommodate memory expansion cards that will likely integrate general\hyp{} and special\hyp{}purpose processing units for \ac{ndp}~\cite{skhyinx-cms}, as in \upmem{}~\cite{devaux2019true} \ac{pim}, where such processing units directly access the same memory as the main \ac{cpu}.

\paragraph{Programming emerging platforms.}
Classic heterogeneous computing runs an application on the \ac{cpu} and offloads a specific part of it to special\hyp{}purpose processing units.
However, when multiple \acp{cpu} of diverse \acp{isa} lie on the same platform, thread migration has been shown to be more beneficial than offloading, enabling decisions at runtime rather than statically deciding on the function to offload at compile time~\cite{devuyst2012,barbalace2015,barbalace2017,8063925}.

While earlier works on heterogeneous\hyp{}\ac{isa} migration~\cite{smith1998a,jul1988} require the transformation of the entire application state before execution on another \ac{isa}, recent approaches transform only part of the state (e.g., the registers and stack), guided by metadata derived during compilation.
Despite that, the transformation step and related metadata still incur execution time and binary size overheads, respectively, which are mostly linear to the number and size of active stack frames when migrating~\cite{devuyst2012,barbalace2017}.
Both overheads impact migration time, potentially hindering its benefits.
Lastly, although outside the scope of this work, state transformation approaches constitute a potential attack surface for the binaries~\cite{rave-rerandom}, e.g., by exposing the return addresses of functions that can be leveraged to create \ac{rop} gadgets.

\paragraph{\tool{}.}
Motivated by the emerging heterogeneous\hyp{}\ac{isa} platforms with shared memory, with the goal of making programmability as simple as homogeneous\hyp{}\ac{isa} platforms,
and removing execution time and code size overheads of state transformation, we propose \tool{}.
\tool{} is a compilation technique that generates multi\hyp{}\ac{isa} binaries with a unified address space layout and application state (including stack, heap, thread\hyp{}local storage, etc.) across different \acp{isa}, enabling thread migration without transformations. 
We achieve this by rethinking how compiler backends generate code, extending the code generation passes that impact the application state to adhere to a common set of rules, without the need to graft any metadata in the binary.

We prototyped \tool{} targeting the 64\hyp{}bit versions of the \armarch{} and \pcarch{} \acp{isa}, and studied its efficacy on different benchmarks.
Comparing against binaries generated using previous heterogeneous\hyp{}\ac{isa} \ac{cpu} migration projects, we demonstrate that \tool{} adds on average no more than $6\%$ execution time overheads and no more than $10\%$ code size increases.
We envision \tool{} integrated in existing compiler frameworks, and used in modern heterogeneous hardware platforms.

We make the following \emph{contributions}:
\begin{itemize}
  \item A compilation technique, \tool{}, that enables thread migration amongst heterogeneous\hyp{}\ac{isa} \acp{cpu} \emph{without state transformation}, removing its initialization, runtime, and code overheads.
  \item We prototype \tool{} targeting the \pcback{} and \armback{} \llvm{} backends, and validate migration on the \acf{npb} suite, utilizing the \criu{} checkpoint and restore software.
    \tool{} is released as \ac{oss} at 
    \url{https://github.com/systems-nuts/unifico}.
  \item An evaluation across different benchmarks and microarchitectures demonstrating, on average, no more than $10\%$ binary size overhead, no more than $6\%$ overhead on execution time (without migration) for high-end processors, and no more than $10\%$ overhead for low-power ones.
  \item A comprehensive performance analysis of \tool{}\hyp{}compiled binaries identifying key features for a common stack layout and proposing improvements to alleviate performance bottlenecks.
\end{itemize}


%% file: 020_background_motivation.tex
\section{Background and Motivation}\label{sec:background-motivation}

We provide background on software migration along with its main issues motivating our approach.

\subsection{Heterogeneous-ISA Architectures}\label{subsec:heterogeneous-isa-architectures}

This work is motivated by the emerging compute heterogeneity, i.e., heterogeneous \acp{isa}~\cite{devuyst2012, venkat2014, barbalace2015, lee2017, barbalace2017, balkind2020, cho2020flick}, coupled with next\-/generation memory architectures materialized by new interconnect technologies~\cite{ccix-webpage, tamimi2022, genz-webpage, opencapi-webpage, kyriazis2012, drucker2020}, e.g., \cxl{}~\cite{cxl-webpage, ahn2022}.
In such configurations, thread migration has been shown to be more advantageous than the typical prevalent offloading techniques~\cite{devuyst2012, barbalace2015}.

In particular, we focus on a combination of \pcarch{} and \armarch{} \acp{cpu} inspired by a family of emerging platforms closely related to \ac{ndp}~\cite{devaux2019true, 10.1145/3460201, cxl-ndp, skhyinx-cms, barbalace2017s}.
These usually accommodate a brawny host processor (e.g., \pcarch{}) plus one or more simpler \ac{risc} processors (usually of different \acp{isa}, e.g., \armarch{}) near the memory, to avoid data movement and increase bandwidth utilization.
Memory\-/intensive applications that exhibit weak locality are well suited for these architectures~\cite{gomez-luna2022},
with offloading being their dominant programming paradigm.

\subsection{Thread Migration Techniques}\label{subsec:thread-migration-techniques}

Dynamic software thread migration is the act of moving a thread's execution context (e.g., register state, stack contents, page mappings, etc.) between different processing units in a system~\cite{attardi1988,dubach1989,smith1998a,jul1988}.
In \ac{smp} systems, this can be achieved through hardware and \ac{os} mechanisms.
However, in the case of heterogeneous\-/\ac{isa} systems, additional compiler and runtime support is needed, since the thread state needs to be transformed in order to match the architecture-specific details of the target processor~\cite{devuyst2012, venkat2014, barbalace2015, barbalace2017, barbalace2020, cho2020flick,xing2022}.

There are three main axes to consider when performing thread migration:
First, \textit{migration granularity}, which denotes where the program is able to migrate (e.g., function boundaries~\cite{vonbank1994, barbalace2017, cho2020flick}).
This property directly affects programmability and performance (\Cref{subsec:motivation-1:-programmability,subsec:motivation-2:-performance-potential}).
Second, \textit{state transformation cost}, which is incurred at runtime before the migration, affects the overall performance of the migration.
Finally, \textit{compiler support}, which describes the changes made to the compiler in order not only to enable migration (i.e., generate code for both \acp{isa}), but also control the granularity of migration and transformation costs.
Unfortunately, trying to improve on two of the axes, means that compromise or more effort should be placed on the third one.
For example, to maximize granularity without modifying extensively the compiler, binary translation can help migrate threads across architectures before reaching a valid migration point, though this approach introduces additional execution overhead~\cite{vonbank1994, devuyst2012}.

We believe that in order to maximize programmability and performance in heterogeneous-\ac{isa} systems, more weight should be put in the compiler support, as described in the following sections.

\subsection{The Burden of Programmability}\label{subsec:motivation-1:-programmability}

Utilizing offloading commonly requires modifications at the source code level of an application in order to use a specific supported \ac{api}.
This mainly involves:
\begin{enumerate*}[label=\roman*)]
  \item setup/teardown of communication with the accelerator, and
  \item code segmentation and data movement for offloading computation.
\end{enumerate*}
This approach is inherently at odds with ease of programming, portability and programmer productivity~\cite{10.1145/3378678.3391881}.

\begin{wrapfigure}{r}{0.47\linewidth}
  \centering
\begin{lstlisting}[style=pythonstyle, xleftmargin=10pt]
queue.push(root)
while len(queue) > 0:
  for src in queue:
    for dst in out_edges(src):
    # Can migrate here with UNIFICO
      if parent[dst] == -1:
        parent[dst] = src
        queue.push(dst)
\end{lstlisting}
  \caption{Pseudo\-/code of a queue\-/based BFS application.
  The for\-/loops cannot be used directly as offload kernels without major modifications.}\label{fig:bfs}
\end{wrapfigure}

\Cref{fig:bfs} shows the main part of a small fragment of the \ac{bfs} algorithm, as taken from~\cite{hein2018near}.
Porting the algorithm to an offloading-based programming model, e.g., the recently introduced \upmem{} \ac{pim} architecture~\cite{devaux2019true}, requires substantial modifications to the code and replicating the data structures among all processing units~\cite{gomez-luna2022, prim-bfs}.
However, with thread migration, there is no need of replicating data structures because threads can migrate where the data is, and \tool{} also enables more fine-grained migration, beyond just function boundaries.

\subsection{The Overhead of State Transformation}\label{subsec:motivation-2:-performance-potential}

\begin{figure}[t]
  \centering
  \captionsetup[sub]{aboveskip=1pt,belowskip=1pt}
  \begin{subfigure}[t]{.5\linewidth}
      \centering
    \includegraphics[width=\textwidth,clip]{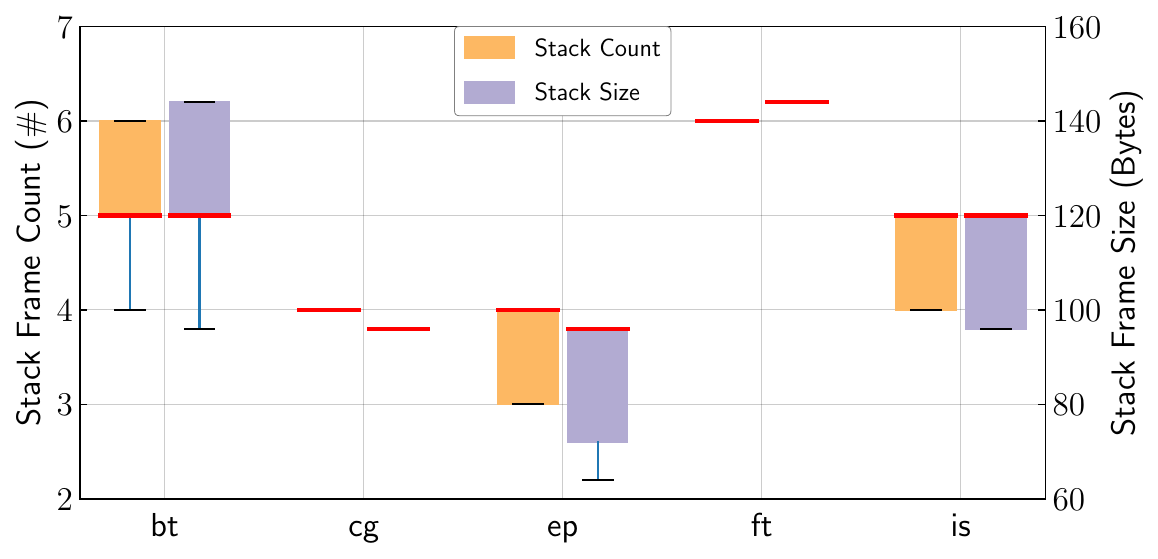}
    \caption{\ac{npb} benchmarks.}\label{fig:npb-stack-stats}\label{fig:stack-npb-combined}
  \end{subfigure}%
  \begin{subfigure}[t]{.5\linewidth}
      \centering
    \includegraphics[width=\textwidth,clip]{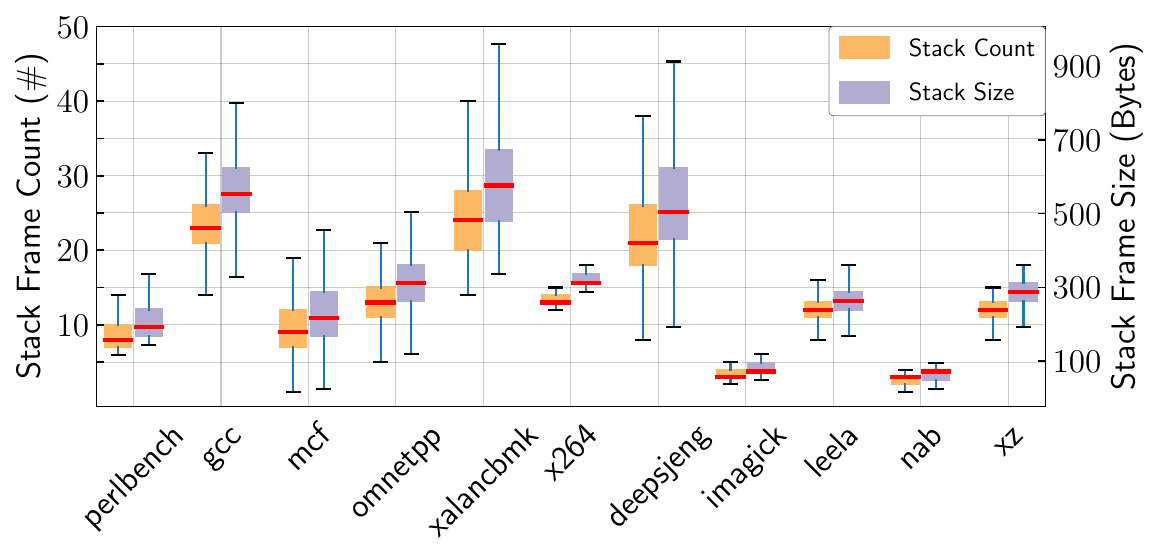}
    \caption{\spec{} benchmarks.}\label{fig:spec-stack-stats}\label{fig:stack-spec-combined}
  \end{subfigure}
  \caption{Stack frame size and count distribution for benchmarks on \armarch{}.
  Larger, more complex applications like those in \spec{} also have higher frame sizes and frame counts and will benefit more from the elimination of stack transformation during migration by \tool{}.}\label{fig:stack-stats}
\end{figure}

Related work shows that for a set of applications from the \ac{npb}, the cost of stack transformation is not significant~\cite{barbalace2017}.
However, since state transformation requires traversing all stack frames for migration between \acp{isa}, the cost depends on the number of variables, stack frames, and their sizes, potentially resulting in higher overheads for other applications.
Thus, we examine stack sizes and stack frame counts of a wide range of \ac{hpc} and compute\-/intensive benchmarks from the \ac{npb}~\cite{bailey2000} and \spec{}~\cite{bucek2018} suites (\Cref{fig:stack-stats}), showing their 5\-/point summary (min/max, lower/upper quartile, median values).

We observe that larger and more complex applications also have higher frame sizes and frame counts on average, e.g, up to 30 times more when comparing \spec{} against \ac{npb} applications.
For example, \textsc{xalancbmk} can have up to 40 stack frames with an average stack frame size of $500$ bytes, amounting to a stack size of $\sim$$20$KB.
If we conservatively assume $8$ byte variables, the runtime stack transformation will need to convert the state of 2500 variables.
This is an order of magnitude larger when compared to the worst case of \textsc{bt} in \ac{npb} with only 6 frames, $120$ bytes stack size, and 90 variables on average.
Moreover, we see that there is a direct correspondence between the number of stack frames and the frame size.
Lastly, frame count and size variability depend on the application and there is no correlation between programs even within the same benchmark suite.
We conclude that larger applications can incur a higher overhead in execution time during migration.
Finally, stack transformation also increases binary size due to additional metadata required (\Cref{subsec:size-comparison}).
Both costs are completely avoided by \tool{}.

%% file: 030_example.tex
\section{The Stack Layout Problem}\label{sec:example}

\input{031_figure_example}

Consider the top left\-/hand side of \Cref{fig:example}, where a simple loop in \cproglang{} accumulates the result of calls to function \code{hot\_func}, using as input local variable \code{x} passed by reference.
We assume an \armarch{} and \pcarch{} hardware setup which can facilitate migration and that \code{hot\_func} represents a computationally intensive series of operations of a program which typically resides on the low\-/power \armarch{} processor since it is not demanding outside this function.
Then, the function call boundary represents a natural point for migrating the long\-/running computation to the more powerful \pcarch{} processor which would significantly reduce the overall execution time of this program~\cite{bhat2015}.

Despite its simplicity, this code snippet results in different stack layouts on \pcarch{} and \armarch{} \acp{isa} that would inhibit heterogeneous process migration.
\Cref{fig:example-x86-asm,fig:example-arm-asm} show, respectively, the simplified \pcarch{} and \armarch{} pseudo\-/assembly generated for~\Cref{fig:example-source} by the \llvm{} compiler backend code generators.
The main difference between the two versions relates to the treatment of the pass\-/by\-/reference parameter.
On \pcarch{}, the compiler hoists the load operation outside the loop (\Cref{fig:example-x86-asm}, line~\ref{example_x86_hoist_load}) since it is expensive to perform at every iteration, based on the compiler cost model for this \ac{isa}.
This is achieved by using a callee\-/saved register since the value of \code{x} needs to be preserved across calls to \code{hot\_func}, which in turn forces its prior contents to be spilled on the stack (\Cref{fig:example-x86-asm}, line~\ref{example_x86_spill}).
On \armarch{}, the compiler decides the opposite; the address is deemed cheap and, hence, kept in the loop and recalculated (i.e., rematerialized) using the frame pointer \asm{FP} at every iteration (\Cref{fig:example-arm-asm}, line~\ref{example_arm_remat}).

\Cref{sub@fig:example-stack-unaligned-aligned} shows a snapshot of the register file state and the top\-/most stack frame for each process when execution reaches the call to \code{hot\_func}.
It also shows the bookkeeping information required by the \acl{sota} \popcorn{} compiler~\cite{barbalace2017} to perform heterogeneous migration.
The main differences are highlighted in red.
Comparing the two \ac{isa} execution states, we note that:
\begin{enumerate}[topsep=1em,label=\roman*)]
  \item the stack slot contents differ,
  \item the registers assigned to program variables differ, and
  \item the \acl{sota} \popcorn{} compiler~\cite{barbalace2017} requires metadata embedded in the binaries to track differences for correct transformation during migration.
\end{enumerate}
\popcorn{} \linux{} uses a runtime library to transform the stack, guided by the metadata embedded in the binaries during their compilation.
Conversely, \tool{} generates code that preserves the stack layout across \acp{isa} (\Cref{fig:example-stack-unaligned-aligned}).
Hence, a binary compiled with \tool{} does not require any metadata to account for the differences during heterogeneous migration.
This obviates the stack transformation costs by significantly reducing binary size and state transformation overheads.

%% file: 031_figure_example.tex
\begin{figure*}
  \centering
  \begin{subfigure}[t]{0.295\columnwidth}
    \begin{adjustwidth}{0.0cm}{}
    \centering
\begin{lstlisting}[style=cstyle]
 int i;
 double x = 0, sum = 0;

 for (i = 0; i < 16; ++i)
  (*@\colorbox{lightgray}{sum += hot\_func(\&x);}@*)

 // x also accessed here
\end{lstlisting}
    \end{adjustwidth}
    \caption{Code snippet in \cproglang{}.}\label{fig:example-source}
  \end{subfigure}
  \hspace*{\fill}
  \begin{subfigure}[t]{0.330\columnwidth}
    \begin{adjustwidth}{0.0cm}{}
    \centering
\begin{lstlisting}[style=asmstyle]
 push Rcs //spill|\label{example_x86_spill}|
 load Rcs, xaddr //hoisted|\label{example_x86_hoist_load}|
 Rtmp1 = 1
 jmp end if Rtmp1 > 15
loop:
 |\colorbox{lightgray}{call hot\_func()}| //arg:Rcs
 Rtmp1++
 jmp loop if Rtmp1 < 16
end:
\end{lstlisting}
    \end{adjustwidth}
    \caption{\pcarch{} pseudo\hyp{}assembly for~\ref{sub@fig:example-source}.}\label{fig:example-x86-asm}
  \end{subfigure}
  \hspace*{\fill}
  \begin{subfigure}[t]{0.330\columnwidth}
    \begin{adjustwidth}{0.0cm}{}
    \centering
\begin{lstlisting}[style=asmstyle]
 Rtmp1 = 1
 jmp end if Rtmp1 > 15
loop:
 //address recalculation
 load Rtmp2, FP + 3|\label{example_arm_remat}|
 |\colorbox{lightgray}{call hot\_func()}| //arg:Rtmp2
 Rtmp1++
 jmp loop if Rtmp1 < 16
end:
\end{lstlisting}
    \end{adjustwidth}
    \caption{\armarch{} pseudo\hyp{}assembly for~\ref{sub@fig:example-source}.}\label{fig:example-arm-asm}
  \end{subfigure}
  \vfill
  \begin{subfigure}[b]{\textwidth}
    \centering
    \includegraphics[clip,width=\linewidth]{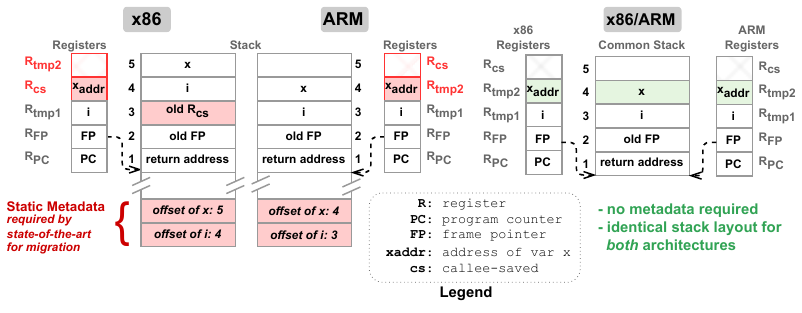}
    \caption{Unaligned stack layout (left) and \tool{} aligned stack layout for~\ref{sub@fig:example-source}.}\label{fig:example-stack-unaligned-aligned}
  \end{subfigure}
  \caption{Execution snapshot of the stack layout of a simple loop~\ref{sub@fig:example-source} on \pcarch{}~\ref{sub@fig:example-x86-asm} and \armarch{}~\ref{sub@fig:example-arm-asm} \acp{isa} just before a function call.
  Red highlighted areas (bottom left) show the differences and their required bookkeeping to enable migration using a \acl{sota} technique~\cite{barbalace2017}.
  \tool{} (this work) enforces a common layout (bottom right) with minimal overhead for heterogeneous migration.}\label{fig:example}
\end{figure*}

%% file: 040_design.tex
\section{Design of \tool{}}\label{sec:design}

Our goal is to provide the automatic generation of a unified memory stack layout that will simplify programming and allow faster heterogeneous migration without the associated overheads of metadata\-/dependent approaches.
To this end, we have developed \tool{}, a compiler backend technique which imposes a uniform stack layout between the targeted architectures and eliminates the need for stack transformation during migration.

We chose the 64\-/bit \pcarch{} and \armarch{} \acp{isa} as targets, being the most widespread architectures.
Moreover, their primary \ac{abi} properties partially overlap (e.g., alignment, register width, pointer size, endianness, etc.), simplifying an initial prototype.

\begin{figure}[t]
  \includegraphics[clip,width=\textwidth]{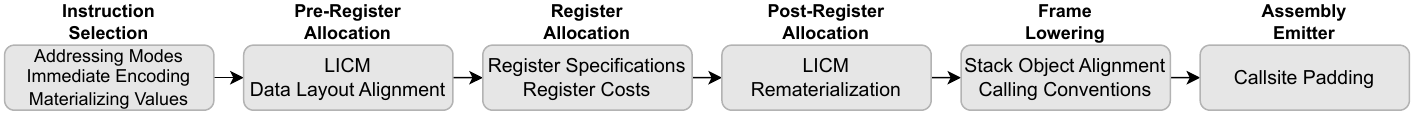}
  \caption{High\-/level design of \tool{} showing code generation passes with our extensions in each backend compilation stage for both architectures.}\label{fig:design-diagram}
\end{figure}

A high\-/level overview of our approach is shown in~\Cref{fig:design-diagram}, where a modern modular compiler structure is assumed.
The \ac{ir} of the program is given as input to both compiler backends, in order to be lowered to the target assembly.
By extending each of the backend passes, we mitigate the differences in the final stack layout.
Code is lowered in every stage of the backend, so that the final binary (one per architecture), when executed, will have the same stack layout for both architectures.
Hence, no metadata is required, as we do not need any stack transformation.

The rest of the section presents the main design challenges of \tool{}.
First, we decompose the variations related to each architecture's \acp{abi}.
These are mostly straightforward to fix, since they are clearly documented as specifications and solving them first will also ease mitigating the rest of the differences.
Then, we examine the differences that arise from the different instructions offered by each \ac{isa}.
Finally, in every compiler implementation, certain decisions have gone into it that reflect tacit knowledge based on practice and experience, which is challenging to identify.

\subsection{\ac{abi} Treatment}\label{subsec:abi_treatment}

The first fundamental factor that causes stack layout mismatch is the register file.
Registers are finite, so any values that they cannot hold, need to be stored (spilled) on the stack.
Also, if a register is \textit{callee\-/saved}, it must be pushed to the stack before being used by a function, affecting the stack layout (contrary to a \textit{caller\-/saved} or \textit{temporary}).

\input{041_table_regmap}

One of \tool{}'s guiding principles is to use the same number of registers between the architectures, while mapping the registers to the same functionality (we assume that such mapping exists, e.g., if one architecture supports 256-bit vector registers, the other one should support them too).
However, removing registers that have a dedicated purpose in the \ac{isa} is challenging.
For instance, in \armarch{}, the registers \asm{r16} and \asm{r17} can be used to support long branches, and the register \asm{r18} can be used to carry inter-procedural state (e.g., the thread context)~\cite{aarch64abi-webpage}.
Therefore, we give priority to those registers and keep them in the \ac{isa}, whereas we can freely remove others with no special use, e.g., simple temporary registers.

For the calling conventions~\cite{lu2021,aarch64abi-webpage}, we need to maintain:
\begin{enumerate*}[label=\roman*)]
  \item the same number of registers for function arguments since the extra arguments are passed through the stack,
  \item the same return value registers, and
  \item match the callee\-/saved registers which are also pushed on the stack.
\end{enumerate*}
We apply this to both general\-/purpose and \ac{fp} registers.

\Cref{tab:mapping} shows the register sets of \pcarch{} and \armarch{} after our mapping.
Overall, we reduced the number of the \armarch{} general\-/purpose and \ac{fp} registers from 32 to 16, per category, which is compatible with \pcarch{}, and modified the usage of a series of registers.
For the callee\-/saved registers, we also ensured that they are saved in the same order on the stack by the called functions.

Moreover, if a register has a special use in one architecture (e.g., \asm{r16}--\asm{r18} in \armarch{}), we preserve it and map the register to its counterpart in the other architecture, as long as there are no incompatibilities between the registers (e.g., if they both need to be reserved for a different purpose).



\subsection{Instruction Treatment}\label{subsec:instdiff}

Another contributing factor to stack layout mismatch stems from the instructions offered by each architecture.
For example, loading a constant from memory in \pcback{} and \armback{} requires different number of instructions (\Cref{fig:load-asm}).
\pcarch{} calculates the address of the constant and loads the value from memory in one instruction.
On the other hand, \armarch{} calculates the address using a separate instruction and then loads the value in a subsequent instruction.
Therefore, \armarch{} requires an additional register and may spill one more value to the stack compared to \pcback{}.
In addition, \armarch{} uses two instructions to complete this behavior, which affects the register allocation regions and, hence, indirectly the stack layout.

\begin{wrapfigure}{r}{0.47\linewidth}
  \centering
  \begin{subfigure}[t]{0.40\linewidth}
\begin{lstlisting}[style=asmstyle]
R0 = constant
||
\end{lstlisting}
    \caption{}\label{fig:load-asm-x86}
  \end{subfigure}
  \begin{subfigure}[t]{0.55\linewidth}
\begin{lstlisting}[style=asmstyle, numbers=none]
R0 = addressof(constant)
R1 = load R0
\end{lstlisting}
    \caption{}\label{fig:load-asm-arm}
  \end{subfigure}
  \caption{Pseudo\-/assembly for loading a constant value on \pcarch{} (left) and \armarch{} (right).
  \armarch{} requires an additional instruction for the same high-level operation.}\label{fig:load-asm}
\end{wrapfigure}

These differences are tightly coupled with two important compiler problems: register allocation and instruction selection.
Register allocation determines which values reside in registers and which in memory, and for each architecture it may utilize a different number of registers for a specific operation.
Instruction selection lowers the IR code into machine instructions and may choose different instructions for the same high\-/level operation, depending on the architecture or implementation\-/specific decisions (\Cref{subsec:implementation-specific-differences,sec:implementation}).

\subsection{Compiler Backend Treatment}\label{subsec:implementation-specific-differences}

Finally, independently of the \ac{abi} or the instructions of an \ac{isa}, there are some differences caused by the manner the code is generated and optimized in the compiler.
As shown in~\Cref{fig:example}, the code motion caused by the compiler reflects specific implementation decisions in the backend, affecting the stack layout even when the \ac{abi} differences between \acp{isa} have been bridged.

These differences comprise alignment decisions (apart from the \ac{abi} specifications), register allocation decisions, and a series of optimizations, like rematerialization~\cite{briggs1992} and code motion (\Cref{sec:implementation}).
We make up for the differences described in the last two sections by modifying the instruction selection (to select instructions with similar behavior), register allocation, and other phases (e.g., optimizations of constants).

%% file: 041_table_regmap.tex
\begin{wraptable}{r}{0.47\linewidth}
  \scriptsize
  \centering
  \caption{\tool{}'s mapping between \pcback{} and \armback{} register files ensures a low execution\hyp{}time impact.
  We show the callee-saved, return value, parameter passing, and temporary registers.
This applies for both general\hyp{}purpose and \acs{fp}/vector registers.}\label{tab:mapping}
  \begin{tabularx}{\linewidth}{ l l @{\hskip 4pt} l }
    \toprule
    \multicolumn{2}{c}{\textbf{Registers}} & \textbf{Usage} \\
    \cmidrule(lr){1-2}
    \textbf{\pcarch} & \textbf{\armarch} & \\
    \midrule
    \multicolumn{2}{c}{\textit{Callee\hyp{}saved}} & \\
    \cmidrule(lr){1-2}
    \asm{rsp} & \asm{SP} & Stack pointer \\
    {--} & \asm{r30} & Link register \\
    \asm{rbp} & \asm{r29} & Frame pointer \\
    \asm{rbx}, \asm{r15} & \asm{r19}, \asm{r20} & General purpose \\
    \addlinespace
    \multicolumn{2}{c}{\textit{Caller\hyp{}saved}} & \\
    \cmidrule(lr){1-2}
    \asm{rax}, \asm{rdx} & \asm{r8}, \asm{r2} & Return \\
    \asm{rdi} & \asm{r0} & Arg \#1/return\\
    \asm{rsi,rdx,rcx,r8,r9} & \asm{r1}--\asm{r5} & Args \#2--\#6 \\
    \asm{r10}--\asm{r14} & \asm{r6}, \asm{r7}, \asm{r16}--\asm{r18} & Temp registers \\
    \asm{xmm0}--\asm{xmm1} & \asm{v0}--\asm{v1} & \ac{fp} args/return \\
    \asm{xmm2}--\asm{xmm7} & \asm{v2}--\asm{v7} & \Ac{fp} args \\
    \asm{xmm8}--\asm{xmm15} & \asm{v8}--\asm{v15} & Temp \ac{fp} registers \\
    \bottomrule
  \end{tabularx}%
\end{wraptable}

%% file: 050_implementation.tex
\section{Implementation}\label{sec:implementation}\label{subsec:llvm_changes}

\subsection{The \tool{} Backend}

\input{052_table_llvm_changes}

We implemented \tool{} by extending the \armarch{} and \pcarch{} backends of \llvm{}.
Having demonstrated how code generation can affect the stack layout,~\Cref{tab:tool_compiler_transforms} enumerates how we implemented \tool{} based on the \llvm{} backend infrastructure, in order to achieve a unified address space layout and application state.
We describe the various backend parts based on their functionality.

\subsubsection{Alignment}\label{subsubsec:alignment}

\paragraph{Symbol Alignment.}

Similarly to related work~\cite{devuyst2012, barbalace2017}, the symbols of the program, namely the functions and the data, need to lie on the same virtual addresses for both architectures.
This way, accesses to global data will be consistent between the architectures, and the functions will be aliased to the same address, which is necessary when copying the memory images between \acp{isa}.
Therefore, we align the symbols in the code and data sections, by having one symbol per section and by adding padding between these sections during linking.
The result is shown for \asm{main} in the first lines in the snippets of~\Cref{fig:callsites1,fig:callsites2}, which are all aligned to the address \asm{1000}.

\paragraph{Callsite Alignment.}

When calling a function, the address of the instruction after the call is pushed to the stack (at least for the calling conventions in question).
Typically, the return addresses pushed will differ for the two architectures, so if a migration happens inside a function
the destination processor will return to a wrong address after the function returns.
As shown in~\Cref{fig:callsites1,fig:callsites2}, in the original version of the assembly, the instructions that follow the call to \code{hot\_func} differ by three bytes (offset \asm{1104} vs offset \asm{1107}).

To address this issue, we align the callsites similarly to our approach for the program symbols, using \asm{nop} instructions.
Before the call instructions, we emit \asm{nop} instructions to pad the return addresses (\Cref{fig:callsites3,fig:callsites4}).

\input{053_figure_callsite_align}

\paragraph{Stack Object Alignment.}

We ameliorate the alignment differences for the objects in the stack of the two architectures.
These include the use of an emergency spill slot and the alignment of local stack objects and callee-saved registers, as listed in~\Cref{tab:tool_compiler_transforms}.
To elaborate on the emergency spill slot, the \armback{} backend will scavenge an extra register in case it needs to materialize large stack offsets (i.e., more than $255$ bytes), which do not fit in one instruction.
If a register cannot be found, a special spill slot is reserved.
For simplicity, \tool{} conservatively reserves this slot and places it after the callee\hyp{}saved registers for both architectures.

\subsubsection{Addressing Modes}\label{subsubsec:addr_modes}

\armback{} does not support addressing modes of the form \textit{[base + scaled register + offset]} for indexing arrays.
When the \pcback{} backend uses this mode, the \armback{} backend needs to reserve an extra register for the same operation.
Usually, the \armback{} backend keeps the address of the array, i.e., \textit{[base + offset]}, in a separate callee\hyp{}saved register, to be able to reuse it for indexing the array multiple times.
This usage of one extra callee\hyp{}saved register may introduce extra spills, hence, we disable the former complex addressing mode in \pcback{}, to get the same behaviour in \armback{}.

\subsubsection{Immediate Encoding}\label{subsubsec:imm_encoding}

The two \acp{isa} do not support the same set of immediates.
Due to smaller instruction size, \armback{} instructions can encode explicitly up to 21 bits of immediates for \texttt{pc}\hyp{}relative addressing~\cite{arm-v8-arch-ref-webpage}, 12 bits (with an optional shift of 12) for arithmetic operations~\cite{arm-v8-arith-imm-webpage}, and up to 64\hyp{}bit logical immediates~\cite{arm-v8-logical-imm-webpage}.
Finally, the \ac{isa} allows moving up to 16\hyp{}bit immediates, optionally shifted, to registers.
\tool{} keeps the same immediate encoding in \pcback{}.

\subsubsection{Register Allocation}\label{subsubsec:regalloc}

To limit possible overheads, we have kept the default \textit{greedy} register allocator~\cite{llvm-ti-codegen-ref-webpage}, which uses global live range splitting, minimizing the cost of spilled code.
However, we need to make sure that the allocator will take the same decisions when assigning registers for the two backends, despite the heuristics it uses internally.
Expanding on the first two corresponding entries of the table, we first assign the same order of preference to the registers.
For example, \asm{r15} is more expensive on \pcback{} than \asm{rbx}, due to encoding reasons, so we do the same for \armarch{}'s \asm{x20} and \asm{x19}, even though their cost is the same, to achieve similar allocation.
Also, since \armback{} has a dedicated zero register that is not callee\hyp{}saved, we avoid using callee\hyp{}saved registers for \pcback{} to hold the zero constant, as this would require extra spills in memory.

\subsubsection{Rematerialization and Code Motion}\label{subsubsec:remat_code_motion}

\begin{wrapfigure}{r}{0.54\linewidth}
  \centering
  \centering
  \vspace{0pt}
  \begin{subfigure}[t]{0.48\linewidth}
\begin{lstlisting}[style=asmstyle]
  |\tikzmark{align_x86_remat_before_start1}|push Rcs|\tikzmark{align_x86_remat_before_end1}|
  |\tikzmark{align_x86_remat_before_start2}|load Rcs, xaddr|\tikzmark{align_x86_remat_before_end2}|
  Rtmp1 = 1
  jmp end if Rtmp1 > 15
loop:

  // arg: Rcs
  call hot_func()
  jmp loop if Rtmp1 < 16
\end{lstlisting}
    \caption{}\label{fig:x86remat1}
  \end{subfigure}
  \hspace*{\fill}
  \begin{subfigure}[t]{0.48\linewidth}
\begin{lstlisting}[style=asmstyle, numbers=none]
||
||
  Rtmp1 = 1
  jmp end if Rtmp1 > 15
loop:
  |\tikzmark{align_x86_remat_after_start1}|load Rtmp2, xaddr|\tikzmark{align_x86_remat_after_end1}|
  // arg: Rtmp2
  call hot_func()
  jmp loop if Rtmp1 < 16
\end{lstlisting}
    \caption{}\label{fig:x86remat2}
  \end{subfigure}
  \caption{\pcback{} pseudo\hyp{}assembly from~\Cref{fig:example} \emph{with} (left) and \emph{without} (right) rematerialization of \code{\&x} using \asm{load}.
  In~\subref{fig:x86remat1} the address calculation is hoisted out of the loop (highlighted), occupying a callee\hyp{}saved register whose contents are spilled to the stack.
  In~\subref{fig:x86remat2}, \tool{} does not hoist the calculation, keeping the stacks among \acp{isa} aligned.}\label{fig:x86remat}
  \begin{tikzpicture}[remember picture,overlay]
    \begin{scope}
      \tikzset{hilite_unaligned node/.style 2 args = {draw=none, fill=carnationpink, inner sep=0pt, fill opacity=0.3, yshift=2pt, minimum height=8pt, fit=(#1) (#2)}}
      \tikzset{hilite_aligned node/.style 2 args = {draw=none, fill=applegreen, inner sep=0pt, fill opacity=0.3, yshift=2pt, minimum height=8pt, fit=(#1) (#2)}}
      \node[hilite_unaligned node={pic cs:align_x86_remat_before_start1}{pic cs:align_x86_remat_before_end1}] {};
      \node[hilite_unaligned node={pic cs:align_x86_remat_before_start2}{pic cs:align_x86_remat_before_end2}] {};
      \node[hilite_aligned node={pic cs:align_x86_remat_after_start1}{pic cs:align_x86_remat_after_end1}] {};
    \end{scope}
  \end{tikzpicture}
\end{wrapfigure}

Most changes in this category are related to how the compiler generates code for getting the values or the addresses of variables.
As discussed in \Cref{sec:example} and shown in~\Cref{fig:x86remat}, we instruct the compiler to rematerialize the \asm{load} instructions, leading to the code in \Cref{fig:x86remat2}, where \asm{load} is kept inside the loop without the need for a callee\hyp{}saved register.
This way the address is calculated at every iteration (line 6) and passed to the first argument register (line 7).
There are eight different cases (\Cref{tab:var_access}).

\paragraph{Getting values of variables.}

Getting the value of local variables is done similarly in both architectures, with simple loads from the stack frame.
For global variables, \pcback{} needs only one instruction (a \asm{rip}\hyp{}relative \asm{mov}), whereas \armback{} requires two instructions (one to calculate the global address and one to load from it).
Since \llvm{} cannot currently rematerialize multiple instructions, when \armback{} needs to reuse a global variable it will spill it in the stack, while for \pcback{} it suffices to recompute the \asm{rip}\hyp{}relative \asm{mov}.
Regardless, we ensure that \pcback{} will also spill the value in this case.

\paragraph{Getting references of variables.}

Getting the reference (i.e., address) of a local variable in \armback{} uses simple arithmetic (\asm{add}/\asm{sub}) with the frame pointer (\Cref{fig:example}), while \asm{lea} instructions (termed \asm{load} in our pseudo\hyp{}assembly) are used in \pcback{} (\Cref{fig:example,fig:x86remat}).
The \asm{add}/\asm{sub} instructions are usually not hoisted out of loops by the compiler, whereas the \asm{lea} instruction is, since it is a little more expensive in the general case~\cite{instruction-tables-fog-webpage,arm-cortex-a7-timings-webpage}.
Hoisting the instruction may consume a callee\hyp{}saved register so, instead, our compiler rematerializes these instructions if it needs to reuse their result.
Since we are not hoisting \asm{lea} instructions, however, we need to do the same for the \asm{adrp} instructions (the \armarch{} counterpart), to also have an aligned behavior for the case of global variables (\asm{adrp} vs rip-relative \asm{lea}).
Therefore, we additionally rematerialize \asm{adrp} instructions.

\begin{wrapfigure}{r}{0.40\linewidth}
    \input{043_table_variable_access}
\end{wrapfigure}

\paragraph{Reusing constants.}

When the same constant is used multiple times in \armback{}, the compiler tries to place its value in a register.
Instead, \pcback{} tries to materialize the constant again by encoding it separately for every instruction.
\tool{} emulates the first behavior for \pcback{}.

\subsubsection{Verifying Stack Layout}

To verify that the stack layout between an \pcarch{} and an \armarch{} binary is the same and, hence, that migration will be successful, we use the \llvm{} StackMaps~\cite{llvm-stackmaps} to track the location of values in the stack.
These are inserted in a separate \ac{elf} section, which is cross\hyp{}checked after compilation for both binaries and then discarded.
Compilation correctness is ensured by the \llvm{} backend tests, after they are ported to the new \tool{} changes.

\subsection{Other Considerations and Limitations}

\paragraph{Libraries.}
We currently only support static linking for the libraries, to simplify the engineering effort required to align all the symbols, and to create consistent memory images of the binaries between the machines.
Moreover, we are not migrating inside library calls, so they do not need a consistent layout.
However, some functions in the \musl{} library (e.g., \code{setjmp}/\code{longjmp}) use inline assembly when handling signals/exceptions, so we modified their assembly to obey our new \acp{abi} for compatibility.

\paragraph{Interaction with the \ac{os}.}

We assume a replicated\hyp{}kernel \ac{os}, where there is a kernel per core and each kernel loads the address space of the respective binary~\cite{barbalace2017}.
The address space of each binary has an identical layout, but the \code{.text} section is natively compiled, and then aligned (\Cref{subsubsec:alignment}), so that each function in the two \acp{isa} will have the same virtual address.
The \ac{os}\hyp{}specific details (e.g., page mapping, process scheduling, view of the \ac{os} by different processors, etc.) are out of scope and can be found in related work~\cite{barbalace2015}.

\paragraph{Backend and optimization flags.}
Regarding the backend infrastructure, we are not using the machine instruction scheduler since it can lead to instruction reorderings that invert the order of spilling. 
Identifying a good compromise between allowing machine scheduling and getting a predictable order of spilled values is left for future work.
In addition, although the ARMv8 architecture is bi\hyp{}endian, we keep the (default) little\hyp{}endian setting to match with \pcarch{}.
The technical effort to support different endianness is outside the scope of this work.
Finally, migration is supported currently for up to the \code{-O1} flag.
An open issue in the \llvm{} code generator\footnote{https://github.com/llvm/llvm-project/issues/56880} makes the allocator run out of registers for most of the \ac{npb} in \code{-O2}/\code{-O3}, when using stackmaps.
Therefore, we cannot verify \tool{} for the full suite in \code{-O2} or higher, although we are working on an \llvm{} patch to fix that.
However, since the \llvm{} backend enables optimizations for any optimization level other than \code{-O0}, having supported \code{-O1} should cover from the outset many programs compiled with higher flags.

\paragraph{Applicability to other architectures.}

Even though the details described in this section are specific to \pcarch{} and \armarch{}, our technique provides a blueprint and the high\hyp{}level ideas for supporting other combinations of general-purpose processors (e.g., \pcarch{} and \riscv{}).
Much of the target-specific details, e.g., unifying the \acp{abi}, the instruction formats, the register allocation costs, the rematerialization properties of values, etc., are usually encoded easily as a backend specification, e.g., in \llvm{}'s \tablegen{}~\cite{tablegen} or \gcc{}'s Machine Description~\cite{gcc-machine-description}.
More elaborate implementation, e.g., which address immediates are legal, can be guided using the insights gleaned from this section and our \ac{oss} artifact.
For target features not covered in this work, these can be detected through the stackmap machinery, e.g., a constant optimized specifically in one architecture will appear as a different architecture\hyp{}specific value (or values) in the stackmaps.

\paragraph{Multithreading and memory consistency.}

Our evaluation is for a single thread, however \tool{}'s design is orthogonal to multithreaded execution.
Since \tool{} is based on a unified address space between the architectures, and the evaluated architectures support different memory consistency models~\cite{owens2009better, chong2008reasoning}, an inter\hyp{}device coherence protocol is assumed (e.g., \cxl{}~\cite{cxl-webpage}), along with a fused memory model (e.g., compound models~\cite{goens2023compound}) between the two architectures.
In our setup, we assume that upon migration all threads are stopped and buffers/caches are flushed to memory, therefore migration points act as memory barriers.
We will investigate multithreaded execution in future work.

\subsection{Migration}

There is no platform available today with \pcarch{} and \armarch{} that share memory.
Therefore, to validate migration under our approach, we are using \criu{}~\cite{criu-webpage} for a prototype.
\criu{} offers a checkpoint\hyp{}restore mechanism in user space, by dumping a multi\hyp{}file image of the application when pausing it, and restoring the state later, continuing the execution.
However, in our prototype, we are leveraging multi\hyp{}\ac{isa} binaries, so we dump the state of the binary in the starting processor, rewrite the necessary \criu{} images to be consistent with the target machine, transfer the images (via SSH), and continue execution of the other binary by restoring the rewritten state, similarly to related work~\cite{xing2022, bapat2024dapper}.

Contrary to related work~\cite{venkat2014,barbalace2017,xing2022,bapat2024dapper}, we do not transform the state at all, but only rewrite images like \textit{core}~\cite{criu-webpage}, which contain core process and architecture-specific state information, e.g., the registers.
For example, based on our register mapping (\Cref{tab:mapping}), we need to simply copy the value of the callee\hyp{}saved \texttt{rbx} from the source image to the value of \texttt{r19} in the destination image, if we are migrating from \pcarch{} to \armarch{}.
Rewriting happens through a script invoked by a simple runtime, but in our ideal use\hyp{}case scenario, e.g., an \pcarch{}/\armarch{} machine with coherent shared memory, this is facilitated by a user- or kernel-space service without the need to transfer the state between the machines.


We take some further steps to keep the virtual memory address mappings between the architectures consistent.
First, we disable \ac{aslr} on both machines, otherwise the start address of the stack will be different between the architectures, thus breaking the match of the stack layouts.
We do this for simplicity and leave the effort of assigning the same random stack start address to both architectures as a future extension.
Furthermore, by default, \pcarch{} programs map their stack to addresses starting from \asm{0x800000000000}, whereas in \armarch{}, frames start from the address \asm{0xffffffffffff}~\cite{lu2021,aarch64abi-webpage}.
We modify the startup function from the \cproglang{} library, so that the frame of the \asm{main} function starts from \asm{0x800000000000} in both machines.

%% file: 052_table_llvm_changes.tex
\setlength{\textfloatsep}{10pt}  
\begin{table}[t]
    \footnotesize
    \centering
    \caption{\tool{}'s code generation extensions to the \armarch{} and \pcarch{} backends maintaining a unified stack layout.}
    \label{tab:tool_compiler_transforms}

    \begin{minipage}[t]{0.49\textwidth}
        \centering
        \begin{tabularx}{\linewidth}{l c @{\hskip 4pt} c}
            \toprule
            \textbf{Category \& Description} & \textbf{\pcarch} & \textbf{\armarch} \\
            \midrule

            \multicolumn{2}{l}{\textbf{Alignment} (\S~\ref{subsubsec:alignment})} & \\
            \cmidrule(lr){1-1}
            Align symbols in code \& data sections & \successmark{} & \successmark{} \\
            Align return addresses after callsites & \successmark{} & \successmark{} \\
            Allocate emergency spill slot & \successmark{} & \successmark{} \\
            Align local stack objects to at least 4 bytes & \successmark{} & \\
            Align callee-saved registers to 8 bytes & & \successmark{} \\
            \addlinespace

            \multicolumn{2}{l}{\textbf{Addressing Modes} (\S~\ref{subsubsec:addr_modes})} & \\
            \cmidrule(lr){1-1}
            Do not encode complex addressing & \successmark{} & \\
            Match legal address immediates & \successmark{} & \successmark{} \\
            \addlinespace

            \multicolumn{2}{l}{\textbf{Immediate Encoding} (\S~\ref{subsubsec:imm_encoding})} & \\
            \cmidrule(lr){1-1}
            Do not encode immediates in multiplication & \successmark{} & \\
            Encode same immediates for data-processing & \successmark{} & \\
            Do not materialize non-zero \ac{fp} constants & & \successmark{} \\
            \addlinespace

            \multicolumn{2}{l}{\textbf{Register Allocation} (\S~\ref{subsubsec:regalloc})} & \\
            \cmidrule(lr){1-1}
            Match register cost and allocation order & \successmark{} & \successmark{} \\
            \bottomrule

        \end{tabularx}
    \end{minipage}%
    \hfill
    \begin{minipage}[t]{0.49\textwidth}
        \centering
        \begin{tabularx}{\linewidth}{l c @{\hskip 4pt} c}
            \toprule
            \textbf{Category \& Description} & \textbf{\pcarch} & \textbf{\armarch} \\
            \midrule

            \multicolumn{2}{l}{\textbf{Register Allocation (cont.)} (\S~\ref{subsubsec:regalloc})} & \\
            \cmidrule(lr){1-1}
            Hold the zero constant in temp registers & \successmark{} & \\
            \makecell[tl]{Optimized two-address format \\ for integer instructions} & & \successmark{} \\
            Match instruction input/output operand size & \successmark{} & \successmark{} \\
            \addlinespace

            \multicolumn{2}{l}{\textbf{Rematerialization \& Code Motion} (\S~\ref{subsubsec:remat_code_motion})} & \\
            \cmidrule(lr){1-1}
            Rematerialize local variable loads & & \successmark{} \\
            Do not rematerialize \asm{movss}/\asm{movsd} & \successmark{} & \\
            Rematerialize \asm{lea} & \successmark{} & \\
            Rematerialize \asm{adrp} & & \successmark{} \\
            Reuse constants instead of rematerializing & \successmark{} & \\
            \addlinespace

            \multicolumn{2}{l}{\textbf{Other Optimizations}} & \\
            \cmidrule(lr){1-1}
            Disable heuristic for frame object ordering & \successmark{} & \\
            Match optimization of special constants & \successmark{} & \\
            Lower conditional \textit{select} similarly & \successmark{} & \successmark{} \\
            Vectorize pairs of \textit{double} & \successmark{} & \\
            \bottomrule

        \end{tabularx}
    \end{minipage}
\end{table}

%% file: 053_figure_callsite_align.tex
\begin{figure}[t]
  \centering
  \begin{subfigure}[t]{0.48\textwidth}
    \begin{adjustwidth}{0.10cm}{}
    \centering
\begin{lstlisting}[style=asmstyle, xleftmargin=50pt]
1000: main:
1000: SP = SP - 32
...
1108: call hot_func()
|\tikzmark{align_arm_ret_before_start1}|1112|\tikzmark{align_arm_ret_before_end1}|: // after call
\end{lstlisting}
    \end{adjustwidth}
    \caption{}\label{fig:callsites1}
  \end{subfigure}
  \begin{subfigure}[t]{0.48\textwidth}
    \begin{adjustwidth}{0.0cm}{}
    \centering
\begin{lstlisting}[style=asmstyle, xleftmargin=50pt]
1000: main:
1000: push FP
...
1102: call hot_func()
|\tikzmark{align_x86_ret_before_start1}|1107|\tikzmark{align_x86_ret_before_end1}|: // after call
\end{lstlisting}
    \end{adjustwidth}
    \caption{}\label{fig:callsites2}
  \end{subfigure}
  \hfill
  \begin{subfigure}[t]{0.48\textwidth}
    \begin{adjustwidth}{0.10cm}{}
\begin{lstlisting}[style=asmstyle, xleftmargin=50pt]
1000: main:
1000: SP = SP - 32
...
|\tikzmark{align_arm_ret_before_start2}|1100: four_byte_nop|\tikzmark{align_arm_ret_before_end2}|
1104: call hot_func()
|\tikzmark{align_arm_ret_before_start3}|1108|\tikzmark{align_arm_ret_before_end3}|: // after call
\end{lstlisting}
    \end{adjustwidth}
    \caption{}\label{fig:callsites3}
  \end{subfigure}
  \begin{subfigure}[t]{0.48\textwidth}
    \begin{adjustwidth}{0.0cm}{}
\begin{lstlisting}[style=asmstyle, xleftmargin=50pt]
1000: main:
1000: push FP
...
|\tikzmark{align_x86_ret_before_start2}|1102: one_byte_nop|\tikzmark{align_x86_ret_before_end2}|
1103: call hot_func()
|\tikzmark{align_x86_ret_before_start3}|1108|\tikzmark{align_x86_ret_before_end3}|: // after call
\end{lstlisting}
    \end{adjustwidth}
    \caption{}\label{fig:callsites4}
  \end{subfigure}
  \caption{Call site alignment in pseudo\hyp{}assembly before (top) and after (bottom) \tool{}'s operation for \armback{} (left) and \pcback{} (right).
  The \asm{main} symbol is placed at the same address (offset \asm{1000}) for both \acp{isa}.
  Highlighted lines in top figures show the difference in return addresses, while in bottom figures show the code emitted after \tool{} adds \asm{nop} instructions for padding, resulting in same return address.}
  \begin{tikzpicture}[remember picture,overlay]
    \begin{scope}
      \tikzset{hilite_unaligned node/.style 2 args = {draw=none, fill=carnationpink, inner sep=0pt, fill opacity=0.3, yshift=2pt, minimum height=8pt, fit=(#1) (#2)}}
      \tikzset{hilite_aligned node/.style 2 args = {draw=none, fill=applegreen, inner sep=0pt, fill opacity=0.3, yshift=2pt, minimum height=8pt, fit=(#1) (#2)}}
      \node[hilite_unaligned node={pic cs:align_arm_ret_before_start1}{pic cs:align_arm_ret_before_end1}] {};
      \node[hilite_unaligned node={pic cs:align_x86_ret_before_start1}{pic cs:align_x86_ret_before_end1}] {};
      \node[hilite_aligned node={pic cs:align_x86_ret_before_start2}{pic cs:align_x86_ret_before_end2}] {};
      \node[hilite_aligned node={pic cs:align_x86_ret_before_start3}{pic cs:align_x86_ret_before_end3}] {};
      \node[hilite_aligned node={pic cs:align_arm_ret_before_start2}{pic cs:align_arm_ret_before_end2}] {};
      \node[hilite_aligned node={pic cs:align_arm_ret_before_start3}{pic cs:align_arm_ret_before_end3}] {};
    \end{scope}
  \end{tikzpicture}
\end{figure}

%% file: 043_table_variable_access.tex
    \scriptsize
    \centering
    \vspace{0pt}
    \captionof{table}{Accessing values (\code{x}) and references (\code{\&x}) of variables per architecture.}\label{tab:var_access}
    \begin{tabularx}{0.4\columnwidth}{ l l l l }

        \toprule

        & \textbf{Scope} & \textbf{Access \code{x}} & \textbf{Access \code{\&x}} \\
        \midrule

        \multirow{2}{*}{\textbf{\armback}}

        & \textit{Local} & simple load & simple \asm{add}/\asm{sub} \\
        \cmidrule(lr){2-4}

        & \textit{Global} & \asm{adrp} + load & \asm{adrp} \\
        \midrule

        \multirow{2}{*}{\textbf{\pcback}}

        & \textit{Local} & simple load & \asm{lea} \\
        \cmidrule(lr){2-4}

        & \textit{Global} & rip-relative \asm{mov} & rip-relative \asm{lea} \\

        \bottomrule
    \end{tabularx}

%% file: 070_evaluation.tex
\section{Evaluation}

We answer the following questions to evaluate \tool{}:
\begin{enumerate*}[label=\roman*)]
    \item What is the binary size improvement compared to related work (\Cref{subsec:size-comparison})?
    \item What is the overall and per feature impact of \tool{} on execution time (\Cref{subsec:perf-breakdown})?
    \item How to analyze and limit the effect of the most impactful features (\Cref{subsec:callsite-alignment,subsec:reg-pressure,subsec:misched})?
    \item How does performance vary on different microarchitectures (\Cref{subsec:portability})?
    \item How does \tool{} compare in execution time with related work (\Cref{subsec:migration-comparison})?
\end{enumerate*}

\subsection{Setup}\label{subsec:setup}

\paragraph{Hardware.}

We use two hardware configurations for evaluation:
\begin{enumerate*}[label=\roman*)]

      \item (\armarch{}) GIGABYTE\textsuperscript{\textregistered} R181{-}T92{-}00 (SABER SKU), Dual Cavium ThunderX2\textsuperscript{\textregistered} \ac{cpu} CN9980 v2.2 2.20GHz (32 cores/128 threads), and
      \item (\pcarch{}) Dell PowerEdge R440, Intel\textsuperscript{\textregistered} Xeon\textsuperscript{\textregistered} Silver 4110 \ac{cpu} 2.10GHz (8 cores/16 threads).

      For the performance portability experiment (\Cref{subsec:portability}) we additionally use:

      \item (\pcarch{}) Intel\textsuperscript{\textregistered} Xeon\textsuperscript{\textregistered} Gold 6230R \ac{cpu} @ 2.10GHz,
      \item (\pcarch{}) Intel\textsuperscript{\textregistered} Core\textsuperscript{\texttrademark} i9-10900 \ac{cpu} @ 2.80GHz,
      \item (\armarch{}) Neoverse-N1: Ampere\textsuperscript{\textregistered} Altra @ 1.7GHz, and
      \item (\armarch{}) Cortex-A53\textsuperscript{\texttrademark} @ 1.5GHz

\end{enumerate*}

\paragraph{Software.}

\tool{} extends \llvm{} version 9.0.1, consisting of $\sim$3450 \ac{loc} (1520 and 860 \ac{loc} for the \pcback{} and \textsc{AArch64} backends respectively).
The rest of the code relates to target\hyp{}independent functionality, including a few changes on \clang{} version 9.0.1.
We reuse the modified linker and symbol aligner from the \popcorn{} compiler~\cite{popcorn-compiler-gh-webpage} (commit \code{4cc8805}).
Regarding the migration mechanism, we prototype a method using \hetcriu{}\cite{xing2022}, an extension of \criu{}\cite{criu-gh-webpage} at version 3.17.1.
Specifically, we extend the \dapper{} framework~\cite{bapat2024dapper}, a lightweight program state rewriter based on \hetcriu{}, to support heterogeneous checkpoint-restore with and without transformation\footnote{https://github.com/systems-nuts/TransProc}.
Finally, we employ a modified version of \musl{} version 1.1.22, which we statically link to the binaries and release with our \llvm{} modifications.

\paragraph{Benchmarks.}
We focus on compute\hyp{} and memory\hyp{}intensive \cproglang{} benchmarks to explore the impact of \tool{}'s code modifications which may have been hidden otherwise (e.g., by system calls, I/O operations, etc.).
In our experiments, we used a \cproglang{} implementation~\cite{seo2011} of the \ac{npb} suite~\cite{bailey2000} (input classes A, B, C).
All benchmarks are compiled with the \code{-O1} flag.
For the class C of the \textsc{ft} and \textsc{mg} benchmarks, the enormous amount of static data declared lead to a relocation overflow error in \armarch{} (hence we cannot compile also the aligned \pcarch{} binary), so we omit these results.
This is under investigation (e.g., using different code models), but does not affect our overall exploration here.
%

%
\subsection{Size Comparison}\label{subsec:size-comparison}

%
\begin{figure}[t!]
  \centering
  \begin{subfigure}[t]{0.49\linewidth}
    \includegraphics[width=\columnwidth,clip]{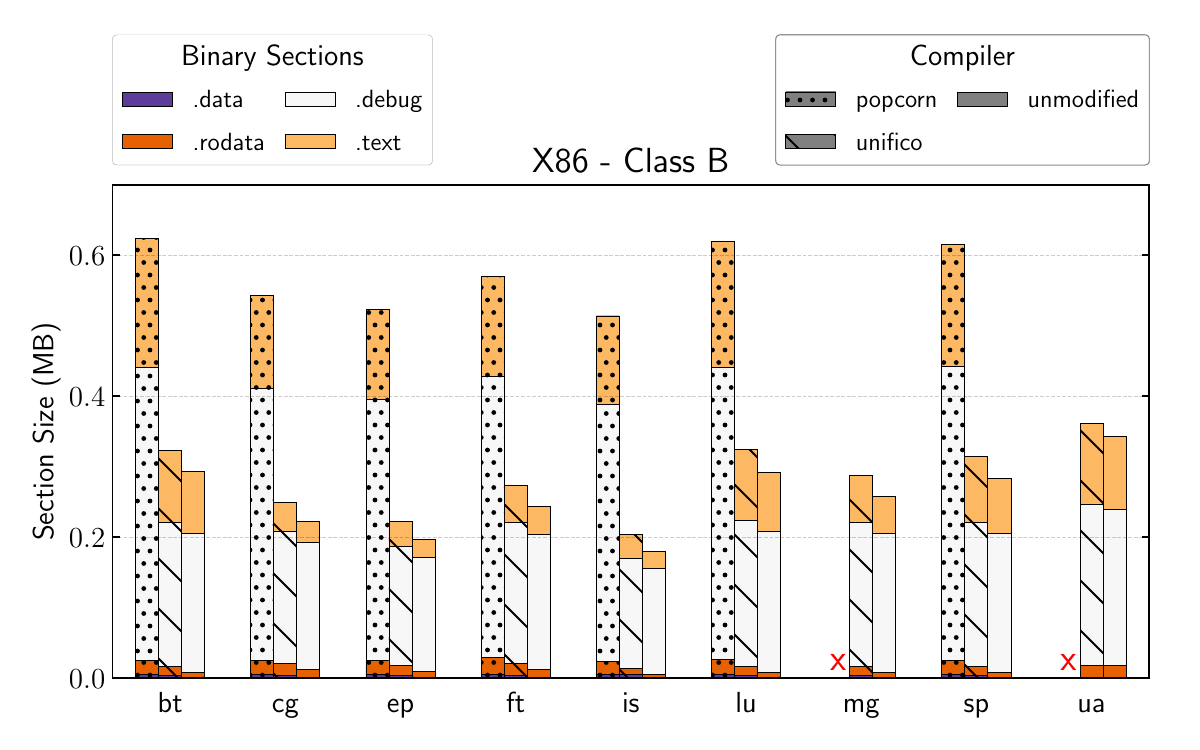}
    \caption{}\label{fig:section-sizes-x86}
  \end{subfigure}
  \begin{subfigure}[t]{0.49\linewidth}
    \includegraphics[width=\columnwidth,clip]{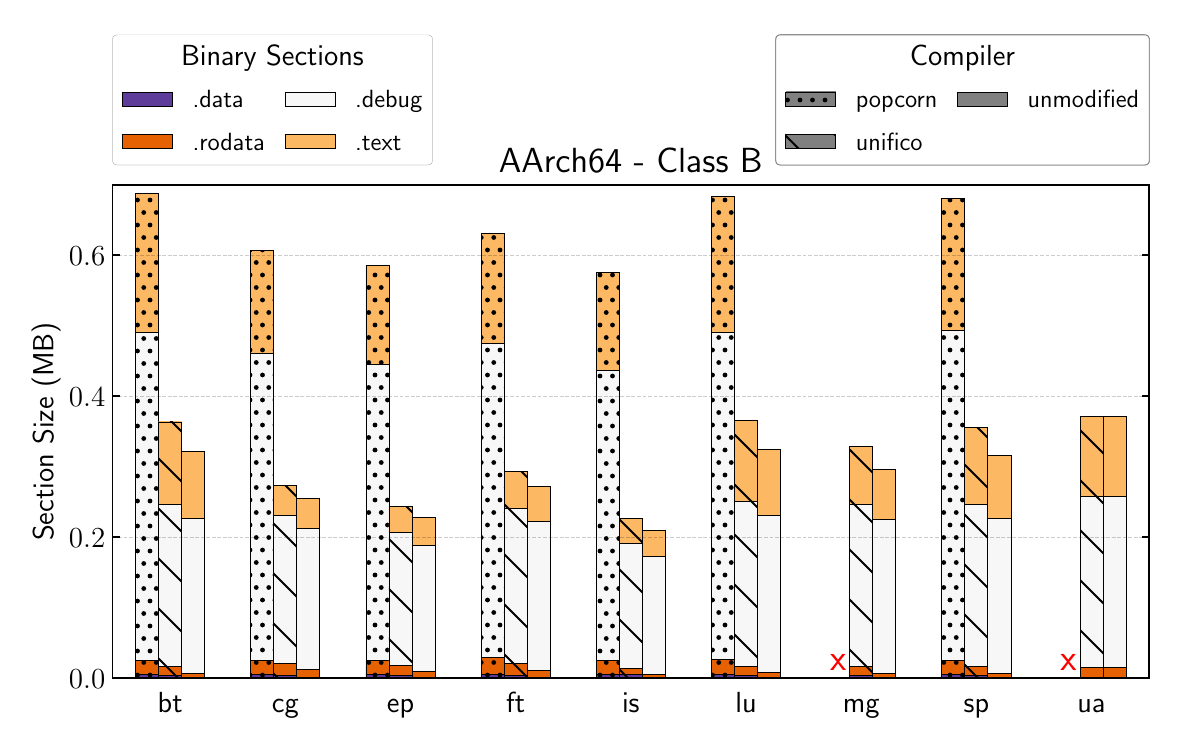}
    \caption{}\label{fig:section-sizes-arm}
  \end{subfigure}
  \caption{Binary size analysis generated with static linking (unmodified \llvm{}), the \popcorn{} compiler, and \tool{}, for \pcarch{} (left) and \armarch{} (right) using the \ac{npb} suite.
  Unlike the \popcorn{} compiler, \tool{} requires no metadata and its size overhead is minimal, within 10$\%$ of statically linked binaries generated with \llvm{}.}\label{fig:binary_sizes}
\end{figure}

We explore the sizes for binaries compiled with the following different methods:
\begin{enumerate*}[label=\roman*)]
  \item an unmodified \clang{}/\llvm{} compiler (statically linked with the \cproglang{} library),
  \item the \popcorn{} compiler toolchain (statically linked with the \cproglang{} library), and
  \item \tool{} (statically linked with the \cproglang{} library).
\end{enumerate*}
We include the unmodified compiler in our evaluation to show the impact of our modifications to the code section size (\Cref{fig:binary_sizes}).\footnote{\popcorn{} cannot compile \textsc{mg} and \textsc{ua} in \code{-O1} due to register allocation issues with stackmaps, but it does not affect our analysis.}
The binaries are for the class B of \ac{npb}, but the trends are similar for the other classes.

We make three observations.
First, in all compilation categories, the \pcarch{} binaries are smaller than the \armarch{} binaries.
This can be attributed to the \pcarch{} being a \ac{cisc} \ac{isa}, encoding more complex instructions that can lead to compact binaries.
Second, in both architectures, the \popcorn{} binaries are up to $2x$ the size of the statically compiled binaries of the unmodified compiler.
This is caused by additional metadata, debug information, as well as libraries (in the \code{.text} section) that are used to facilitate migration and state transformation.
Finally, we observe that in both architectures \tool{} leads to much smaller binaries relatively to \popcorn{}, and within $10\%$ of the unmodified compiler.
The size of the \tool{} binaries is slightly larger than those compiled with the standard \clang{} due to the constraints we imposed to the code generation, visible in the code and data sections.
Overall, we reduced the binary size overhead from ${\sim}200\%$ to ${\sim}10\%$.

\subsection{Analysis of Impact per Architecture}\label{subsec:perf-breakdown}

\begin{figure}[t]
    \begin{subfigure}[b]{\textwidth}
        \centering
        \includegraphics[width=0.8\linewidth]{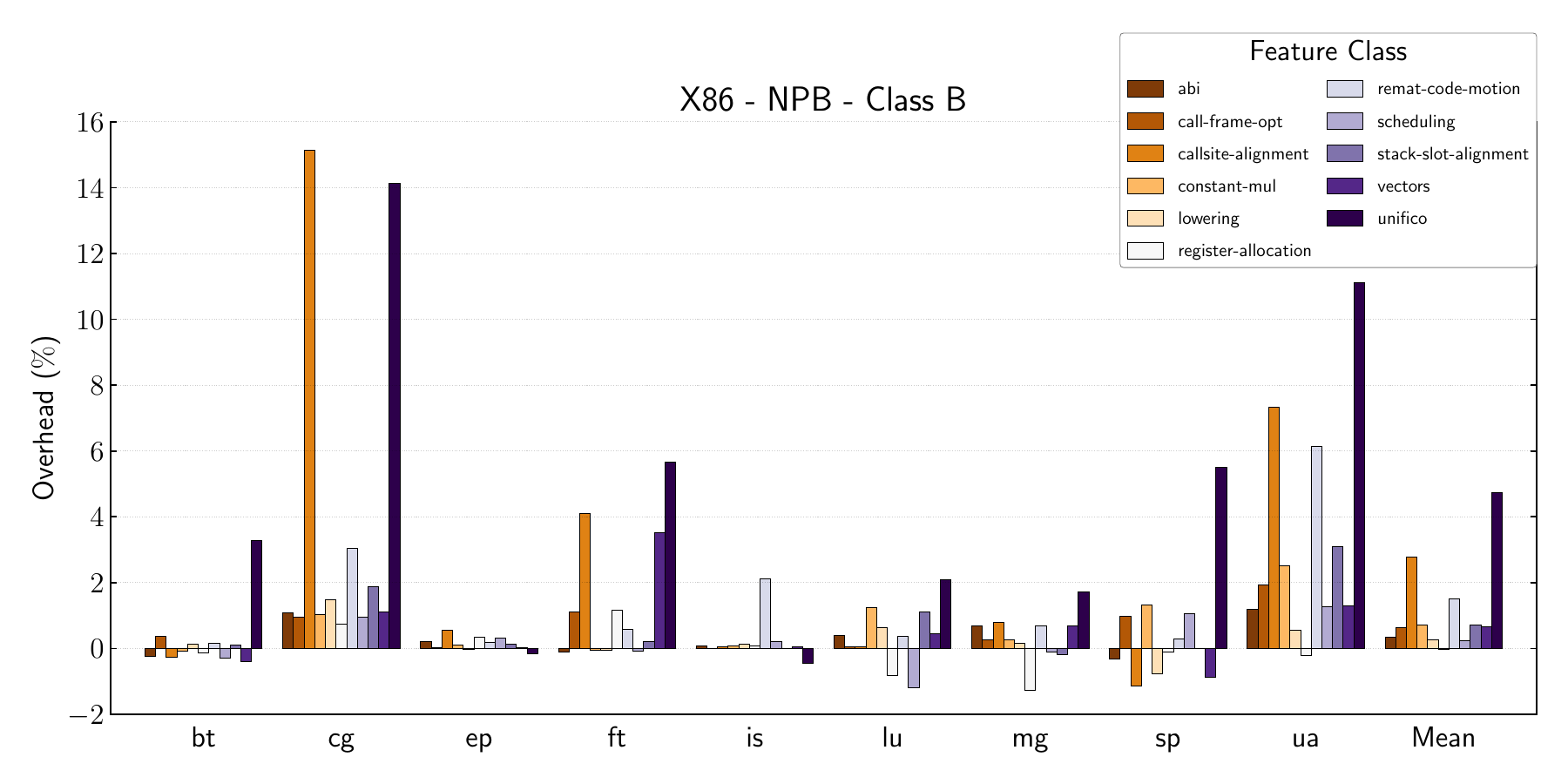}
        \caption{\pcarch{}}
        \label{fig:x86-breakdown}
    \end{subfigure}
    \begin{subfigure}[b]{\textwidth}
        \centering
        \includegraphics[width=0.8\linewidth]{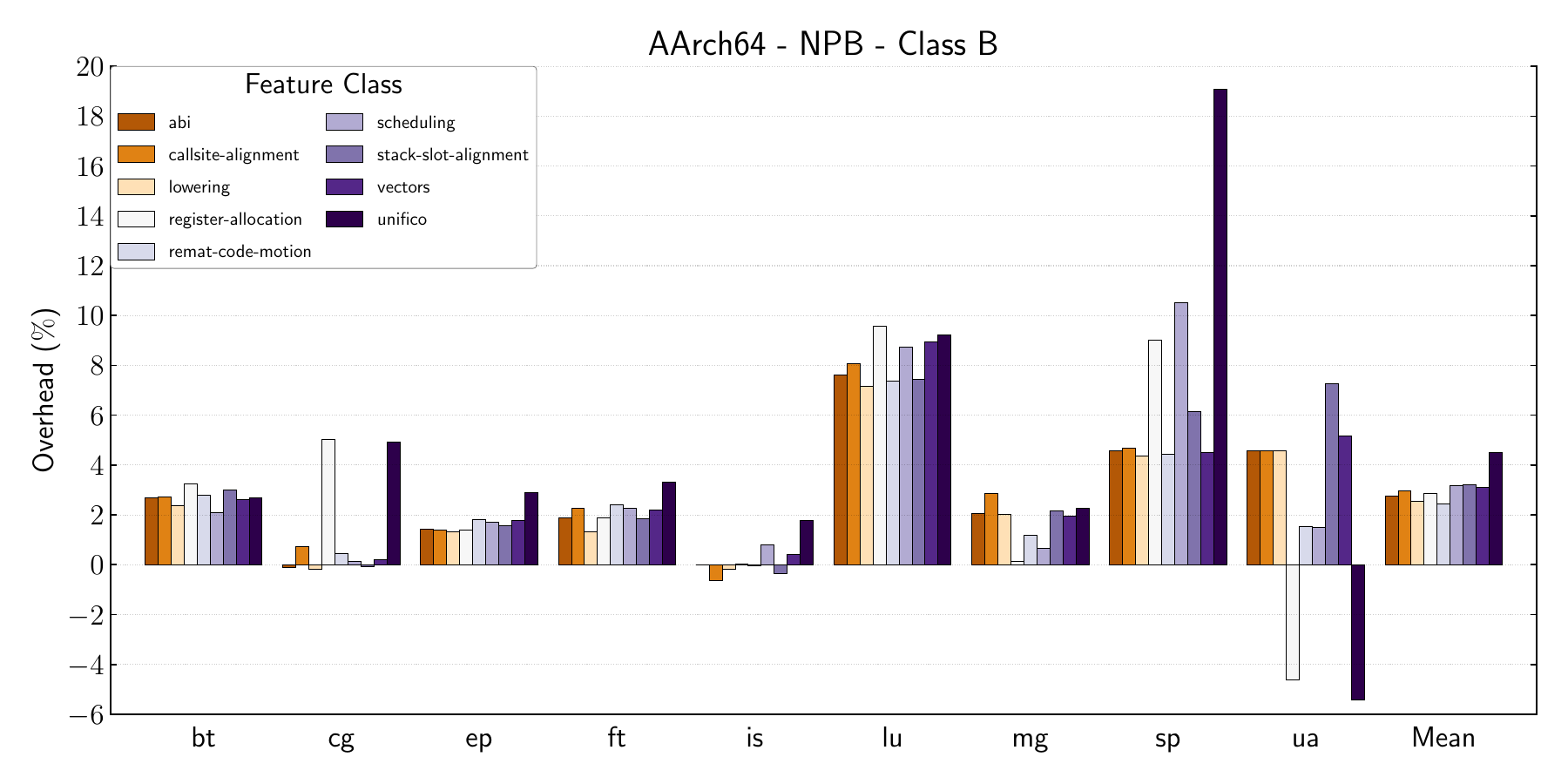}
        \caption{\armarch{}}
        \label{fig:arm-breakdown}
    \end{subfigure}
    \caption{Percentage overhead of \tool{} (last feature) and analysis of the overhead per feature category for \ac{npb} class B.}
\end{figure}

For each architecture, we evaluate the impact of \tool{}, relative to the unmodified compilation.
We run each benchmark three times and all standard deviations were below 1\%.
We compare the overhead introduced by \tool{} against an unmodified baseline.

Since \tool{} comprises different feature changes in the backend, to understand and potentially mitigate their overhead, we perform a comprehensive evaluation analysis for all the modification categories seen in~\Cref{tab:tool_compiler_transforms}.
To this end, we only apply each time the changes of one category and compare it with the baseline.
For a fair comparison, all categories are applied on top of the \ac{abi} changes (\Cref{subsec:abi_treatment}), since their true impact can only be highlighted when the final set of registers is available to them.
For example, the changes in the register allocation category might show minimal impact for \armarch{} if the full set of registers is available, but if we proactively examine their impact on the reduced set of registers, we will likely get more representative results.
We do this for \pcarch{} as well, although the changes in its \ac{abi} are less consequential.

%

Results are shown in~\Cref{fig:x86-breakdown} for \pcarch{} and~\Cref{fig:arm-breakdown} for \armarch{}, for the class B of the \ac{npb}.
For each benchmark, we show the overhead of all the categories (on top of the \ac{abi} changes, namely \textit{abi} in the legend) versus a baseline unmodified \llvm{} compiler.
We make five observations.
First, in the case of \pcarch{}, almost all benchmarks compiled with \tool{} demonstrate less than 5\% difference from the unmodified compiler.
Whereas in the case of \armarch{}, the performance has not noticeably degraded, except for the cases of \textsc{lu} and \textsc{sp}.
Second, in \pcarch{}, the total overhead (\textit{unifico}), seems to be an accumulation of the individual overheads from the categories.
Contrary, in \armarch{}, the individual overheads (which are on top of the \textit{abi} changes), are within a large portion of the total overhead, although combined they don't lead to an exorbitant performance change.
Third, in \pcarch{}, \textit{callsite-alignment} constitutes the most significant overhead factor, for \textsc{cg}, \textsc{ft}, and \textsc{ua}.
Then, \textit{remat-code-motion} follows.
Fourth, in \armarch{}, we observe more overhead in \textit{register-allocation} and \textit{scheduling}.
The other categories do not show much difference compared to the \textit{abi} category (which they already encapsulate), suggesting that their effect is not significant.
Finally, we observe a few speedups compared to the baseline.

In the following sections, we analyze the above observations, focusing on the aforementioned categories with the most impact, we explain the results, and propose solutions.

%
\input{071_callsite_alignment}

%
\input{073_register_pressure}

%
\input{075_mi_scheduling}

%
\input{076_overhead_classes}

%
\input{077_perf_portability}

\paragraph{Discussion.}
Further tuning cross\hyp{}architectural tradeoffs is left as future work.
For example, instead of imposing the two\hyp{}address format for \ac{fp} instructions in \armarch{}, which was needed because the architectures had misaligned spill slots with \ac{fp} values, we saw that padding the spill slots after the register allocation was sufficient, and caused much less overhead for \armarch{}.
Nevertheless, we envision end\hyp{}to\hyp{}end benefits at the system level enabled by finer migration granularity with no metadata and state transformation.
Finally, if an overhead is not acceptable, techniques like \ac{melf} binaries~\cite{tollner2023melf} allow the runtime to pick the right compile\hyp{}time variant of a function, exploiting the tradeoff between migration granularity and performance impact.

%
\input{078_sota_comparison}

%% file: 071_callsite_alignment.tex
\subsection{Callsite Alignment}\label{subsec:callsite-alignment}

\begin{figure}[t]
    \begin{subfigure}[h]{0.45\columnwidth}
        \includegraphics[width=0.9\columnwidth]{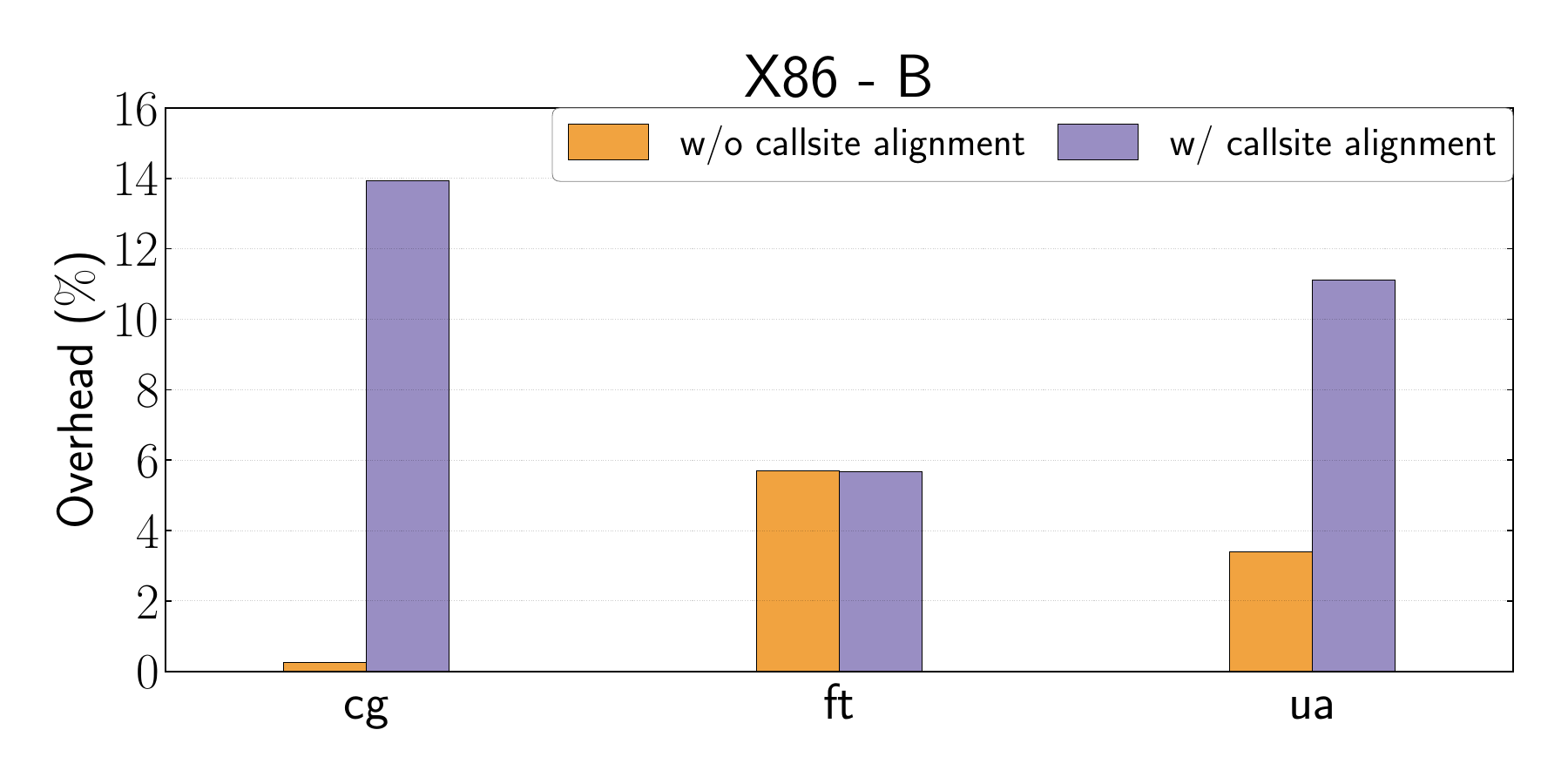}
        \caption{Execution time overhead with\slash{}without the callsite alignment in \tool{} for \pcarch{}.}
        \label{fig:x86-unifico-minus-callsite}
    \end{subfigure}
    \begin{subfigure}[h]{0.42\columnwidth}
        \includegraphics[width=0.9\columnwidth]{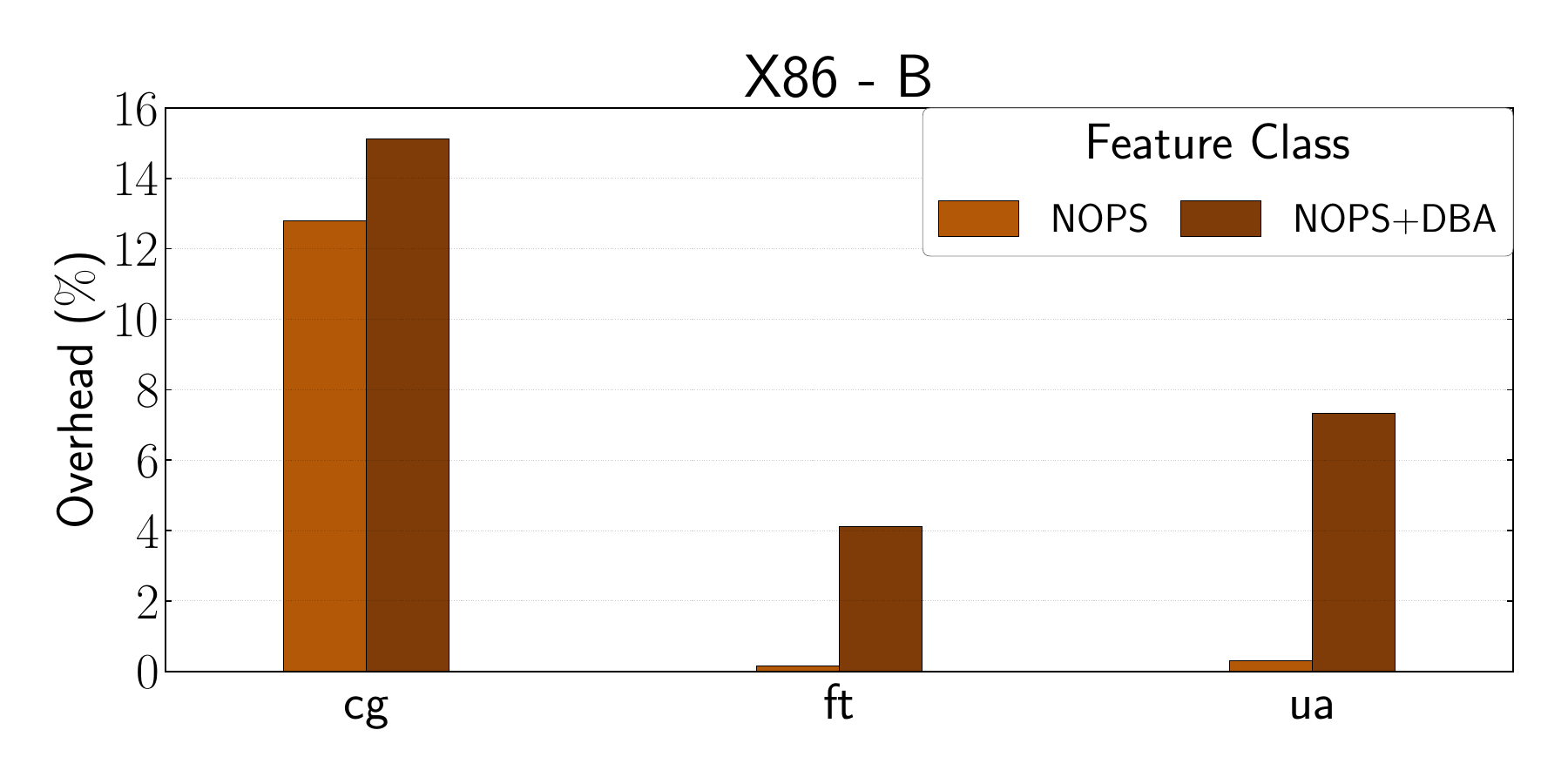}
        \caption{Analysis of the callsite alignment overhead for the most impacted benchmarks on x86.}
        \label{fig:callsite-breakdown}
    \end{subfigure}
    \caption{Impact of callsite alignment to \textsc{cg}, \textsc{ft}, and \textsc{ua}.}\label{fig:callsite-alignment}
\end{figure}

Considering~\Cref{fig:x86-breakdown}, we observe that \textsc{cg}, \textsc{ft}, and \textsc{ua} show the most notable overheads because of the alignment of callsites.
First, we check how much the callsite alignment is contributing to the total \tool{} impact.
\Cref{fig:x86-unifico-minus-callsite} compares the execution time overhead before and after aligning the callsites of the \tool{}\hyp{}compiled binaries.
We observe that the callsite alignment explains most of the total overhead for \textsc{cg}, and a large portion for \textsc{ua}.

\paragraph{Analysis of callsite alignment.}

To further pinpoint and mitigate the impact of callsite alignment, we examine the impact of all its components:
\begin{enumerate*}[label=\roman*)]
    \item insertion of \asm{nop} padding (\textit{NOPS}), and
    \item disabling the block alignment restrictions (\textit{DBA}).
\end{enumerate*}
The \asm{nops} are needed to insert the padding.
Disabling block alignment is needed because after calculating and inserting the padding, the compiler would potentially add additional padding to satisfy the alignment requirements of blocks.
This would break again the alignment between \pcarch{} and \armarch{} since it would require recalculating and inserting the padding, which could cause another padding change for the branch alignment, and so on.
We later show how to fix the last issue using more than two compilations.

\Cref{fig:callsite-breakdown} shows the impact of each component in the most affected \pcarch{} benchmarks.
First, we observe that the \asm{nop} insertion (NOPS) has significant impact on \textsc{cg}.
Second, disabling the alignment constraints on top of the \asm{nop} insertion (NOPS+DBA), causes the performance to further degrade in all benchmarks.
The first observation is explained partly by the number of \asm{nops} inserted in \textsc{cg}.
In \textit{conj\_grad}, its hottest function, there are 76 \asm{nops} inserted, meaning that they introduce significant more stalls to the processor.
Finally, the following section uses performance counters to better understand the impact the alignment.

\subsubsection{Top-Down Microarchitecture Analysis}

To better understand the overhead and how to mitigate it, we perform a \ac{tma}~\cite{yasin2014top, IntelVTune2024}.
\ac{tma} combines performance counters to evaluate how the \ac{cpu} pipeline slots are being used, classifying bottlenecks into four high-level categories (described below), and then optionally drilling deeper into each category.
To run \ac{tma} and access the deeper levels, we use \textit{toplev}, a \textsc{python} tool from the \textit{pmu-tools} suite~\cite{pmu-tools} invoking \textsc{perf}~\cite{linux_perf_2024} under the hood.

We first examine the top level of the analysis, estimating how much time the \ac{cpu} pipeline spends:
\begin{enumerate}[label=\roman*)]
    \item Stalling to issue new instructions ready to be executed (\textbf{Frontend Bound}).
    \item Stalling to recover from mispredicted instruction issues (\textbf{Bad Speculation}).
    \item Stalling to execute issued instructions and access the memory system (\textbf{Backend Bound}).
    \item Retiring instructions and committing their results (\textbf{Retiring}).
\end{enumerate}
Then, we can identify one node of interest and drill down using node-specific metrics into the subcomponents that are actually under more pressure inside the node itself.

\begin{figure}[htbp]
    \centering
    \begin{subfigure}[b]{0.49\textwidth}
        \centering
        \includegraphics[width=0.85\textwidth]{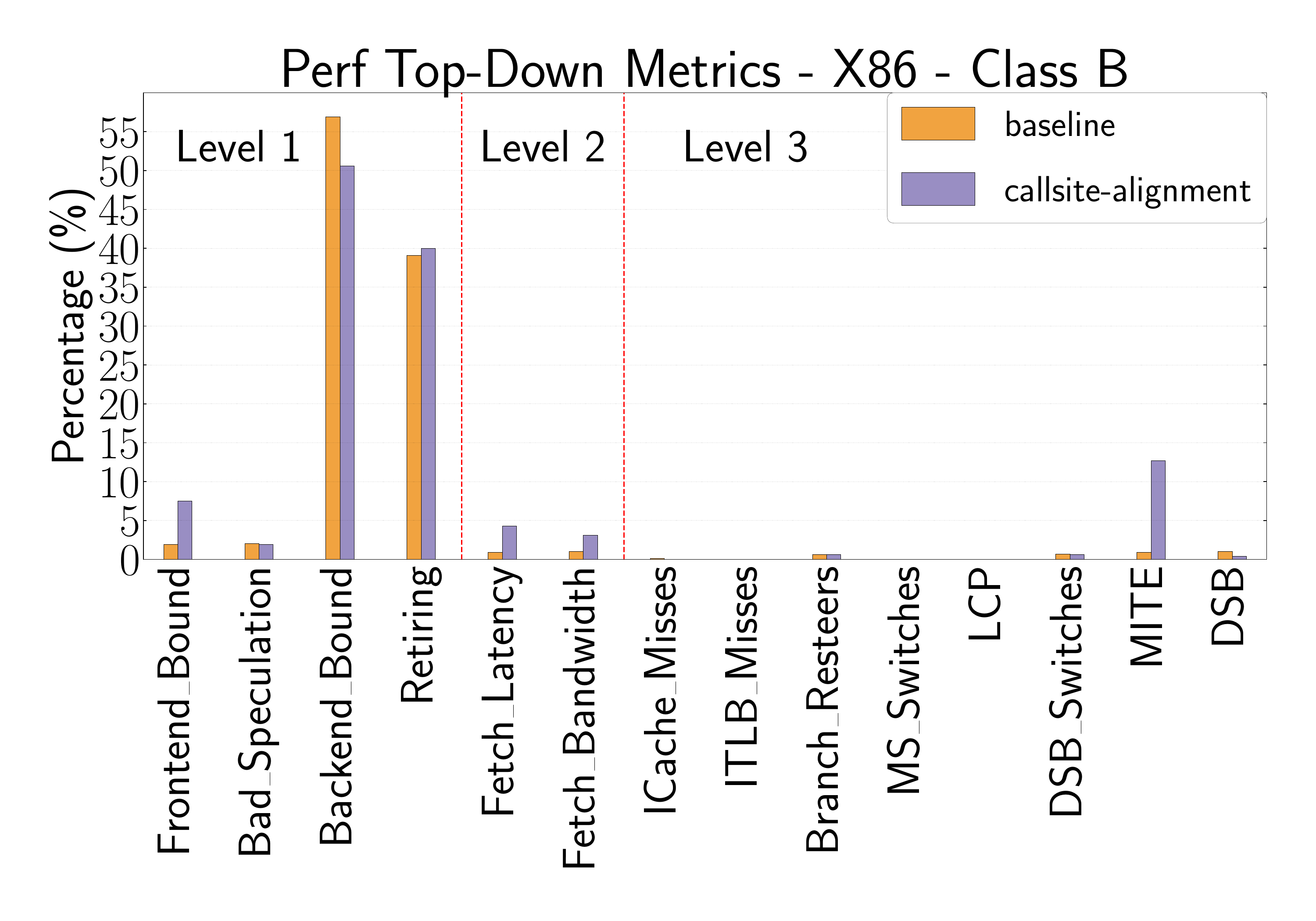}
        \caption{CG}
        \label{fig:figures/tma/cc_vs_vanilla_tma_cg}
    \end{subfigure}
    \begin{subfigure}[b]{0.49\textwidth}
        \centering
        \includegraphics[width=0.85\textwidth]{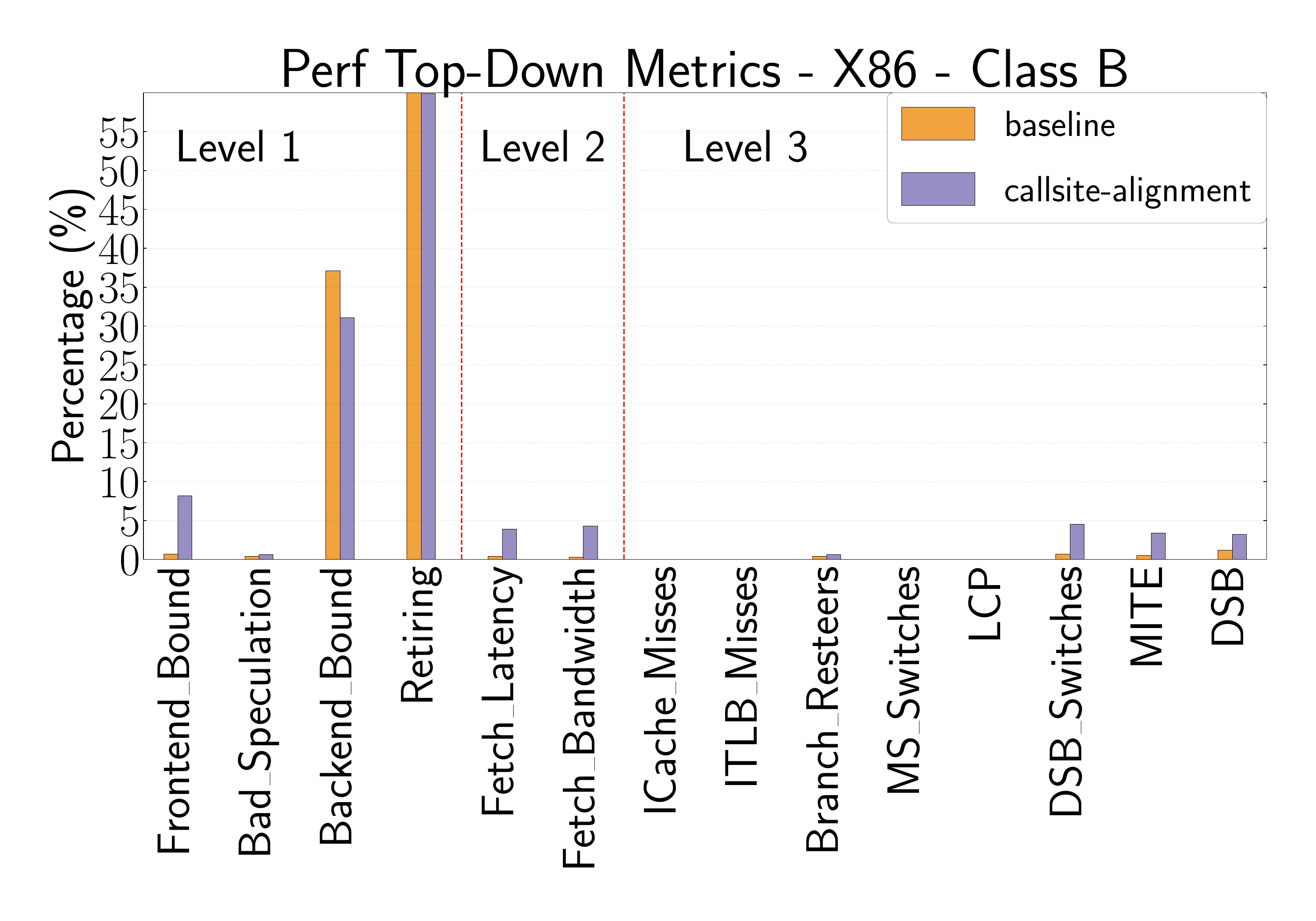}
        \caption{FT}
        \label{fig:figures/tma/cc_vs_vanilla_tma_ft}
    \end{subfigure}
    \begin{subfigure}[b]{0.49\textwidth}
        \centering
        \includegraphics[width=0.85\textwidth]{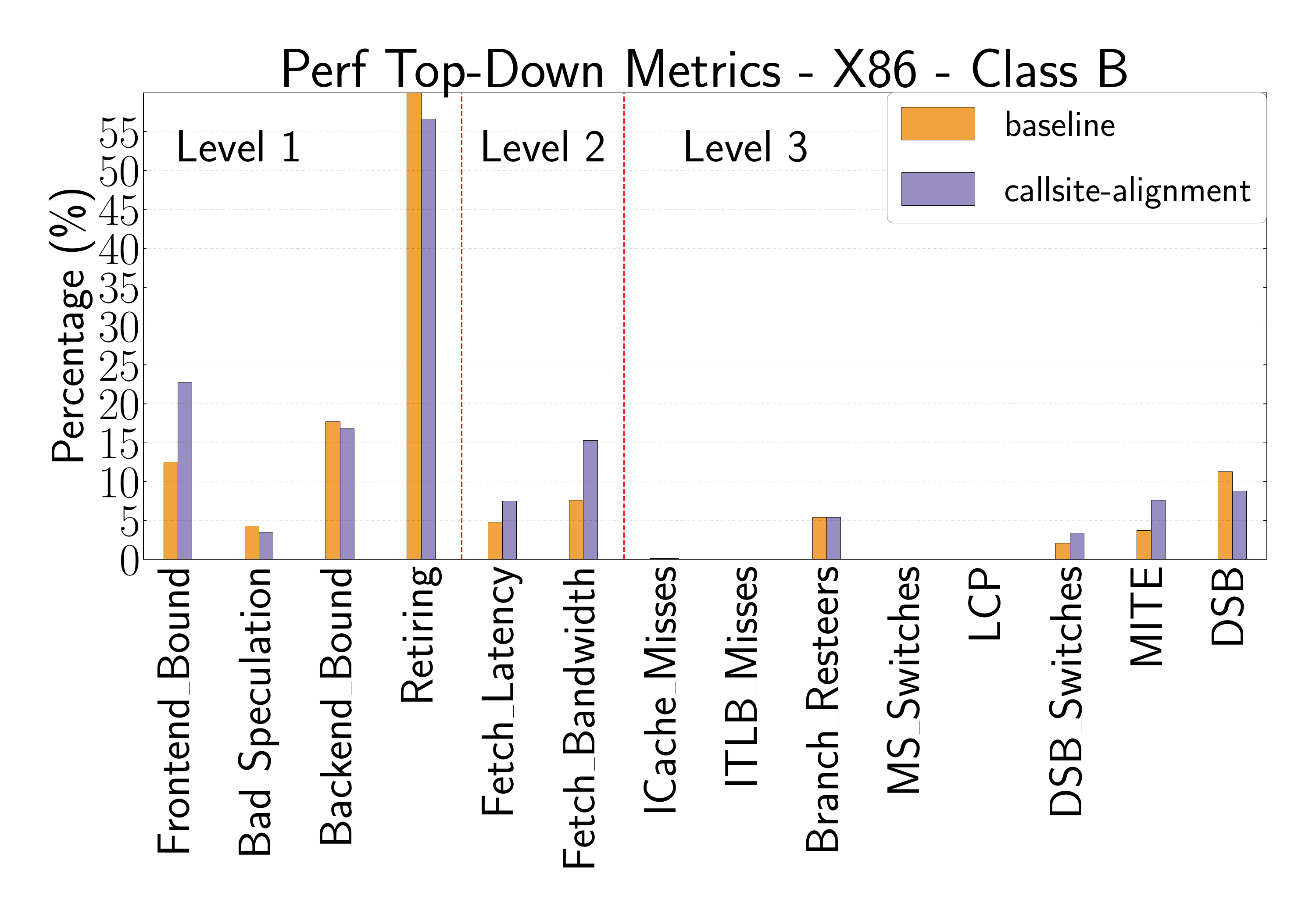}
        \caption{UA}
        \label{fig:figures/tma/cc_vs_vanilla_tma_ua}
    \end{subfigure}
    \begin{subfigure}[b]{0.49\textwidth}
        \centering
        \includegraphics[width=0.85\textwidth]{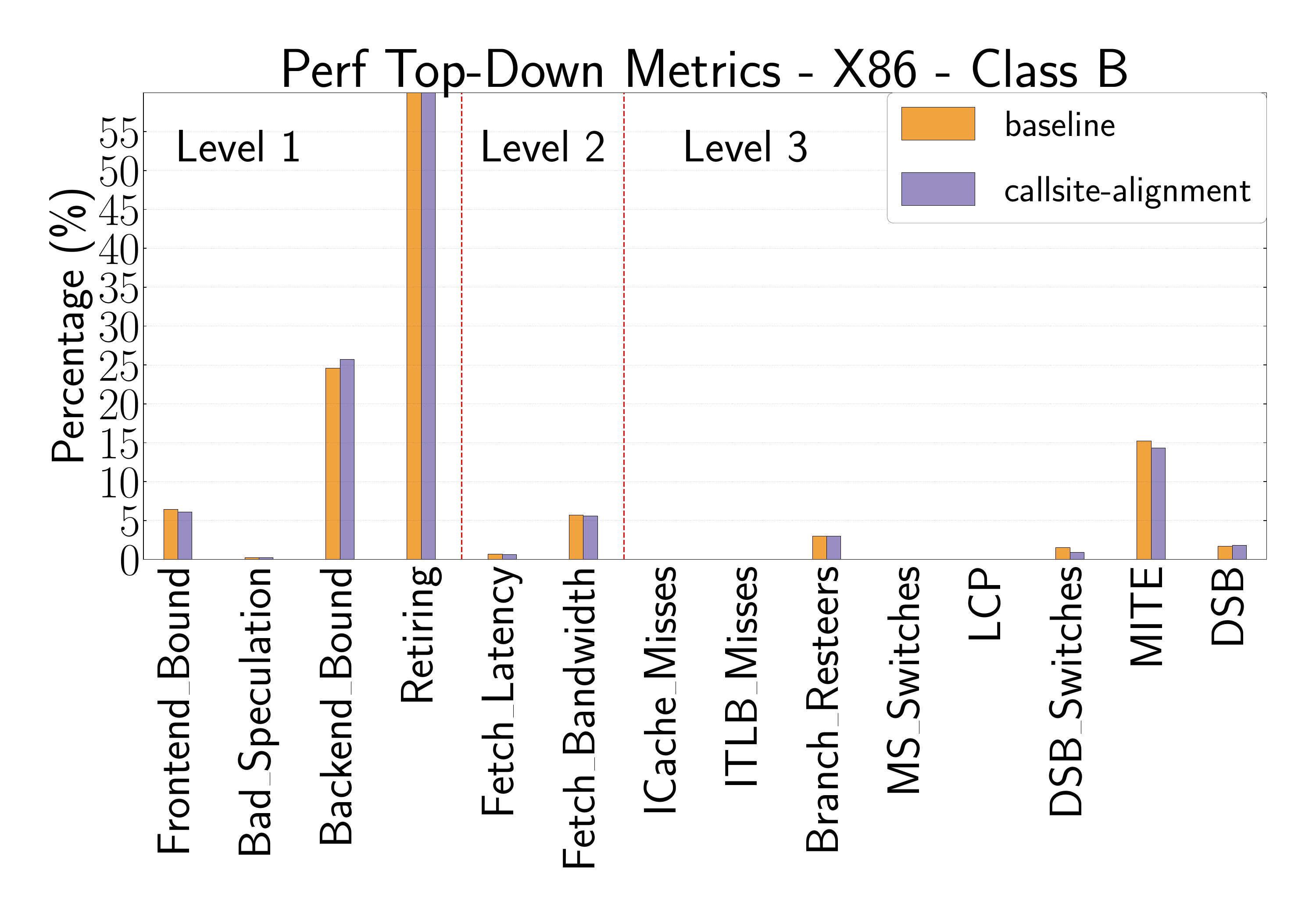}
        \caption{BT}
        \label{fig:figures/tma/cc_vs_vanilla_tma_bt}
    \end{subfigure}

    \caption{Comparison of \ac{tma} metrics, between the callsite alignment and a baseline compiler, for the most affected \textsc{cg}, \textsc{ft}, and \textsc{ua}, versus the unaffected \textsc{bt}.
    For the \ac{tma}, we show level 1 (columns 1--4), the frontend-related metrics of level 2 (columns 5--6), and their expansion in  level 3 (rest of the columns).}
    \label{fig:cc_vs_vanilla_tma}
\end{figure}

The results of the \ac{tma} analysis are shown in~\Cref{fig:cc_vs_vanilla_tma}, where we compare the \ac{tma} metrics between the callsite alignment and a baseline compiler, for the most affected benchmarks of~\Cref{fig:x86-unifico-minus-callsite} against a relatively unaffected benchmark (\textsc{bt}).
We make three observations.
First, in level 1 (columns 1--4), we observe that the first three benchmarks become more frontend-bound after applying the callsite alignment, whereas \textsc{BT} is not impacted notably.
Therefore, we expand only the frontend-related metrics of level 2, fetch latency and fetch bandwidth (columns 5--6), in level 3 (columns 7--14), ignoring the rest nodes of level 1.
Second, in level 3, \textsc{cg}, \textsc{ft}, and \textsc{ua} show higher pressure on the \ac{mite} and less on the \ac{dsb} Switches, suggesting that fewer opcodes are being cached during decoding in the \ac{dsb} and need to switch to \ac{mite}~\cite{intel2023_vtune_cpu_metrics}.
Third, \textsc{ft} demonstrates increased pressure for both \ac{dsb} and \ac{mite}.
Finally, the affected benchmarks do not show differences in the other frontend metrics (like icache misses, iTLB misses, etc.).

The increased pressure in the \ac{dsb} Switches and in \ac{mite} can be attributed to the alignment issues created by the extra \asm{nops} in tandem with the relaxation of alignment constraints.
For example, if the code of a hot loop is not aligned to the recommended boundaries by the architecture, then control may flow out of the region cached by the \ac{dsb}, leading to thrashing and requiring more work from \ac{mite}~\cite{intel2023_vtune_cpu_metrics,bakhvalov2018code}.
Overall, we conclude that if adding the \asm{nops} and not abiding to the alignment constraints decrease the fetch bandwidth in the pipeline, then performance will be impacted (at least for \pcarch{} which has more alignment constraints).

\subsubsection{Improving the performance}

Based on the above analysis, we apply the following fixes during code generation of padding.

\paragraph{Jump over padding}

\input{figure_callsite_jump}

The \asm{nop} instructions inserted might be keeping the processor idle, introducing stalls in the pipeline.
To reduce their impact, we transform the code so that before the \asm{nops} there is an unconditional jump instruction that skips over them.
We enable this in both architectures, i.e., by inserting a \asm{b} instruction for \armarch{} and a \asm{jmp} instruction for  \pcarch{} respectively.
The \asm{b} instruction is an unconditional branch with a $\pm128$MB offset~\cite{holdings2022arm}, while the \asm{jmp} we use is a near, unconditional, relative, 4-byte memory jump~\cite{guide2011intel}.
Since these instructions constitute padding by themselves, we take them into account when calculating the padding, if the compiler decides to emit them.
%
%
The compiler only emits them if the \asm{nop} number is larger than the branch instruction size itself (4 bytes for \armarch{} and 5 for \pcarch{}).
In practice, this means that there will never be a case where both \acp{isa} will need an unconditional jump.

An example where the compiler emits a \asm{nop} instruction in \pcarch{}, is shown in~\Cref{fig:callsite_jmp}.
In~\Cref{fig:callsite_jmps2}, \pcarch{} needed a relatively large padding of 17 \asm{nops} to align with \armarch{} in~\Cref{fig:callsite_jmps1}.
This is not uncommon, since \pcarch{}, being a \ac{cisc} architecture, usually has a more condensed code section than \armarch{}, thus needing more padding in many cases to align its callsites.
To limit the overheads caused by \asm{nops}, in~\Cref{fig:callsite_jmps4} the compiler inserts a \asm{jmp} instruction (5 bytes) before an ensemble of 12 bytes of \asm{nops}, instead of emitting 17 byte \asm{nops}.
This change improved \textsc{cg} by $2\%$, while improving the whole suite slightly on average.

\paragraph{Accumulated padding}

To satisfy both the alignment constraints recommended by the \ac{cpu} vendors, and the callsite alignment needed by \tool{}, we can apply a multi-pass compilation technique.
%
%
The key idea is to cache the calculated \asm{nop} padding for the callsites in every compilation, then allow the compiler to emit further padding to satisfy the block alignment and, finally, repeat another padding calculation on top of the previous one.
Applying this enough times converges to the right padding, at the expense of a slower compilation.
In our experiments, this technique improved the performance by an average of $1\%-2\%$, converging after 7 iterations on average, and with only slightly increased \asm{nop} insertion for the best padding.
Nevertheless, we keep it off by default and exclude it in our total results in~\Cref{subsubsec:final-overhead}, to avoid the compilation overhead, but it can be activated optionally when compile time can be traded off for a lower alignment impact.

%% file: figure_callsite_jump.tex
\begin{figure}[t]
  \centering
  \begin{subfigure}[t]{0.48\columnwidth}
    \begin{adjustwidth}{0.10cm}{}
    \centering
\begin{lstlisting}[style=asmstyle]
1000: main:
1000: SP = SP - 32
...
1120: call hot_func()
|\tikzmark{align_arm_ret_before_start4}|1124|\tikzmark{align_arm_ret_before_end4}|: // after call
\end{lstlisting}
    \end{adjustwidth}
    \caption{}\label{fig:callsite_jmps1}
  \end{subfigure}
  \begin{subfigure}[t]{0.48\columnwidth}
    \begin{adjustwidth}{0.0cm}{}
    \centering
\begin{lstlisting}[style=asmstyle, numbers=none]
1000: main:
1000: push FP
...
1102: call hot_func()
|\tikzmark{align_x86_ret_before_start4}|1107|\tikzmark{align_x86_ret_before_end4}|: // after call
\end{lstlisting}
    \end{adjustwidth}
    \caption{}\label{fig:callsite_jmps2}
  \end{subfigure}
  \hfill
  \begin{subfigure}[t]{0.48\columnwidth}
    \begin{adjustwidth}{0.10cm}{}
\begin{lstlisting}[style=asmstyle]
1000: main:
1000: SP = SP - 32
...

1118: call hot_func()
|\tikzmark{align_arm_ret_before_start5}|1124|\tikzmark{align_arm_ret_before_end5}|: // after call
\end{lstlisting}
    \end{adjustwidth}
    \caption{}\label{fig:callsite_jmps3}
  \end{subfigure}
  \begin{subfigure}[t]{0.48\columnwidth}
    \begin{adjustwidth}{0.0cm}{}
\begin{lstlisting}[style=asmstyle, numbers=none]
1000: main:
1000: push FP
...
|\tikzmark{align_x86_ret_before_start5}|1102: jmp 1119|\tikzmark{align_x86_ret_before_end5}|
|\tikzmark{align_x86_ret_before_start6}|1107: twelve_byte_nop|\tikzmark{align_x86_ret_before_end6}|
1119: call hot_func()
|\tikzmark{align_x86_ret_before_start7}|1124|\tikzmark{align_x86_ret_before_end7}|: // after call
\end{lstlisting}
    \end{adjustwidth}
    \caption{}\label{fig:callsite_jmps4}
  \end{subfigure}
  \caption{Callsite alignment with jump in pseudo\hyp{}assembly before (top) and after (bottom) \tool{}'s operation for \armback{} (left) and \pcback{} (right).
  \pcarch{} needed a 17-byte padding so, instead, the compiler emits a 5-byte \asm{jmp} to the \asm{call} before a 12-byte padding.
  %
  %
  }
  \label{fig:callsite_jmp}
  \begin{tikzpicture}[remember picture,overlay]
    \begin{scope}
      \tikzset{hilite_unaligned node/.style 2 args = {draw=none, fill=carnationpink, inner sep=0pt, fill opacity=0.3, yshift=2pt, minimum height=8pt, fit=(#1) (#2)}}
      \tikzset{hilite_aligned node/.style 2 args = {draw=none, fill=applegreen, inner sep=0pt, fill opacity=0.3, yshift=2pt, minimum height=8pt, fit=(#1) (#2)}}
      \node[hilite_unaligned node={pic cs:align_arm_ret_before_start4}{pic cs:align_arm_ret_before_end4}] {};
      \node[hilite_unaligned node={pic cs:align_x86_ret_before_start4}{pic cs:align_x86_ret_before_end4}] {};
      \node[hilite_aligned node={pic cs:align_x86_ret_before_start5}{pic cs:align_x86_ret_before_end5}] {};
      \node[hilite_aligned node={pic cs:align_x86_ret_before_start6}{pic cs:align_x86_ret_before_end6}] {};
      \node[hilite_aligned node={pic cs:align_x86_ret_before_start7}{pic cs:align_x86_ret_before_end7}] {};
      \node[hilite_aligned node={pic cs:align_arm_ret_before_start5}{pic cs:align_arm_ret_before_end5}] {};
    \end{scope}
  \end{tikzpicture}
\end{figure}

%% file: 073_register_pressure.tex
\subsection{Register Pressure and Register Allocation}\label{subsec:reg-pressure}

We examine the impact of limiting the number of general-purpose and floating-point registers available during code generation, specifically in \armarch{}.

\paragraph{Register Pressure and Rematerialization.}
We define \textit{register pressure} as the number of simultaneous live values in registers at a specific instruction, and we use the number of spilled values in the stack as a proxy to estimate it~\cite{chaitin1981register,briggs1994improvements,touati2002register}.
This makes register pressure a local instruction property.
For example, spilling a value means keeping a value in memory instead of a register, which reduces register pressure.
Thus, if altering code generation leads to more spills and refills to or from the stack, we say that register pressure was decreased (and vice-versa).
Finally, \textit{rematerialization} (\Cref{subsubsec:remat_code_motion}) generally increases register pressure, which further compounds the tradeoff between choosing to apply it and increase register pressure (with potentially more future spills), or allow directly the spill to the stack (increasing the memory pressure).

\subsubsection{Impact of reducing the registers}

As a first step, we measure the impact of only reducing the registers in \armarch{}, versus a baseline \llvm{} compiler, without other \tool{} features.
We only include the remaining \ac{abi} changes, such as which registers are callee-saved or temporary, since these are directly relevant to register usage and spilling.
We use the dynamic instruction distribution as a proxy to changes in register pressure.
There are five \armarch{} opcode families affecting register pressure:
\begin{enumerate}[label=\roman*)]
    \item \textbf{Load instructions} indicate either a load from a spill slot in the stack or a load of a global symbol or of a constant (e.g., \asm{ldp}/\asm{ldr}).
    \item \textbf{Store instructions} indicate a store to a spill slot.
    \item \textbf{Arithmetic instructions} to compute load/store addresses or other values (e.g., \asm{add}).
    \item \textbf{Move immediate instructions} used for moving immediates to registers (e.g., \asm{movk}).
    \item \textbf{Address calculation instructions} used for materializing addresses (e.g., \asm{adrp}).
\end{enumerate}

Note that these instructions serve other purposes as well, but we focus on the ones related to register pressure.
Load and store instructions are relevant because they implement spills and refills.
Arithmetic instructions are relevant because of their role in spilling and rematerialization.
For example, more spills increase arithmetic instructions since more offsets to the stack pointer will be needed.
Also, arithmetic instructions take part in rematerialization to recalculate values.
Move immediate instructions are used in rematerialization of global values or constants.
Similarly, address calculation instructions are used in rematerialization of memory addresses (that are not stack related).

\begin{figure}[t]
    \centering
    \begin{subfigure}[b]{0.32\textwidth}
        \centering
        \includegraphics[width=\textwidth]{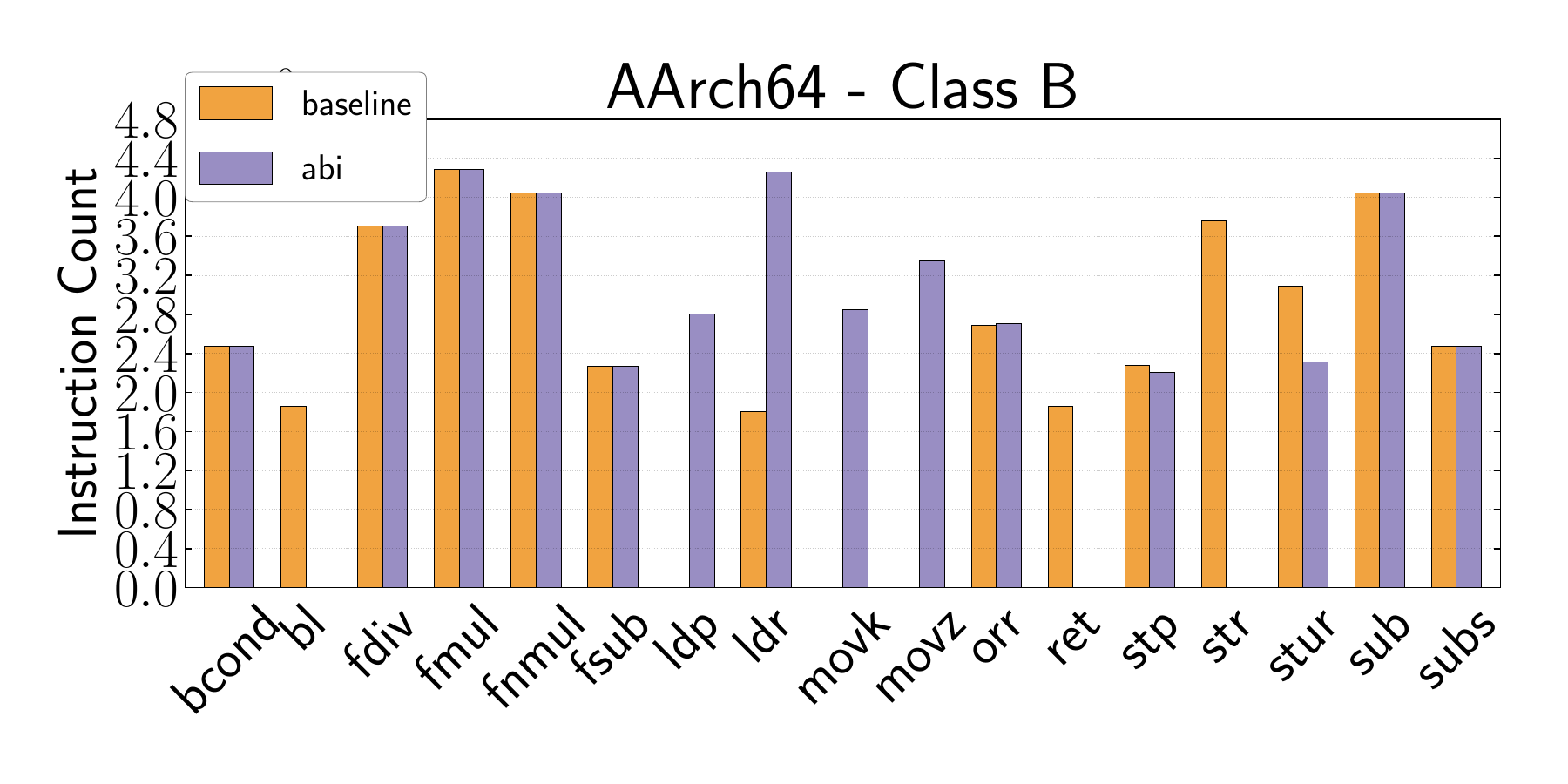}
        \caption{BT}
        \label{fig:figures/cc_vs_vanilla_opcodes_bt}
    \end{subfigure}
    \hfill
    \begin{subfigure}[b]{0.32\textwidth}
        \centering
        \includegraphics[width=\textwidth]{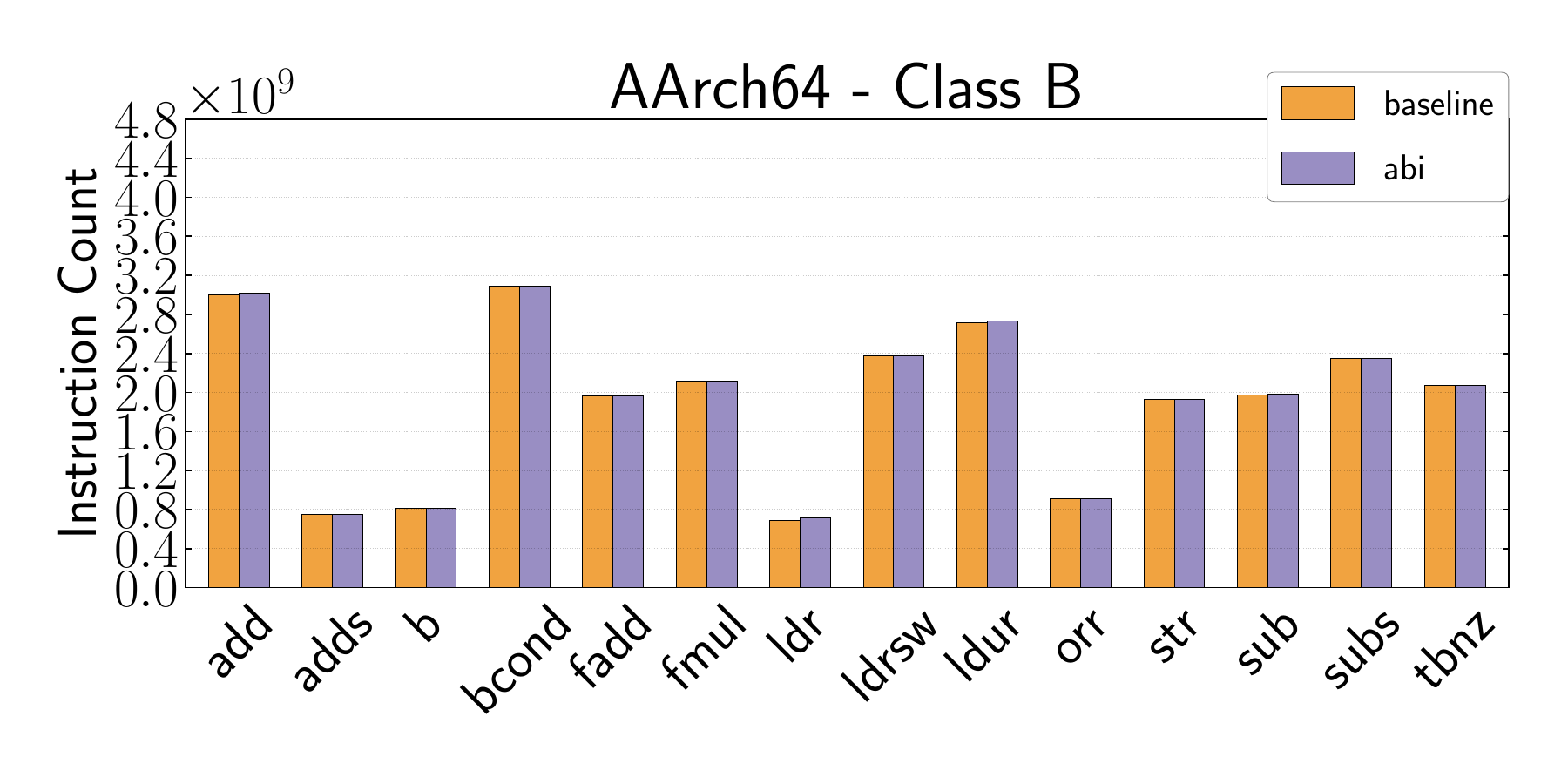}
        \caption{CG}
        \label{fig:figures/cc_vs_vanilla_opcodes_cg}
    \end{subfigure}
    \hfill
    \begin{subfigure}[b]{0.32\textwidth}
        \centering
        \includegraphics[width=\textwidth]{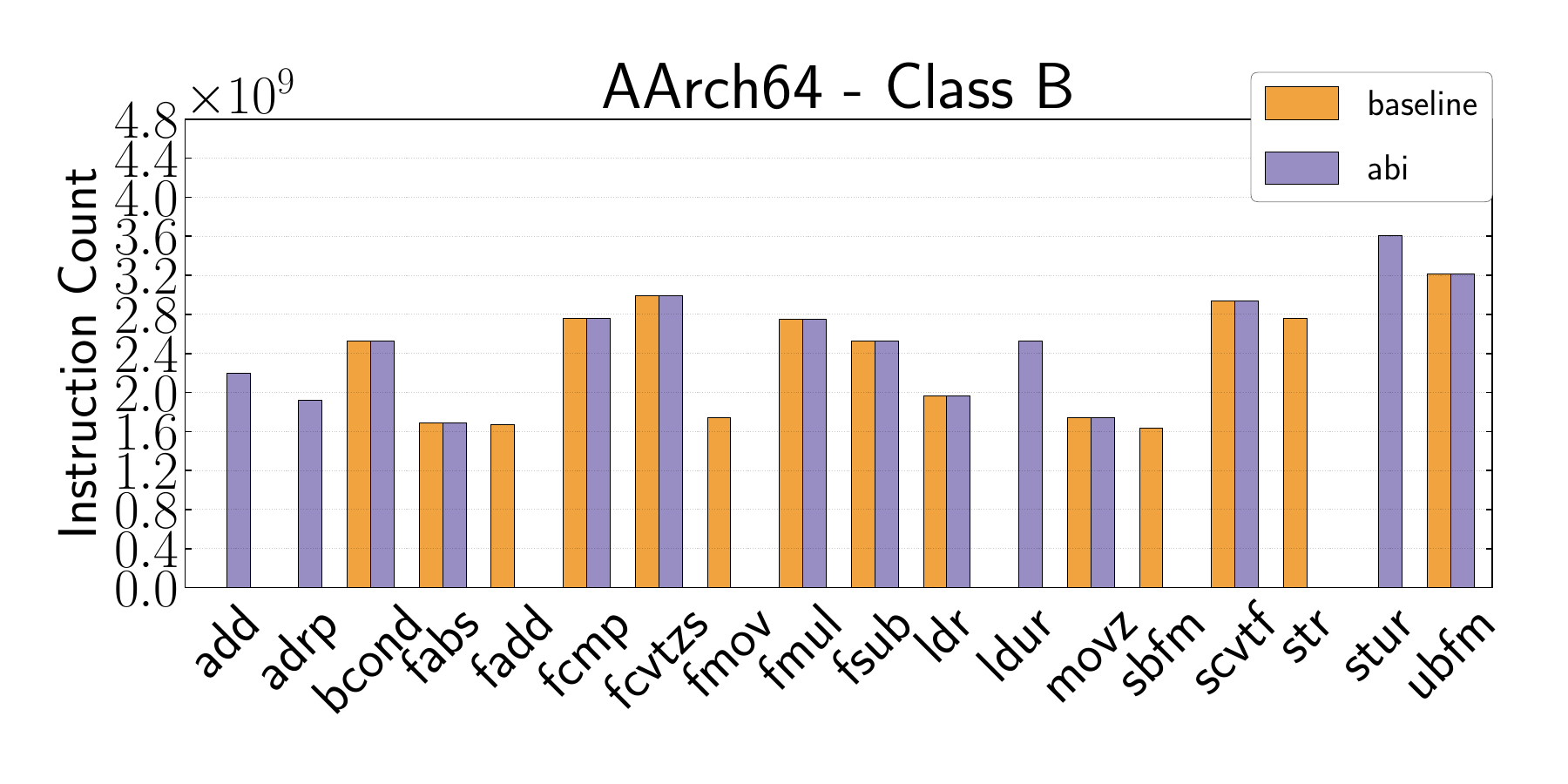}
        \caption{EP}
        \label{fig:figures/cc_vs_vanilla_opcodes_ep}
    \end{subfigure}

    \vspace{1em}
    \begin{subfigure}[b]{0.32\textwidth}
        \centering
        \includegraphics[width=\textwidth]{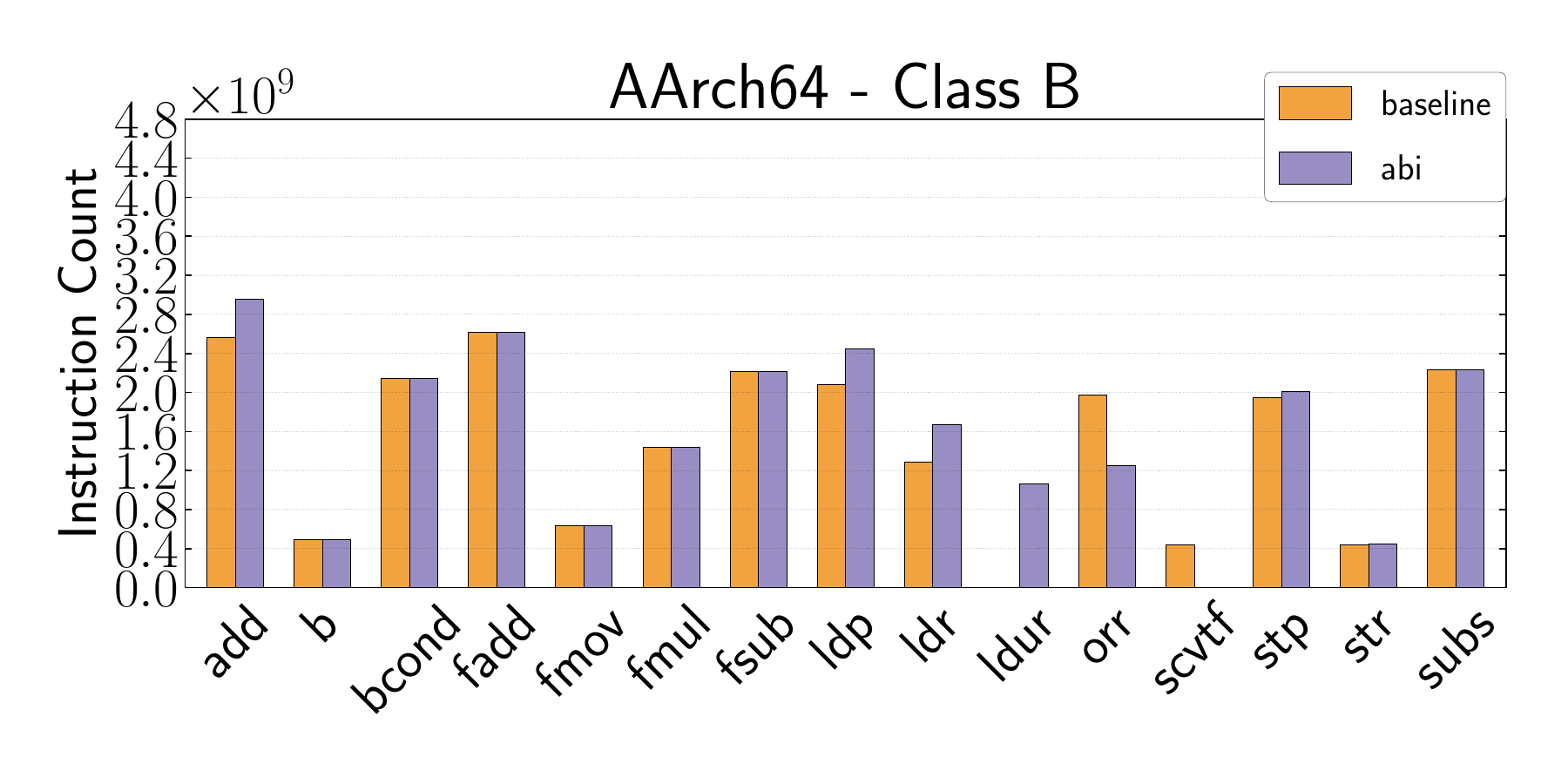}
        \caption{FT}
        \label{fig:figures/cc_vs_vanilla_opcodes_ft}
    \end{subfigure}
    \hfill
    \begin{subfigure}[b]{0.32\textwidth}
        \centering
        \includegraphics[width=\textwidth]{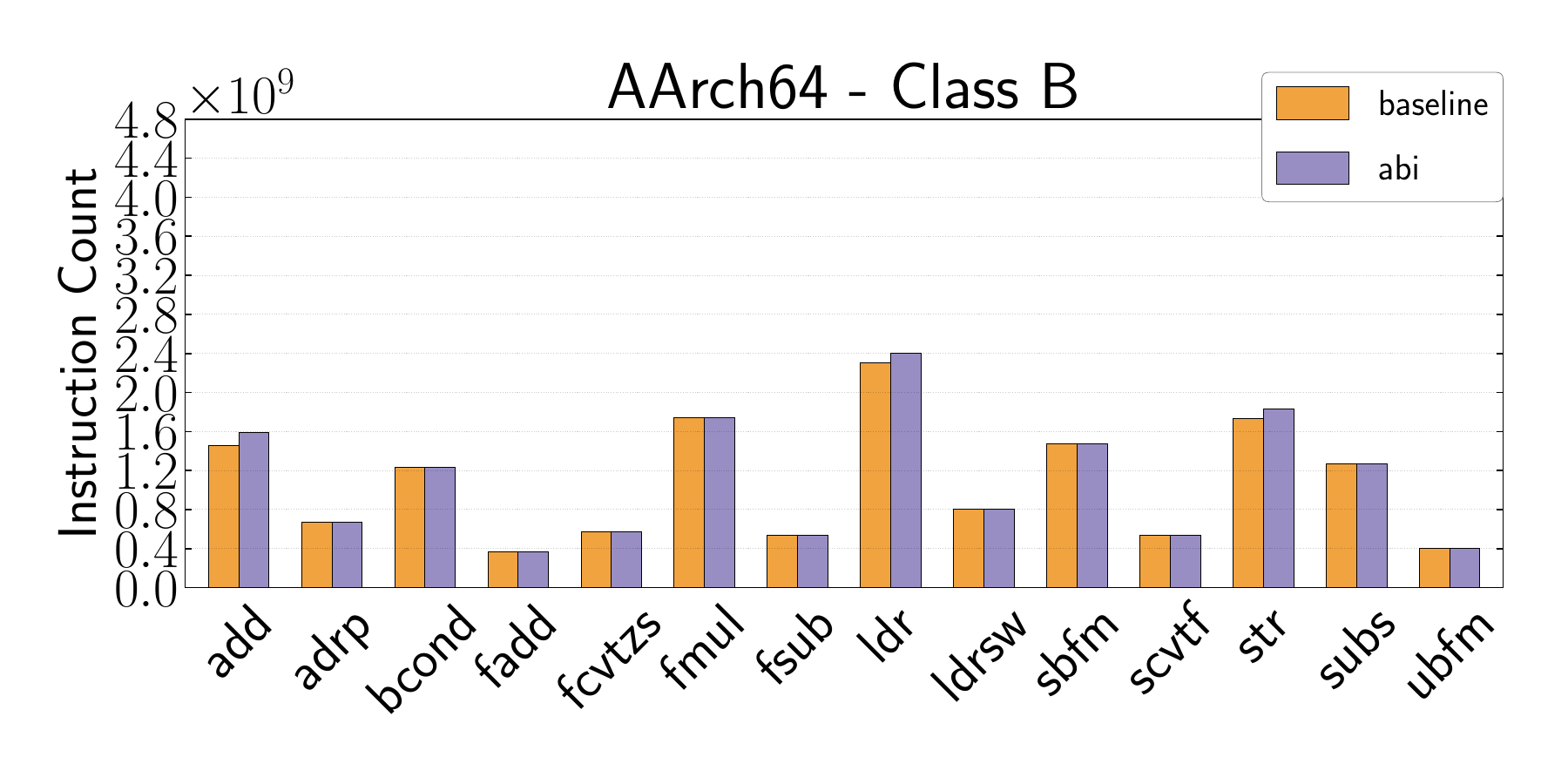}
        \caption{IS}
        \label{fig:figures/cc_vs_vanilla_opcodes_is}
    \end{subfigure}
    \hfill
    \begin{subfigure}[b]{0.32\textwidth}
        \centering
        \includegraphics[width=\textwidth]{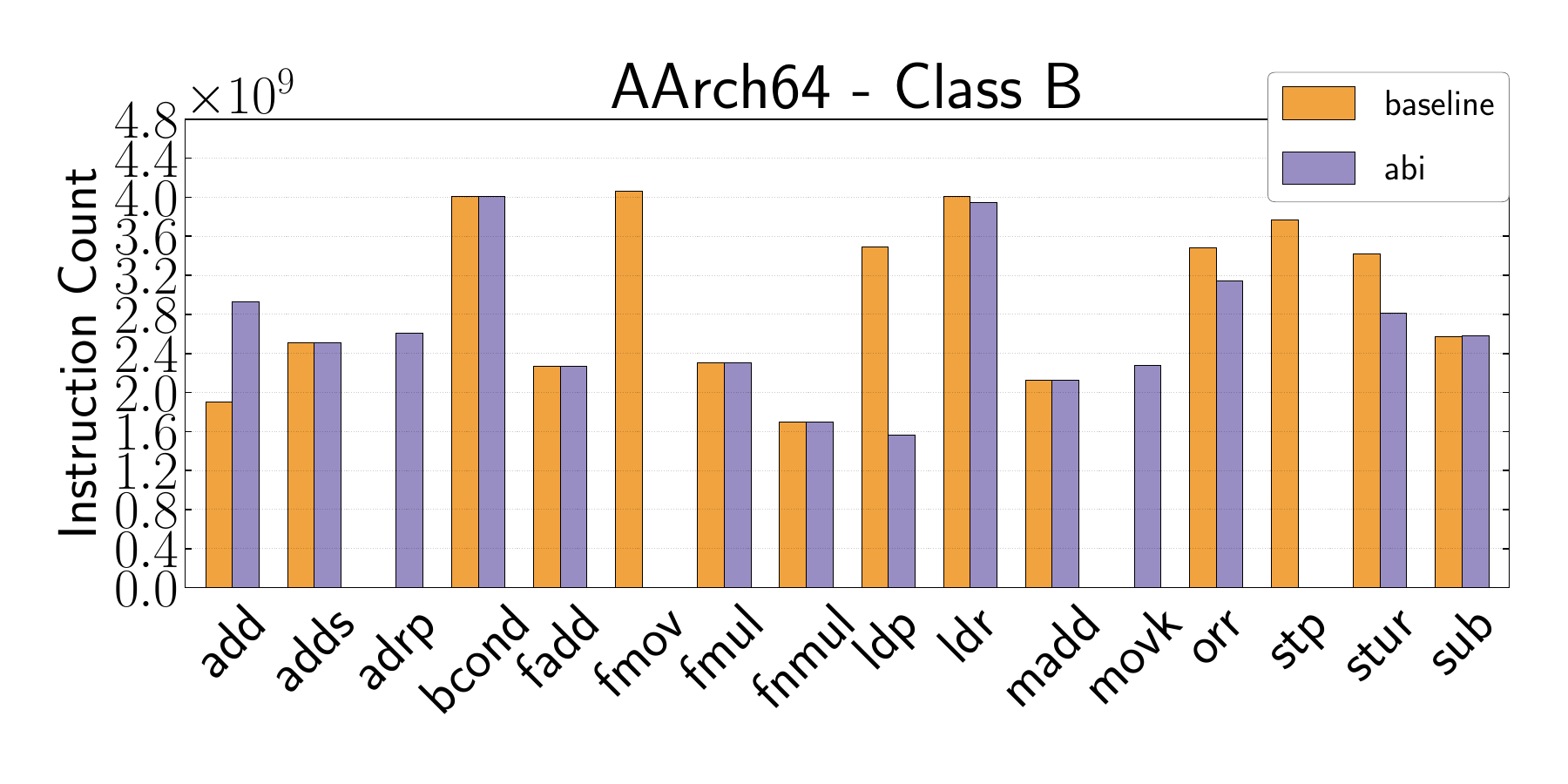}
        \caption{LU}
        \label{fig:figures/cc_vs_vanilla_opcodes_lu}
    \end{subfigure}

    \vspace{1em}
    \begin{subfigure}[b]{0.32\textwidth}
        \centering
        \includegraphics[width=\textwidth]{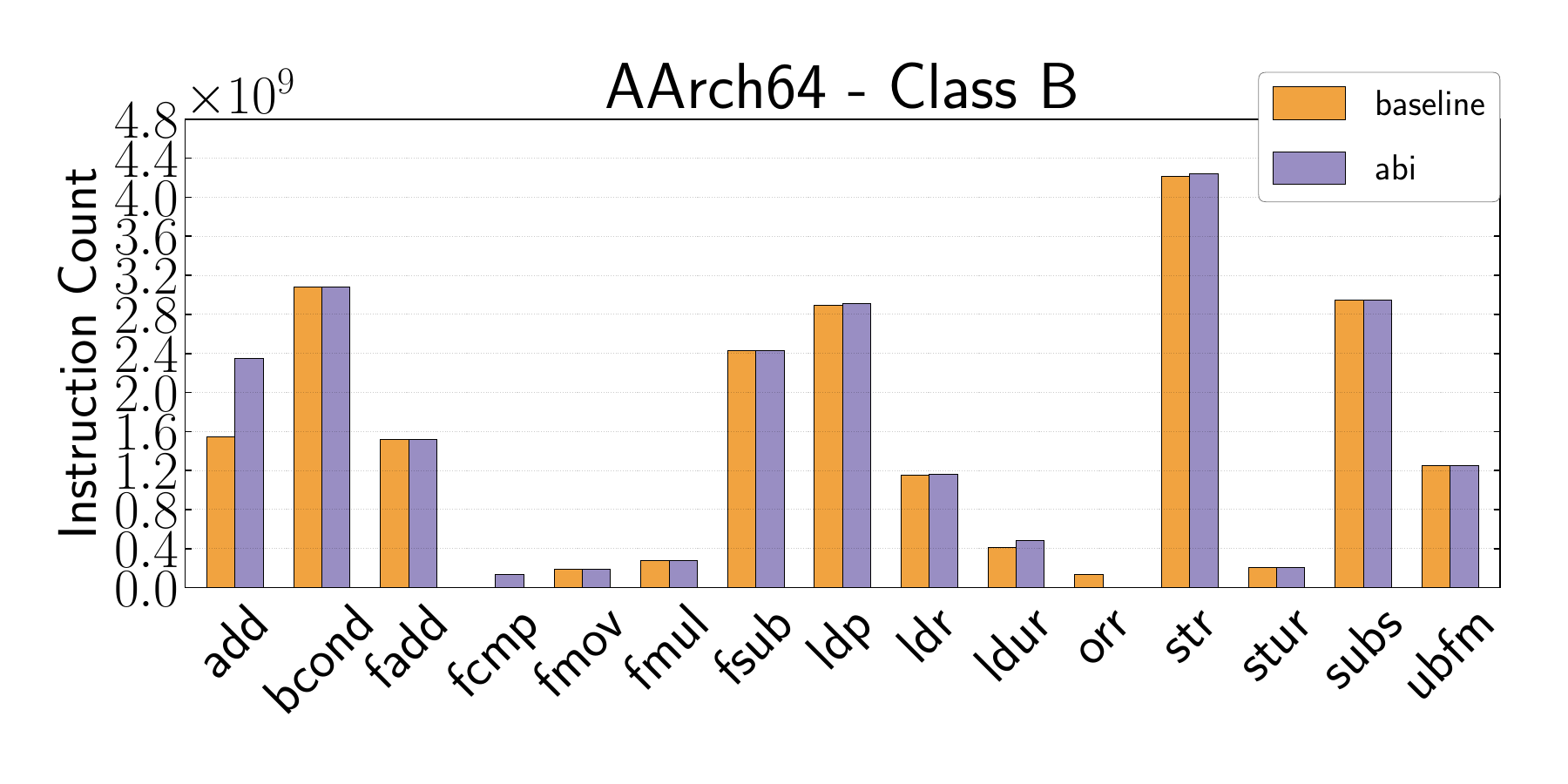}
        \caption{MG}
        \label{fig:figures/cc_vs_vanilla_opcodes_mg}
    \end{subfigure}
    \hfill
    \begin{subfigure}[b]{0.32\textwidth}
        \centering
        \includegraphics[width=\textwidth]{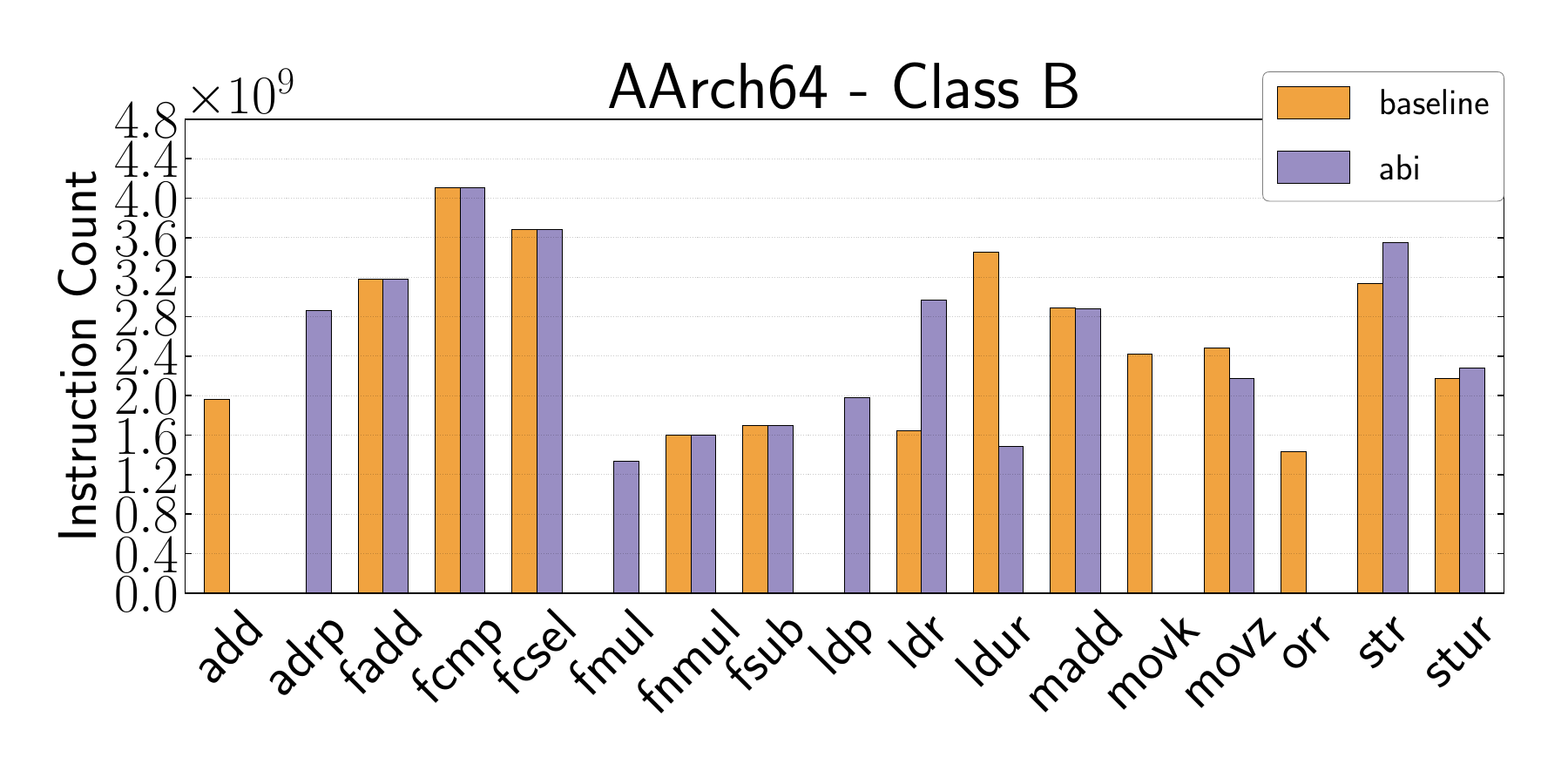}
        \caption{SP}
        \label{fig:figures/cc_vs_vanilla_opcodes_sp}
    \end{subfigure}
    \hfill
    \begin{subfigure}[b]{0.32\textwidth}
        \centering
        \includegraphics[width=\textwidth]{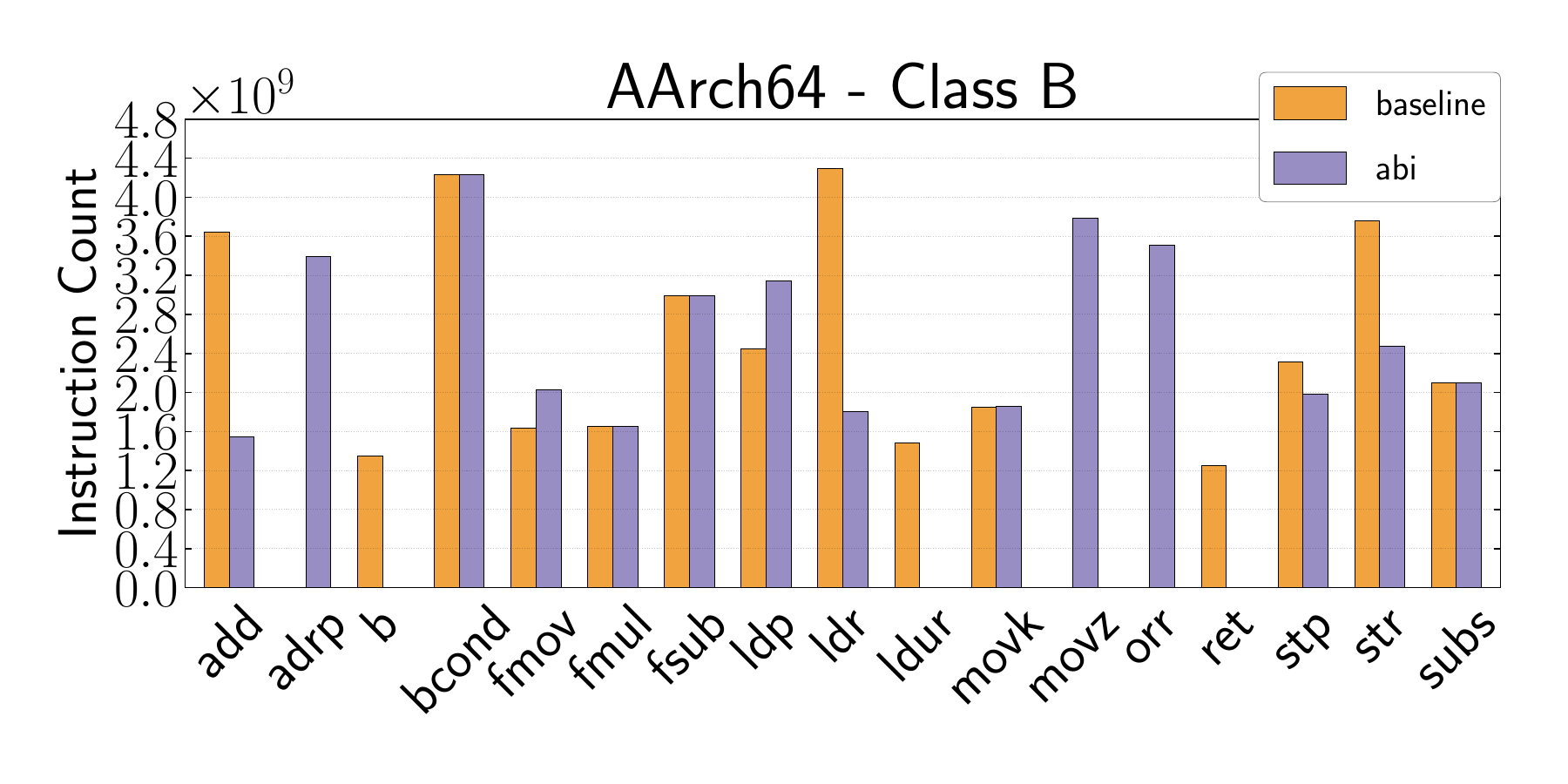}
        \caption{UA}
        \label{fig:figures/cc_vs_vanilla_opcodes_ua}
    \end{subfigure}

    \caption{Dynamic instruction distribution analysis showing the effect of applying the ABI from \tool{} (\textit{abi}), which includes reducing the available registers in \armarch{}, versus \textit{baseline}.
    Only the top 15 instructions are shown in each case (a bar pair has only one bar if the instruction is top-15 in one version, but not the other).
    }
    \label{fig:cc_vs_vanilla}
\end{figure}

We track the changes in dynamic instruction count of the above categories after applying the \tool{} \ac{abi} using \dynamorio~\cite{bruening2004efficient}.
Specifically, we use the \textit{opcodemix} tool similarly to benchmark characterization methodologies~\cite{singh2019memory}.
The reduction of registers applies only to \armarch{}, so we only show results for this architecture.

\Cref{fig:cc_vs_vanilla} shows the dynamic instruction distribution for all \ac{npb} workloads, compiled for \armarch{}, for class B\@.
We show the top 15 instructions per workload, in terms of dynamic count, for the baseline compilation (\textit{baseline}), which is an unmodified \llvm{} compiler, and the feature tested (\textit{abi}), which is applying all the \tool{} \ac{abi} rules, including reducing the registers and modifying the calling conventions.

We make three observations.
First, \textsc{cg} and \textsc{is} exhibit almost no change in their instruction distributions, without changing the mix in the other opcodes.
Second, \textsc{bt}, \textsc{ep}, \textsc{lu}, \textsc{mg}, \textsc{sp}, and \textsc{ua} exhibit increased arithmetic and address materialization instructions.
Finally, \textsc{ep}, \textsc{ft}, and \textsc{sp} exhibit increased load and store instructions.

The results for \textsc{cg} and \textsc{is} are consistent with their performance impact shown in~\Cref{fig:arm-breakdown} (\textit{abi}), which is minimal.
On the other hand, all the benchmarks where the spills and/or rematerialization activity is increased, exhibit notable performance impact in the \textit{abi} category, mostly on \textsc{lu}, \textsc{sp}, and \textsc{ua}.

\input{074_register_pressure_sp_profile}

\subsubsection{Impact of register allocation}

\begin{figure}[t!]
    \centering
    \begin{subfigure}[b]{0.49\textwidth}
        \centering
        \includegraphics[width=0.9\columnwidth]{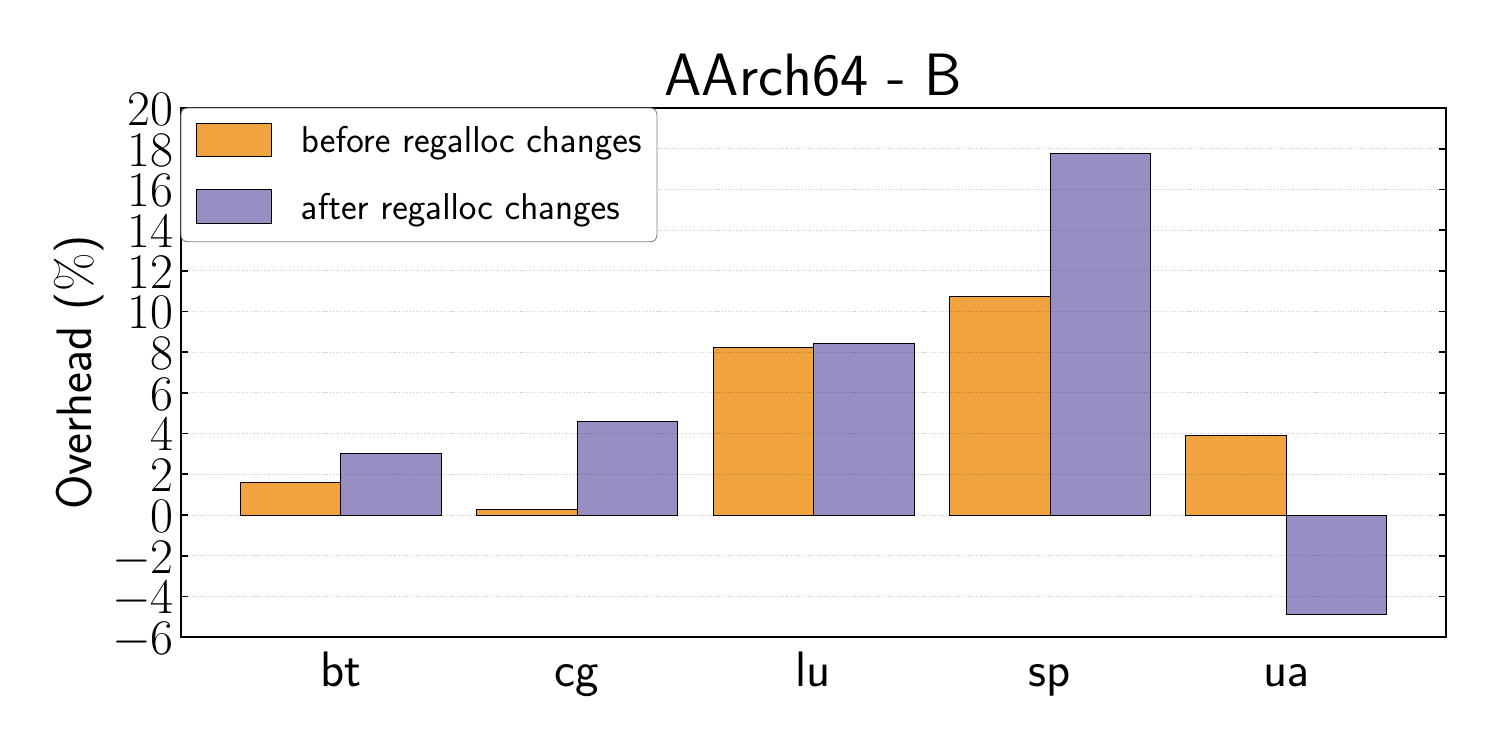}
        \caption{\tool{} \armarch{} binaries with\slash{}without the extensions affecting register allocation.}
        \label{fig:arm-no-regalloc}
    \end{subfigure}
    \hfill
    \begin{subfigure}[b]{0.42\textwidth}
        \includegraphics[width=0.8\columnwidth]{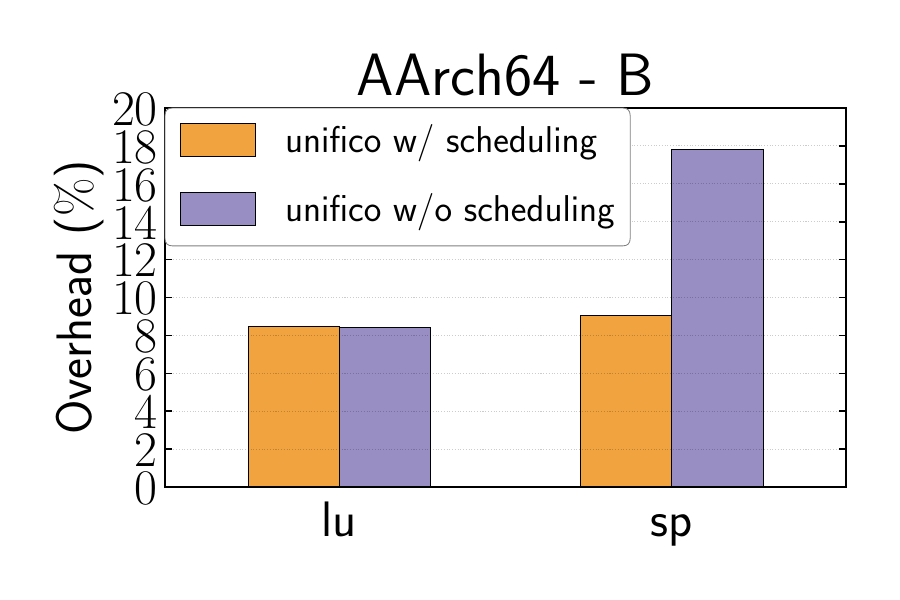}
        \caption{\tool{} \armarch{} binaries with\slash{}without the machine scheduler enabled.}
        \label{fig:arm-no-misched}
    \end{subfigure}
        \caption{Ablation studies for \armarch{}.}
    \label{fig:ablation-plots}
\end{figure}


We also debug the performance of our \armarch{} applications after applying the \textit{register allocation} category changes.
We compare the execution time overhead for the most impacted benchmarks before and after applying the register allocation changes in \tool{}\hyp{}compiled binaries (\Cref{fig:arm-no-regalloc}).\footnote{We omit a \ac{tma} analysis in this section due to lack of detailed \ac{pmu} \textit{core bound} events in our ThunderX2 \armarch{} \ac{cpu}.}
We observe that these changes constitute a large portion of the overhead for \textsc{cg} and \textsc{sp}, and lead to a small speedup in the case of \textsc{ua}.

Regarding the overhead, we found that the most impactful change is the application of the two\hyp{}address format for integer arithmetic in \armarch{} (\Cref{tab:tool_compiler_transforms}).
Despite using similar heuristics with \pcarch{} to bring back the three\hyp{}address format whenever possible (e.g., when \pcarch{} converts a two-address \asm{add} to a three-address \asm{lea}), it still causes $4\%\textup{--}8\%$ execution time overhead for these two benchmarks.
The remaining register allocation changes (e.g., \armarch{} register costs) had far less impact.
Overall, we conclude that applying the two-address format on integer \armarch{} instructions can cause notable overhead in programs with deeply nested loops like \textsc{sp}.


\paragraph{UA speedup}

An outlier we notice in~\Cref{fig:arm-no-regalloc} and~\Cref{fig:arm-breakdown} is a speedup obtained in \textsc{ua}, after applying the changes in the register allocator.
This speedup is carried over to the total \tool{} performance.
From the register allocation features, we identified that avoiding the use of wide multiply\hyp{}add instructions in \armarch{}, with 32-bit multiplication operands and a 64-bit destination operand, e.g.,  \asm{smaddl}, led to better register allocation and execution time improvement.
However, the speedup did not persist at higher optimization levels, indicating missed optimizations in the baseline (compiled with \code{-O1}), rather than a performance improvement introduced by \tool{}.

%% file: 074_register_pressure_sp_profile.tex
\begin{figure}[t]
  \centering
  \begin{subfigure}[t]{0.48\columnwidth}
    \begin{adjustwidth}{0.10cm}{}
    \centering
\begin{lstlisting}[style=asmstyle]
// Enough registers
// to host addresses
ldr d1, [x26, x13]
ldr d2, [x25, x13]
ldr d3, [x27, x13]
ldr d4, [x28, x13]
\end{lstlisting}
    \end{adjustwidth}
    \caption{}\label{fig:sp-before}
  \end{subfigure}
  \begin{subfigure}[t]{0.48\columnwidth}
    \begin{adjustwidth}{0.0cm}{}
    \centering
\begin{lstlisting}[style=asmstyle, numbers=none]
// Rematerialization in x19
add x19, x2, x5
ldr d3, [x19, #4120]
ldr d2, [x19]
ldr d4, [x4, #3352]
ldr d0, [x3, x5]
\end{lstlisting}
    \end{adjustwidth}
    \caption{}\label{fig:sp-after}
  \end{subfigure}
  \caption{A hot loop snippet in \textsc{sp} compiled with an unmodified \llvm{} (left) and with \tool{} after reducing the available registers for code generation (right).
  }
    \label{fig:register-pressure-sp}
\end{figure}

As an example on how reduced registers can increase spills or rematerialization, consider the code snippet in~\Cref{fig:register-pressure-sp}.
The snippet is extracted from a compute-intensive loop in \textsc{SP}, through \perf{}.
On~\Cref{fig:sp-before}, we observe that there are enough registers to hold the global addresses from which the floating-point values are stored.
However, after reducing the available registers (\Cref{fig:sp-after}), the compiler needs to rematerialize an address, adding one more instruction, and also consume a callee-save register (\asm{x19}) to avoid spilling, increasing the demand for callee-saved registers.

%% file: 075_mi_scheduling.tex
\subsection{Machine Instruction Scheduling}\label{subsec:misched}


Lastly, we disable the machine instruction scheduler.
Although we keep the post-register-allocation scheduler, the scheduling that precedes register allocation can reorder instructions before those spill slot locations are finalized (which happens after register allocation).
This can cause different final layout of spill slots, if there is a reorder mismatch between the architectures.

From~\Cref{fig:arm-breakdown}, we notice that the most impacted benchmarks directly from disabling this scheduling pass are \textsc{lu} and \textsc{sp}.
We validate the impact on the overall performance, by comparing \tool{} with scheduling enabled (default in \llvm{} -O1) and disabled (\Cref{fig:arm-no-misched}), observing that disabling the scheduler contributes significantly to the overall performance of \textsc{sp}.

To understand the source of the overhead, we collect some basic event data using \perf{}, which we omit for brevity.
The results showed that \textsc{lu} and \textsc{sp} became more backend bound, and also \textsc{sp} showed almost $9\%$ more branch \ac{mpki}.
Other benchmarks were not affected.
The increase in the backend-bound metric can be attributed to less \ac{ilp} being available without the scheduler, while the increase in the branch misses for \textsc{sp} likely shows that branch prediction has less context to capitalize on around the conditional instructions.


%% file: 076_overhead_classes.tex
\subsection{Final overhead}\label{subsubsec:final-overhead}

%
\begin{figure}[t!]
  \centering
  \begin{subfigure}[t]{0.49\linewidth}
    \includegraphics[width=\columnwidth,clip]{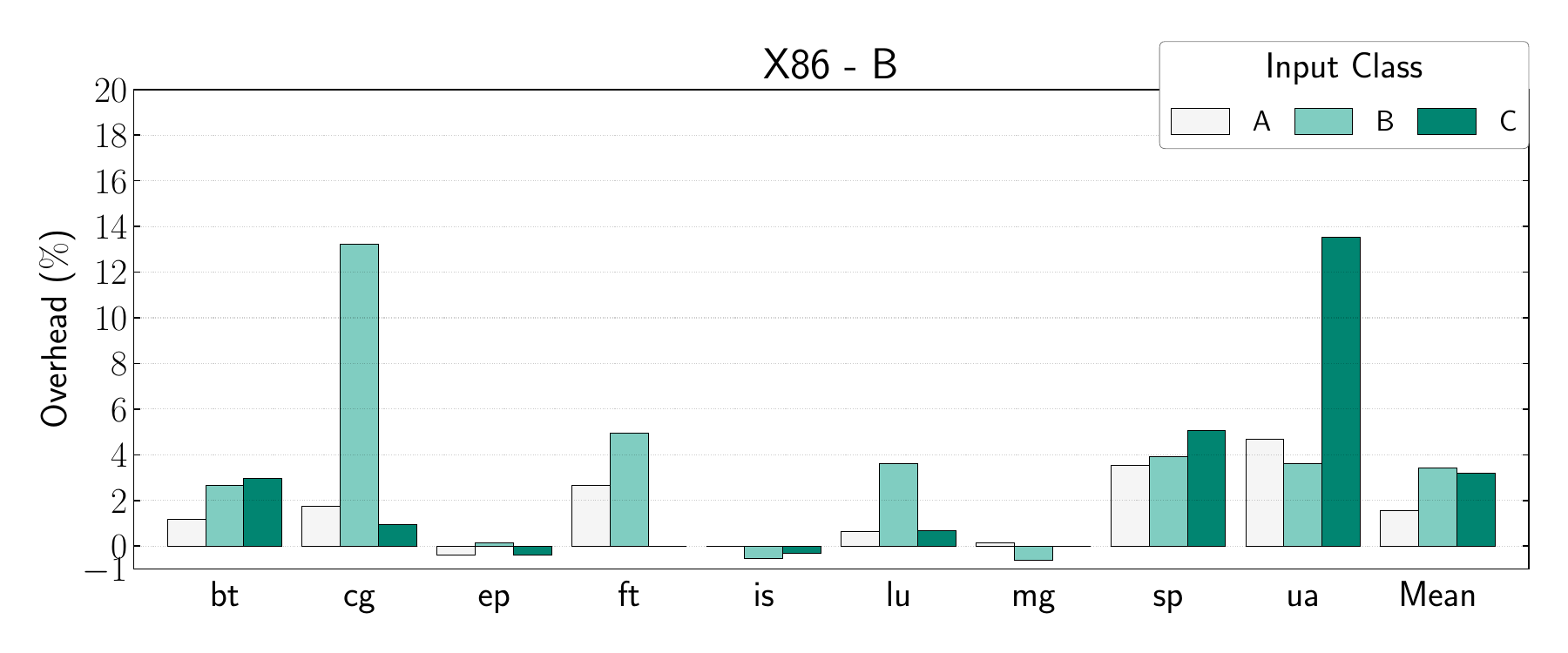}
    \caption{}\label{fig:final-overhead-x86}
  \end{subfigure}
  \begin{subfigure}[t]{0.49\linewidth}
    \includegraphics[width=\columnwidth,clip]{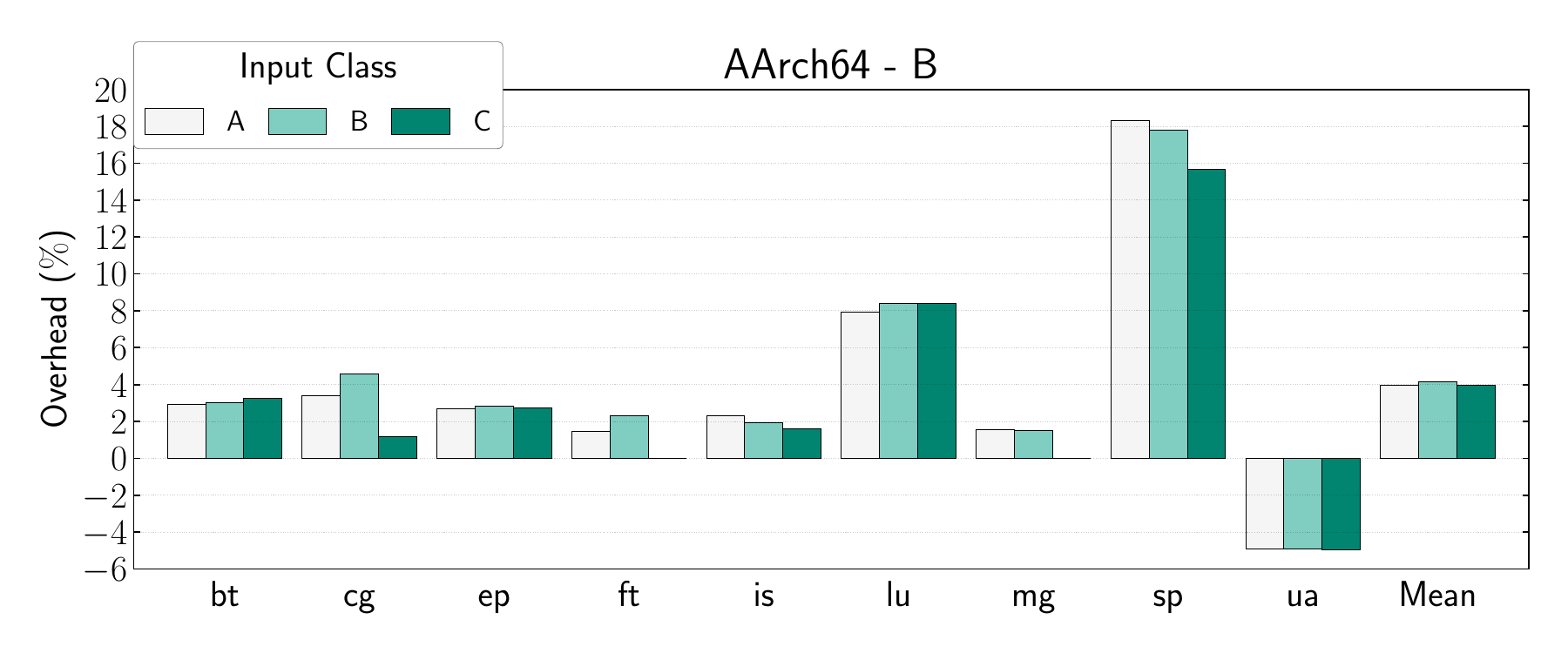}
    \caption{}\label{fig:final-overhead-arm}
  \end{subfigure}
  \caption{Comparison of total overhead with unmodified native execution across all classes.
  \tool{}-compiled binaries perform within $\%6$ of the baseline in almost all benchmarks and classes for both architectures.}\label{fig:final-overhead}
    \Description{Final performance comparison across all classes and architectures.}
\end{figure}

\Cref{fig:final-overhead} shows the total overhead after applying the fixes from the previous sections, for the A, B, and C classes of \ac{npb}.
In \armarch{}, \textsc{sp} has the most overhead, where the code region dominating the runtime is a kernel with a deep\hyp{}nested loop, manipulating 4\hyp{}dimensional global arrays (explained in~\Cref{subsec:reg-pressure,subsec:misched}).
%
In \pcarch{}, the most noticeable outlier is \textsc{cg} for class B, as mentioned in~\Cref{subsec:callsite-alignment}.
Interestingly, for the class C of \textsc{cg}, our investigation concludes that for the much larger array sizes the computation dominates again the alignment issues, making the application mostly backend bound.
Contrary, the simplified addressing, with less available immediates to represent global addresses, causes greater impact in this case (and explains also the overhead for \textsc{ua}).

%% file: 077_perf_portability.tex
\subsection{Performance Portability}\label{subsec:portability}

\begin{figure}[t!]
    \includegraphics[width=0.75\columnwidth]{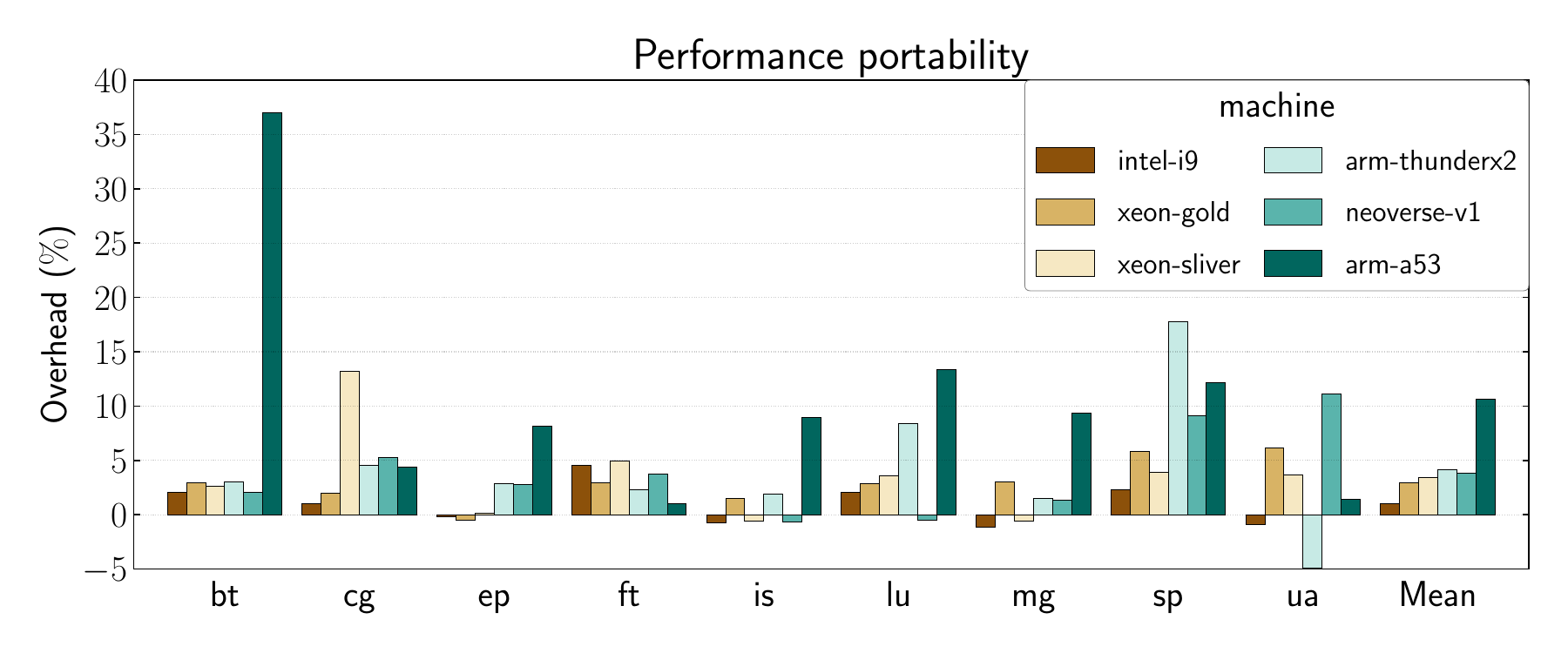}
    \caption{Comparison of the overhead introduced by \tool{} for different \pcarch{} and \armarch{} machines.}
    \Description{Comparison of overheads between different x86 and arm machines of various capabilities.}
    \label{fig:machine-comparison}
\end{figure}

We now compare the performance of \tool{}\hyp{}compiled binaries among machines of different hardware capabilities and microarchitectures.
\Cref{fig:machine-comparison} shows the comparison between three \pcarch{} machines of different base clock frequencies and three \armarch{} machines including two high-end servers and a low-power one.
There are three observations we make.
First, between the \pcarch{} machines, the \textit{Intel i9}, shows less impact on all benchmarks than the \textit{Intel Xeon} counterparts.
Second, the weaker \textit{A53 Cortex}, shows significantly more impact than the high-end \textit{ThunderX2}.
Finally, the \pcarch{} machines overall show less impact than the \armarch{} ones.

For the \pcarch{} case, we conclude that  the higher base clock frequency and the larger caches,  in the case with \textit{Intel i9} compared to \textit{Xeon} machines, diminish the per-architecture impact of the \tool{} backend extensions.
Similarly, features like  out-of-order execution and larger issue width (four versus two), are able to hide some performance degradation in \textit{ThunderX2} compared to the in-order \textit{Cortex A53} machine.
Finally, the \armarch{} architectures exhibit more overhead overall, because only for this \ac{isa} did we use a reduced register set.

%% file: 078_sota_comparison.tex
\subsection{Comparison with \Acl{sota}}\label{subsec:migration-comparison}

To demonstrate the migration, we conduct an experiment using one \armarch{} and one \pcarch{} server, starting execution on one server, getting a \criu{} image dump, and continuing execution on the other server.
We repeat the process and move back to the original server for a number of iterations, depending on the execution time of each benchmark.
We select two representative benchmarks used in similar setups by related work~\cite{barbalace2017,xing2022,bapat2024dapper} and, unlike related work which only shows one migration, we perform multiple trips between the two servers and show an analysis of the execution time.
We compare \tool{} with binaries compiled with the~\popcorn{} toolchain, including their transformation runtime.
Currently, the Popcorn transformation runtime only supports binaries compiled at -O0~\cite{popcorn_compiler_applications}, so we do likewise for this comparison with \tool{}\hyp{}compiled binaries.

\begin{figure}[t!]
  \centering
  \begin{subfigure}[t]{0.27\linewidth}
    \includegraphics[width=1.0\columnwidth,clip]{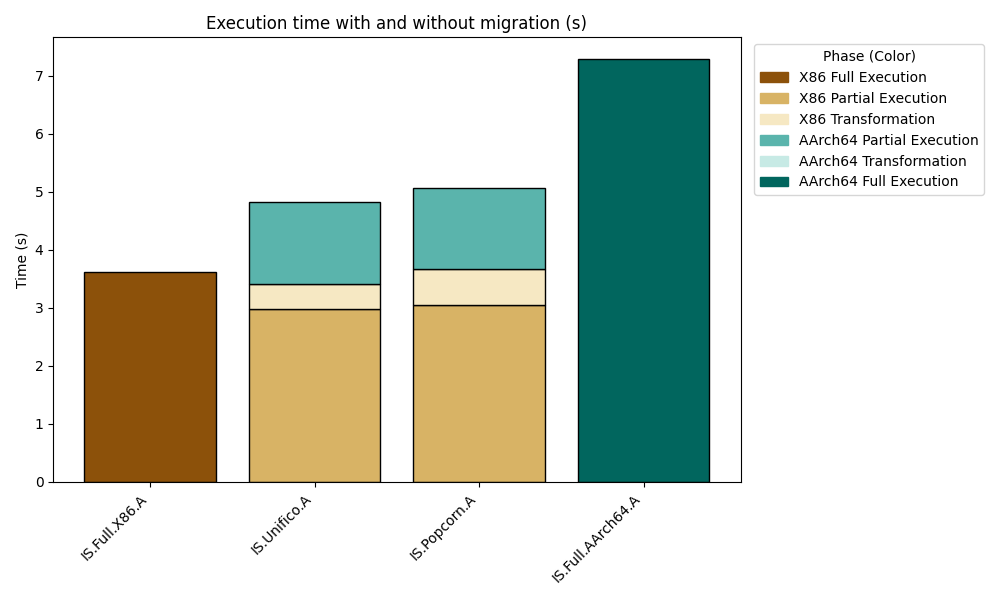}
    \caption{}\label{fig:migration-is-A}
  \end{subfigure}
  \begin{subfigure}[t]{0.22\linewidth}
    \includegraphics[width=0.9\columnwidth,clip]{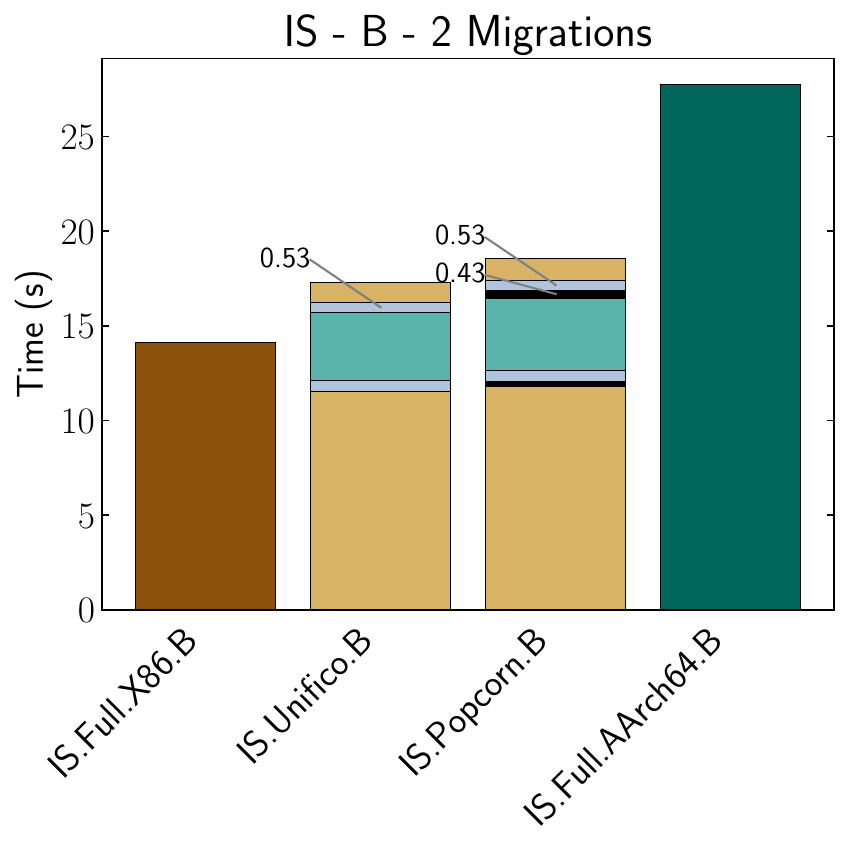}
    \caption{}\label{fig:migration-is-B}
  \end{subfigure}
  \begin{subfigure}[t]{0.22\linewidth}
      \includegraphics[width=0.9\columnwidth,clip]{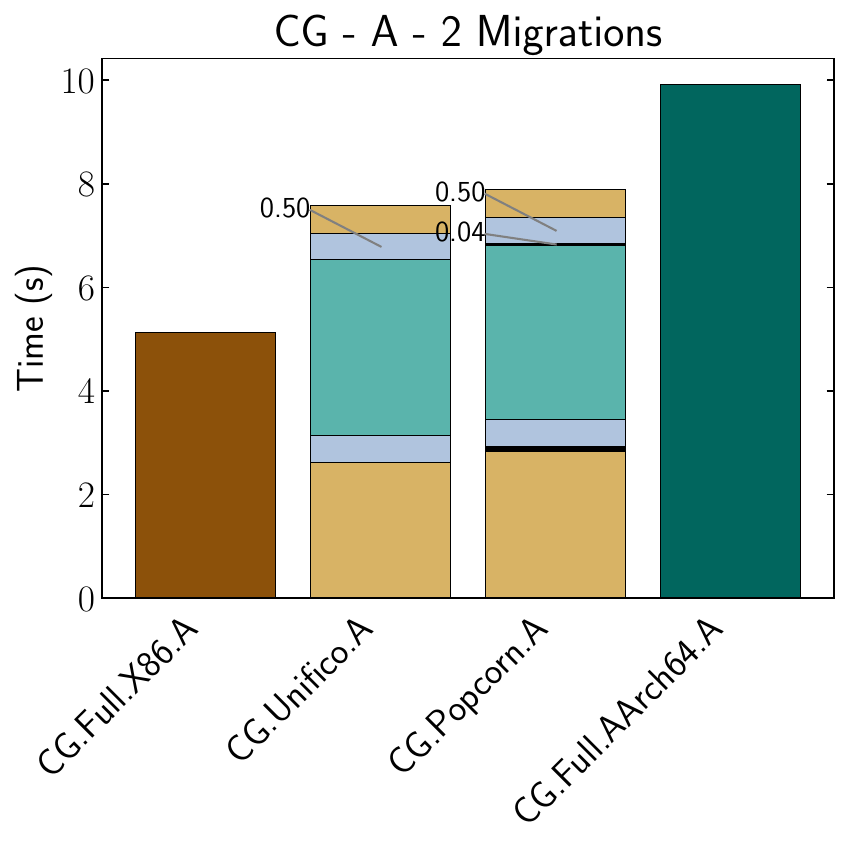}
      \caption{}\label{fig:migration-cg-A}
  \end{subfigure}
  \begin{subfigure}[t]{0.22\linewidth}
      \includegraphics[width=0.90\columnwidth,clip]{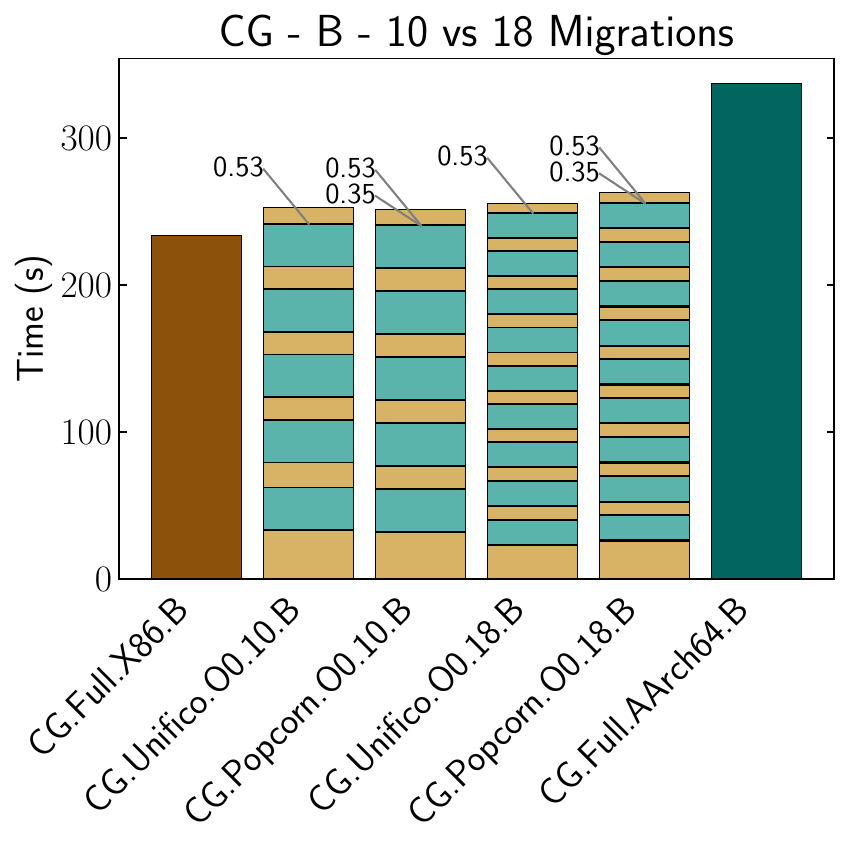}
      \caption{}\label{fig:migration-cg-B}
  \end{subfigure}
  \caption{Analysis of the execution time for \textsc{IS} and \text{CG}, class A and B, and different migration scenarios.}\label{fig:migration}
\end{figure}

Results are shown in~\Cref{fig:migration}, where we also include the standalone executions on each server for unmodified binaries, as a reference.
There are three observations.
First, the cost of transforming the \criu{} images is reduced with \tool{} (shown as \textit{Transformation} in the plots), with the largest increase in class B (up to $45\%$ for \text{IS-B} on \armarch{} - \Cref{fig:migration-is-B}).
Second, there is a baseline cost of dumping the images, which is present both in \tool{} and \popcorn{} (shown as \text{Image Dump}).
Third, for execution times up to $10^{1}s$, the gain of removing the transformation is noticeable (\Cref{fig:migration-cg-B,fig:migration-is-A,fig:migration-is-B}), while for time larger than $10^{2}s$, the benefit appears after a certain number of migrations (18 vs 10 migrations in~\Cref{fig:migration-cg-B}).
We conclude that avoiding the stack transformation can yield modest gains for relatively shorter-lived applications, especially with large input sizes (like \text{IS-B}), or for larger execution times given enough migrations.

%

\paragraph{Discussion.}
In the above setup, we ignore SSH transfer cost of images, since they are present in both compilation toolchains.
However, in a true shared-memory system, works like \popcorn{} \linux{}~\cite{barbalace2015} facilitate transformation through \ac{dsm}, which during migration requires page-level handling even for small stacks, introducing additional overheads like preloading the transformation metadata in memory or pre-allocating and managing the pages of the entire stack (8MB)~\cite{lyerly2019}.
Hence, avoiding the transformation would further reduce memory management overheads, which are hard to model in checkpoint-restore setups.
The transformation gains become more pronounced in multithreaded contexts, where several worker threads undergo transformation before migration.
This is amplified in \ac{ndp} scenarios accommodating thread migrations, described in our motivation (\Cref{sec:background-motivation}), where migrations are more likely to occur due to remote memory accesses.
As future work, we believe \tool{} can enable highly-threaded general-purpose \ac{ndp} architectures of different \acp{isa}, as counterparts to specialized FPGA/ASIC implementations\cite{dysart2016highly, page2022evolution}.

%% file: 080_relatedwork.tex
\section{Related Work}\label{sec:relatedwork}

\paragraph{Dynamic software migration.}
Migration among heterogeneous\hyp{}\ac{isa} compute nodes during execution was initially explored in the context of distributed systems~\cite{attardi1988,dubach1989,smith1998a,jul1988}.
DeVuyst et al.~\cite{devuyst2012}, were the first to examine the feasibility of heterogeneous\hyp{}\ac{isa} migration for multicore CPUs, where state transformation dominates the migration overhead.
Venkat and Tullsen~\cite{venkat2014} showed that \ac{isa} heterogeneity offers $\sim$20\% performance and $\sim$23\% power improvements over the homogeneous\hyp{}\ac{isa} counterpart.
Both works rely on the same compilation techniques, aiming to minimize the state transformation by providing a common data layout as much as possible and using a common ``fat'' binary format for the target architectures.
The evaluation is done through simulation, but artifacts are not publicly available.
The \popcorn{} \linux{} and compiler toolchain~\cite{barbalace2015,barbalace2017}, improve upon these techniques and provide an open source implementation.
The \popcorn{} compiler toolchain~\cite{barbalace2017} avoids ``fat'' binaries by embedding metadata in them to guide the transformation during migration.
Later works based on \popcorn{} \linux{}, examined migration between heterogeneous\hyp{}\ac{isa} systems and FPGAs\cite{horta2021}, and at higher granularity among sets of processes in \linux{} containers~\cite{xing2022}.
\tool{}, as a compilation technique, can be integrated in any of the aforementioned approaches.

\paragraph{Migration granularity.}
Von Bank et al.~\cite{vonbank1994} provide a formal methodology to identify migration points among heterogeneous processes at varied granularity.
A large body of work, along with \tool{}, which operate at the process level, use callsite boundaries as potential migration points~\cite{barbalace2017,barbalace2015,devuyst2012,venkat2014}.
DeVuyst and Venkat~\cite{devuyst2012,venkat2014} propose to allow migration requests at any machine instruction by using \ac{dbt} to fulfill them upon reaching a callsite, when actual migration occurs.
However, \ac{dbt} introduces excessive overhead~\cite{barbalace2017}.
Checkpointing relates to process migration~\cite{plank1995,criu-webpage}, though \hetcriu{}~\cite{xing2022} further blurs any distinctions.


%% file: 090_conclusion.tex
\section{Conclusion}\label{sec:conclusion}

We propose \tool{}, a compilation technique that extends and innovates upon existing compiler support for heterogeneous\hyp{}\ac{isa} \ac{cpu} migration, by removing the need for runtime state transformation.
\tool{} removes state transformation overheads during migration, without creating large binaries and easing programmability, by extending the compiler backend to generate binaries for different architectures with a unique address space and stack layout.
We show that \tool{} does not substantially impact program execution, adding on average no more than $6\%$ ($10\%$) execution time overhead on high-end (low-end) processors and no more than $10\%$ code size increases, compared to the $2x$ size overhead introduced by related work.
We finally analyze the performance of the most impactful features for a common stack layout, and suggest performance improvements.
In future work, we will examine the applicability of our approach to other architectures, the automatic extraction of unification rules for the \ac{abi} and stack layout, as well their encoding to the high\hyp{}level compiler specifications to ease adoption.